\documentclass[aps,prx,reprint,groupedaddress]{revtex4-2}

\usepackage{amsmath,amssymb,mathtools}
\usepackage{graphicx}
\usepackage{dcolumn}
\usepackage{bm}
\usepackage[hidelinks]{hyperref}
\hypersetup{pdfpagemode=UseNone}
\usepackage{siunitx}
\usepackage{physics2}
\usephysicsmodule{braket}

\newcommand{\sSz}{{}^1\mathrm{S}_0}
\newcommand{\sPo}{{}^1\mathrm{P}_1}
\newcommand{\tPz}{{}^3\mathrm{P}_0}
\newcommand{\tPo}{{}^3\mathrm{P}_1}
\newcommand{\tPt}{{}^3\mathrm{P}_2}
\newcommand{\boson}{{}^{174}\mathrm{Yb}}
\newcommand{\fermi}{{}^{171}\mathrm{Yb}}

\allowdisplaybreaks[1]

\begin{document}

\preprint{}

\title{A hybrid atom tweezer array of nuclear spin and optical clock qubits}


\author{Yuma Nakamura}
 \email[]{nakamura.yuma.54c@st.kyoto-u.ac.jp}
 \affiliation{Department of Physics, Graduate School of Science, Kyoto University, Kyoto 606-8502, Japan}

\author{Toshi Kusano}
  \affiliation{Department of Physics, Graduate School of Science, Kyoto University, Kyoto 606-8502, Japan}

\author{Rei Yokoyama}
  \affiliation{Department of Physics, Graduate School of Science, Kyoto University, Kyoto 606-8502, Japan}

\author{Keito Saito}
  \affiliation{Department of Physics, Graduate School of Science, Kyoto University, Kyoto 606-8502, Japan}

\author{Koichiro Higashi}
  \affiliation{Department of Physics, Graduate School of Science, Kyoto University, Kyoto 606-8502, Japan}

\author{Naoya Ozawa}
  \affiliation{Department of Physics, Graduate School of Science, Kyoto University, Kyoto 606-8502, Japan}

\author{Tetsushi Takano}
  \affiliation{Department of Physics, Graduate School of Science, Kyoto University, Kyoto 606-8502, Japan}

\author{Yosuke Takasu}
  \affiliation{Department of Physics, Graduate School of Science, Kyoto University, Kyoto 606-8502, Japan}
  
\author{Yoshiro Takahashi}
  \affiliation{Department of Physics, Graduate School of Science, Kyoto University, Kyoto 606-8502, Japan}


\date{\today}

\begin{abstract}
While data qubits with a long coherence time are essential for the storage of quantum information, ancilla qubits are pivotal in quantum error correction (QEC) for fault-tolerant quantum computing. The recent development of optical tweezer arrays, such as the preparation of large-scale qubit arrays and high-fidelity gate operations, offers the potential for realizing QEC protocols, and one of the important next challenges is to control and detect ancilla qubits while minimizing atom loss and crosstalk. Here, we present the realization of a hybrid system consisting of a dual-isotope ytterbium (Yb) atom array, in which we can utilize a nuclear spin qubit of fermionic $\fermi$ as a data qubit and an optical clock qubit of bosonic $\boson$ as an ancilla qubit with a capacity of non-destructive qubit readout. We evaluate the crosstalk between qubits regarding the impact on the coherence of the nuclear spin qubits from the imaging light for $\boson$. The Hahn-echo sequence with a \qty{399}{nm} probe and \qty{556}{nm} cooling beams for $\boson$, we observe 99.1(1.8)\,\% coherence retained under \qty{20}{ms} exposure, yielding an imaging fidelity of 0.9992 and a survival probability of 0.988. The Ramsey sequence with a \qty{556}{nm} probe beam shows negligible influence on the coherence, suggesting the potential future improvement of low cross-talk measurements. This result highlights the potential of the hybrid-Yb atom array for mid-circuit measurements for ancilla-qubit-based QEC protocols.
\end{abstract}

\maketitle


\section{Introduction}
Ancilla qubits play a central role in quantum error correction (QEC), which is essential for realizing fault-tolerant quantum computing \cite{Gottesman1997-zd}. The efficacy of ancilla-qubit-based QEC protocols has been recently demonstrated in quantum computing architectures with superconducting qubits \cite{Google_Quantum_AI2021-td, Zhao2022-qf,Krinner2022-td} and trapped ion qubits \cite{Egan2021-ai, Da_Silva2024-ek}, supporting their importance and utility in advancing the field. 

Concurrently, significant progress has been made in developing the optical tweezer array system, enabling the creation of large-scale qubit arrays \cite{Huft2022-ke, Tao2023-ri, Pause2024-wk, Norcia2024-zh, Manetsch2024-sp, Pichard2024-ba} and high-fidelity gate operations \cite{Evered2023-nv, Scholl2023-jb, Peper2024-wc, Tsai2024-tu, Radnaev2024-ac}. The attractive features of this system, such as dynamical qubit connection aided by moving tweezers \cite{Bluvstein2022-tf, Bluvstein2024-yx} and error-detecting methods utilizing the rich internal state of atoms \cite{Ma2023-mo}, open up the possibilities for realizing QEC protocols \cite{Wu2022-ze, Cong2022-bm, Xu2023-bt, Sahay2023-eb, Jia2024-ox}.

One of the recent main focuses in the optical tweezer array systems is efficient control and readout of the ancilla qubits while minimizing qubit loss and crosstalk with neighboring data qubits. The difficulties in achieving this arise primarily from the reliance on globally irradiating light for these operations, which can induce heating and decoherence of the atoms \cite{Saffman2016-jh}.

One approach to addressing those challenges is to hide the data qubits spatially or energetically from the controlling light for the ancilla qubits. The former takes advantage of the moving tweezers that allow qubit transport between spatially separated zones designated for a specific operation \cite{ Deist2022-io,Bluvstein2024-yx}. The latter utilizes a local addressing beam, which induces light shifts, sometimes followed by advanced state manipulation, to make data qubit atoms in a dark state \cite{Graham2023-wl, Norcia2023-dt, Lis2023-cb}.

Another approach is to prepare two distinct qubits in a single optical tweezer array, with each qubit acting as a data qubit and an ancilla qubit, respectively. Dual-species \cite{Singh2022-zg, Singh2023-ox} and dual-isotopes \cite{Zeng2017-mi} atom arrays of alkali atoms have successfully demonstrated the capability of qubit readout with negligible crosstalk, owing to their isolated resonant frequencies.

\begin{figure*}[ht]
    \centering
    \includegraphics[width=\textwidth]{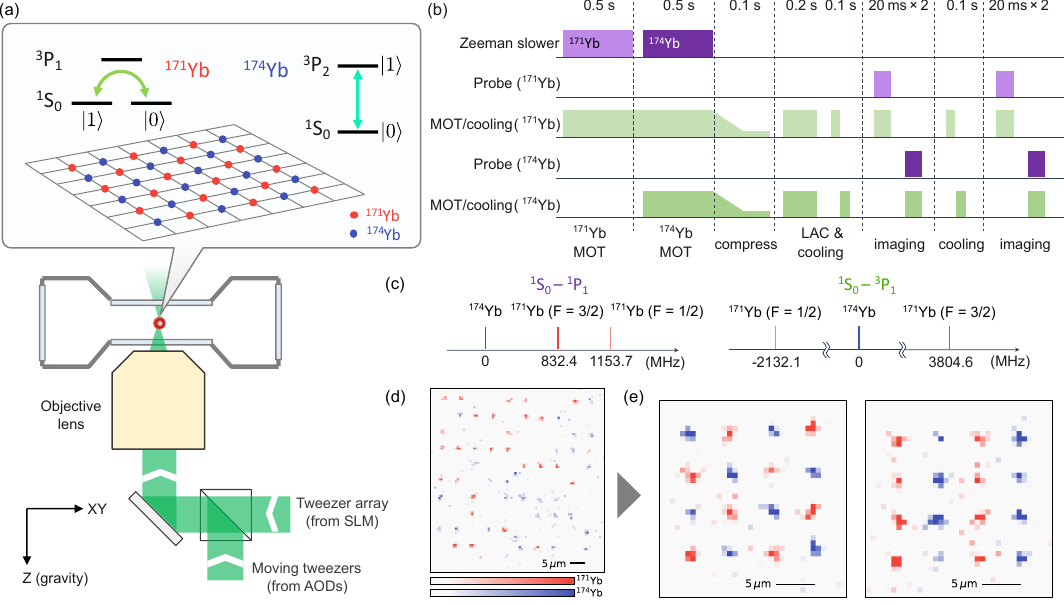}
    \caption{Overview of dual-Ytterbium atom array. (a) The vacuum chamber and the tweezer beam path. The tweezer array beam and moving tweezers are combined with a polarized beam splitter and focused at the center of the chamber through an objective lens with $\mathrm{NA}=0.6$. The fermionic isotope $\fermi$ offers a nuclear spin qubit with a long coherence time and fast qubit control. The qubit states are encoded in the ground state. The bosonic isotope $\boson$ provides an optical clock qubit encoded in the ground state and a metastable excited state, which has the capability of non-destructive qubit readout. (b) Experimental sequence. First, the dual MOT of $\fermi$ and $\boson$ is prepared by switching the frequency of the Zeeman slower. Then, atoms are loaded into the tweezers by compressing the dual-MOT. After LAC and cooling for each isotope, the atoms are sequentially imaged by a \qty{399}{nm} probe and \qty{556}{nm} cooling beams. (c) Isotope shifts of the $\sSz\text{--}\sPo$ ($\lambda=\qty{399}{nm},~\Gamma/(2\pi) = \qty{29}{MHz}$)\cite{Das2005-fx} and $\sSz\text{--}\tPo$ ($\lambda=\qty{556}{nm},~\Gamma/(2\pi) = \qty{182}{kHz}$)\cite{Pandey2009-fp} transitions. We use the $F=3/2$ state of the $\sPo$ and $\tPo$ states for imaging and cooling for $\fermi$. (d) Combined single-shot image of dual-Yb array. The optical tweezer array is $10\times10$ with \qty{5}{\mu m} intervals. The red and blue dots represent single $\fermi$ and $\boson$ atoms, respectively. (e) Rearranged dual-Yb atom array in checkerboard (left) and stripe (right) patterns. The atoms are moved between adjacent rows or columns to avoid the filled sites.}
    \label{fig:overview}
\end{figure*}

A simple system consisting of data qubits with inherently long coherence times and ancilla qubits with non-destructive measurement capability, along with negligible crosstalk between them and no need for additional techniques, would be ideal for an ancilla-based QEC protocol \cite{Fowler2012-sb}. Here, we present the realization of a hybrid system consisting of dual-isotope alkali-earth-like ytterbium (Yb) atom array, where the fermionic isotope $\fermi$ provides a nuclear spin qubit as a data qubit with long coherence and fast qubit control \cite{Jenkins2022-lt, Ma2022-gy}, and the bosonic isotope $\boson$ offers an optical clock qubit as an ancilla qubit with the capacity of non-destructive state-selective qubit readout \cite{Schine2022-sn} (Fig.\,\ref{fig:overview}(a)). By optimizing the light-assisted collision (LAC) parameters and using the heteronuclear photoassociation (PA) process, we obtain a loading rate of \qty{21(5)}{\%} for each isotope while minimizing the probability of dual occupancy, where both isotopes are loaded in the same site simultaneously, to only \qty{0.2(7)}{\%}. The strength of this system is the capability to measure the ancilla qubits with minimal impact on the data qubit coherence without technically demanding local operations, owing to their isotope shifts. We carefully evaluate the crosstalk between the isotopes in terms of the influence on the coherence of the nuclear spin qubits from the imaging light for $\boson$ by the Hahn-echo and Ramsey sequences. The Hahn-echo sequence, where our typical imaging condition of \qty{20}{ms} exposure with \qty{399}{nm} probe and \qty{556}{nm} cooling beams yields an imaging fidelity of 0.9992 and survival probability of 0.988, reveals that the nuclear spin qubits maintain 99.1(1.8)\,\% contrast of the Hahn-echo signal. Moreover, we observe negligible impact on the coherence from a resonant \qty{556}{nm} beam for imaging, suggesting that \qty{556}{nm} imaging could be a promising future enhancement for low cross-talk measurements in this system. These results highlight the potential utility of the dual-Yb atom array for ancilla-based QEC protocols.

\section{Preparation of a dual-Y\lowercase{b} atom array}
Our experimental sequence starts by simultaneously trapping both isotopes in a magneto-optical trap (MOT) (Fig.\,\ref{fig:overview}(b)). Subsequently, the atoms are loaded into the $10\times10$ tweezer array of $k_{\mathrm{B}}\times\qty{0.9}{mK}$ depth with a beam radius of approximately \qty{550}{nm}. The tweezer wavelength is \qty{532}{nm}, corresponding to a near-magic wavelength for both the $\sSz-\tPo$ transition \cite{Yamamoto2016-vk, Saskin2019-xk} and the $\sSz-\tPt$ transition \cite{Tomita2019-ie, Okuno2022-cd} of $\boson$. After applying red-detuned LAC beams of \qty{556}{nm} for each isotope to prepare single atoms in the tweezers, we image them successively for \qty{20}{ms} exposure time by irradiating a single resonant \qty{399}{nm} beam and simultaneously cooling with MOT beams (Fig.\,\ref{fig:overview}(d)). The intensities of the probe and cooling beams are $\num{1e-3}\,I_{\mathrm{s,399}}$ and $10\,I_{\mathrm{s,556}}$, respectively. Here $I_{\mathrm{s,399}} = \qty{59}{mW/cm^2}$ and $I_{\mathrm{s,556}} = \qty{0.14}{mW/cm^2}$ are the saturation intensities of the $\sSz - \sPo$ and $\sSz - \tPo$ transition, respectively. The atom temperatures are \qty{38(2)}{\mu K} for $\fermi$ and \qty{25(1)}{\mu K} for $\boson$, which are relatively higher than the previous work \cite{Jenkins2022-lt, Saskin2019-xk} due to intensity imbalance of the cooling beams. The atom survival probabilities after imaging are $\qty{98.7}{\%}$ for $\fermi$ and \qty{98.8}{\%} for $\boson$. Since the isotope shifts of the relevant transitions between $\fermi$ and $\boson$ are on the order of \qty{1}{GHz} and much larger than the natural linewidth of the $\sSz - \sPo$ ($\sSz - \tPo$) transition of \qty{29}{MHz} (\qty{182}{kHz}) \cite{Das2005-fx, Pandey2009-fp} (Fig.\,\ref{fig:overview}(c)), the probe and cooling beams of one isotope do not cause atom loss of the other. Thus, the survival probabilities are not degraded compared to the single-isotope case. 

\subsection{Dual atom rearrangement}
We use \qty{532}{nm} moving tweezers generated with a pair of acousto-optic deflectors (AODs) to rearrange the atoms into an arbitrary pattern, as shown in Fig.\,\ref{fig:overview}(e). Since the \qty{532}{nm} tweezers can trap both isotopes and the atoms are randomly loaded in the optical tweezer array, there can be misfilled sites where the $\boson$ atoms fill the site designated for $\fermi$ atoms and vice versa. Additionally, we must prevent the atoms from passing through the correctly-filled sites during the transport of the atoms. These difficulties complicate the rearrangement procedure compared with the single-isotope case. Previous research on a dual-isotope atom array has developed protocols that move the atoms along trajectories on the array grid after removing misplaced atoms that may block the paths \cite{Sheng2022-az, Tao2022-pr}. Instead, we adopt trajectories that pass between the rows and columns of the array and extend the parallel sort-and-compression algorithm \cite{Tian2023-xj}, originally designed for single-isotope cases, to apply to a dual-isotope atom array.

\begin{figure}[b]
    \centering
    \includegraphics[width=\linewidth]{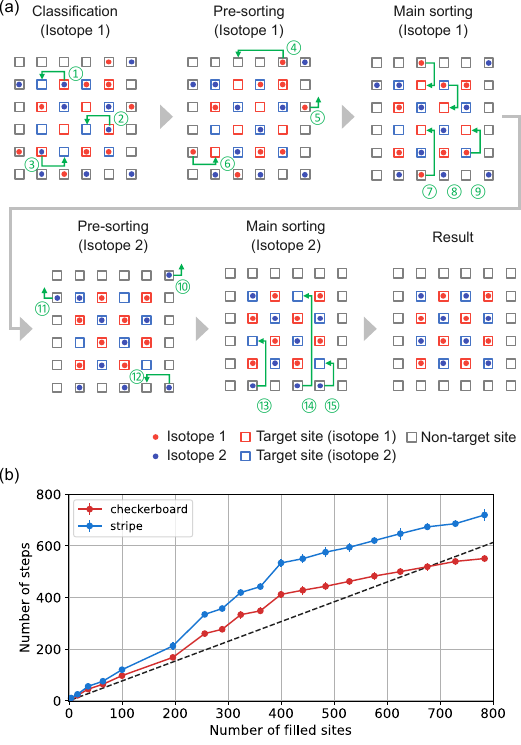}
    \caption{Rearrangement of a dual-isotope atom array. (a) The rearrangement protocol consists of classification, pre-sorting, and main sorting steps for each isotope. In the classification step, the atoms that occupy incorrect target sites are moved to non-target sites or correct target sites. After pre-sorting to prepare a sufficient number of atoms for each column, the atoms are transported to their target sites. In all steps, the moving tweezers pass between the rows or columns to avoid the occupied sites. (b) The scaling of our rearrangement protocol. The efficiency is comparable to that in the pioneering work \cite{Sheng2022-az} represented as a dashed black line. The scaling slope becomes more gradual as the target array size increases, indicating the efficiency may improve for larger arrays due to parallel sorting. }
    \label{fig:rearrange_protocol}
\end{figure}

Our protocol consists of three steps for each isotope: ``classification'', ``pre-sorting'', and ``main sorting'' (Fig.\,\ref{fig:rearrange_protocol}). In the classification step, the isotope 2 atoms that occupy the sites designated for the isotope 1 are moved to correct or non-target sites. Next, to make the number of the isotope 1 atoms in each column equal to or larger than the number of sites designated for the isotope 1, we re-distribute the isotope 1 atoms between columns in the pre-sorting step. The excess atoms are removed from the optical tweezer array during this step if necessary. Finally, the isotope 1 atoms are transported one by one, or simultaneously if possible, to the designated sites column by column in the main sorting step. After rearrangement of the isotope 1, we repeat the same procedure for the isotope 2. Note that, if the number of the isotope 1 atoms is equal to the number of the designated sites for the isotope 1 in each column, all isotope 1 atoms are successfully rearranged to the designated sites, and then the classification step for the isotope 1, scheduled before the pre-sorting step for the isotope 2, will be skipped.  Importantly, the moving tweezers used in all steps pass between the adjacent rows or columns to avoid the filled sites.

The scalability of the rearrangement is primarily determined by the efficiency of the rearrangement algorithm. As shown in Fig.\,\ref{fig:rearrange_protocol}(b), the scaling of our algorithm is comparable to the pioneering work on dual-species rearrangement \cite{Sheng2022-az} for mid-sized arrays. Notably, the milder scaling slope in larger array sizes suggests that our algorithm may perform even more efficiently for large arrays, likely due to the parallel sorting technique demonstrated in \cite{Tian2023-xj}.

It should be noted that the size of the defect-free array shown in Fig.\ref{fig:overview}(e) is currently much smaller than the full $10\times10$ tweezer array. This is primarily due to our unoptimal transport trajectories, where atoms follow parabolic paths as they move between adjacent rows and columns. In this trajectory, the moving tweezers may pass close to other filling sites during transportation, leading to atom loss \cite{Norcia2024-zh}. We have observed that atom loss occurs when the moving tweezer is activated at a distance of approximately \qty{1}{\mu m} from the filling sites. In our setup, the success probability of making a $4\times4$ defect-free array from $10\times10$ array with a \qty{20}{\%} loading of each isotope is only $\sim\qty{1}{\%}$. However, this could be improved by iterative rearrangement, as demonstrated in \cite{Tian2023-xj}. Once we use a proper sweep trajectory that does not pass close to the filling site such as the one used in \cite{Bluvstein2024-yx}, the size of the defect-free array will then be limited primarily by the loading probability of each isotope.

\subsection{Controlling the dual occupancy probability }
\begin{figure}[tbp]
    \centering
    \includegraphics[width=\linewidth]{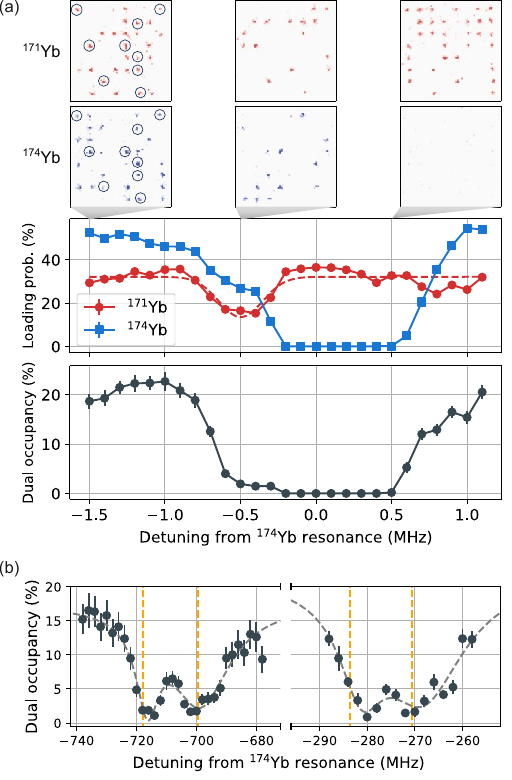}
    \caption{Frequency dependence of dual occupancy probability. (a) Loading probability of each isotope (middle) and dual occupancy (bottom) as a function of the LAC frequency around the $\tPo$ resonance of $\boson$. The dip of $\fermi$ loading probability around \qty{-0.5}{MHz} indicates one-photon heteronuclear PA. The dashed line is a fitting curve by a Gaussian function. The top row shows typical atom array images at the LAC detuning of \qty{-1.5}{MHz}, \qty{-0.5}{MHz}, and \qty{0.5}{MHz}, respectively. The black circle represents dual occupancy. Note that the loading probability of each isotope without applying heteronuclear PA fluctuates in the range of $40\pm\qty{10}{\%}$, depending on the experimental conditions. (b) Heteronuclear PA resonances of deeper bound states around \qty{-710}{MHz} and \qty{-280}{MHz}. The data are fitted by a double Lorentzian function. The yellow dashed lines represent the theoretical PA resonances (see Appendix \ref{sec:PA_calc}). }
    \label{fig:PA}
\end{figure}

The probability of dual occupancy strongly depends on the frequency of the LAC beam for $\boson$. The difference of the dual occupancy rate in the dual-alkali and dual-Yb systems is discussed in Appendix \ref{sec:DO_diff}. Through frequency scanning, we determine the optimal detuning to minimize dual occupancy at \qty{-0.50(2)}{MHz} from the resonance of the $\tPo$ state in the tweezers (Fig.\,\ref{fig:PA}(a)). At this detuning, the probability of the simultaneous loading of both isotopes decreases, indicating that heteronuclear single-photon PA occurs. Consequently, each of the loading probabilities of $\fermi$ and $\boson$ becomes \qty{21(5)}{\%}, with a dual occupancy probability of only \qty{0.2(7)}{\%}. Regarding the system size, the sequential loading with shelving atoms to a metastable state \cite{Shaw2023-gg} and blue-detuned LAC \cite{Grunzweig2010-lo, Brown2019-rr, Jenkins2022-lt} could enhance the loading probabilities, thereby facilitating the creation of larger dual-Yb arrays.

Additionally, we explore other PA lines and identify four resonances at \qty{-268(2)}{MHz}, \qty{-281(1)}{MHz}, \qty{-698.8(8)}{MHz}, and \qty{-716.4(5)}{MHz} (Fig.\,\ref{fig:PA}(b)), consistent with theoretical expectations with a few MHz precision (see Appendix \ref{sec:PA_calc}). These observations offer an important possibility of utilizing these PA lines in stimulated Raman adiabatic passage \cite{Ni2008-kk,Chotia2012-xv} or two-photon PA \cite{Kitagawa2008-fj, Yu2021-wk} to prepare ground-state heteronuclear molecules in the optical tweezers \cite{Liu2018-ec, Liu2019-ho, Zhang2020-up, Yu2021-wk, Anderegg2019-kk, He2020-wd, Ruttley2023-nl}. Such molecules possess multiple degrees of freedom, including rotational and vibrational levels, as well as nuclear spin states decoupled from hyperfine interactions. The presence of nuclear spin states with long coherence times \cite{Park2017-qq, Gregory2021-yx} makes it feasible to realize proposals that leverage molecular degrees of freedom for quantum computation, thereby potentially enhancing the scalability of the system \cite{Sawant2020-qj, Albert2020-bl, Furey2024-bk}. We note that while polar molecules consisting of alkali atoms do possess nuclear spin states in the ${}^1\Sigma$ electronic ground states, preparing such molecules is generally technically demanding \cite{Cairncross2021-gw}. In contrast, the dual-Yb system offers a relatively straightforward method  \cite{Reinaudi2012-il, Stellmer2012-pg, Kato2012-dw} for producing heteronuclear molecules with nuclear spin states. 

It is also worth noting that the cooling beams during imaging do not induce the heteronuclear PA because they are detuned by approximately $-3\Gamma$ from the PA resonance, as shown in the top-left panel of Fig.\,\ref{fig:PA}(a). This allows us to implement various operations while maintaining the dual occupancy sites, such as spin initialization of $\fermi$, cooling of the motional degrees of freedom via Raman sideband cooling \cite{Thompson2013-jp, Jenkins2022-lt} for $\fermi$ and a clock transition \cite{Zhang2022-gp} for $\boson$, as long as the operation lasers are detuned from the PA resonances. Furthermore, it is feasible to image atoms dissociated from a molecule after performing specific operations, allowing the internal state of the molecule to be detected without losing the atoms. These capabilities not only enhance the fidelity of molecule creation, but also simplify the overall process of conducting molecular experiments, making them more efficient and accessible.

\section{Crosstalk measurement}
In the dual-Yb atom array, we can utilize two distinct types of qubits: nuclear spin qubits and optical qubits. The nuclear spin qubit is encoded in the ground state of $\fermi$ as $\ket{0} = \ket{\sSz,m_F=1/2}$ and $\ket{1} = \ket{\sSz,m_F=-1/2}$, which exhibit a second-order coherence time and the capability of MHz-order fast control (\cite{Ma2022-gy, Jenkins2022-lt}, and see Appendix \ref{sec:ns_qubit_property}). The optical qubit, encoded in the ground state and the $m_J=0$ state of an excited metastable state of $\boson$, such as $\tPz$, $\tPt$, and $4f^{13}5d6s^2~(J=2)$ \cite{Ishiyama2023-df, Kawasaki2024-qo}, can be controlled by M2-transition \cite{Kato2016-ie} or magnetic-field induced E1 transition \cite{Taichenachev2006-ta, Barber2006-tu} and imaged non-destructively, as the metastable state naturally serves as a dark state for the probe transition, successfully demonstrated in a recent work using the $\tPz$ state \cite{Lis2023-cb}.

We investigate the crosstalk between them, specifically the ability to image $\boson$ without losing the coherence of the nuclear spin qubits. Since the resonant frequencies of the probe and cooling light for $\boson$ are well-isolated from those for $\fermi$ in GHz order (Fig.\,\ref{fig:overview}(c)), the photon scattering rates of $\fermi$ during the $\boson$ imaging are calculated to be approximately \qty{30}{Hz} for \qty{399}{nm} light and below \qty{0.1}{Hz} for \qty{556}{nm} light, respectively, which are small enough for our imaging conditions of \qty{20}{ms} exposure time. Note that $\boson$ is the best isotope for $\fermi$ in terms of isotope shifts to suppress the crosstalk \cite{Das2005-fx, Pandey2009-fp}. 
\begin{figure}[tbp]
    \centering
    \includegraphics[width=\linewidth]{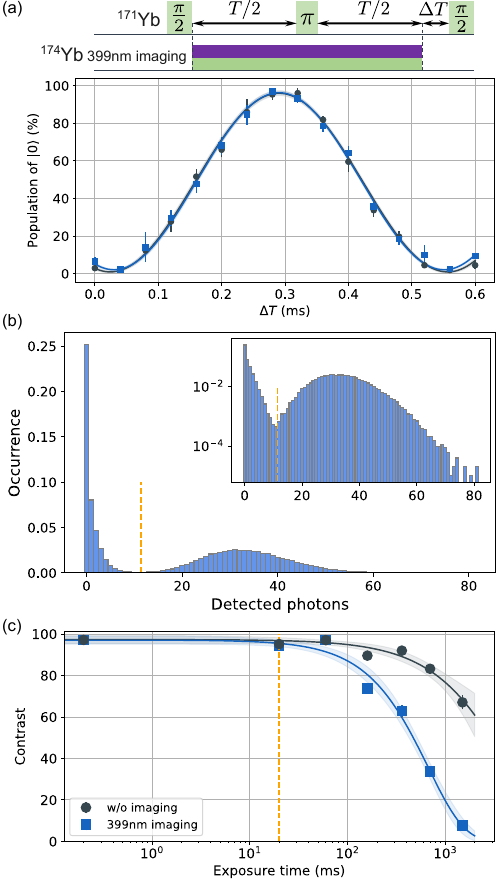}
    \caption{Influence on the coherence of the nuclear spin qubit from the imaging light for $\boson$ in the Hahn-echo sequence. (a) Hahn-echo signal for holding time of $T=\qty{20}{ms}$ with (blue square) and without (black circle) \qty{399}{nm} probe and \qty{556}{nm} cooling beams. The solid lines are the fitting curves by a sinusoidal function. The contrast ratio is 0.991(18), highlighting the possibility of the \qty{399}{nm} imaging without disturbing the coherence of the nuclear spin qubits. (b) The histogram of detected photons from $\boson$ with an exposure time of \qty{20}{ms}. The orange dashed line is the threshold for determining the occupied sites, yielding an imaging fidelity of 0.9992 and a survival probability of 0.988. (c) Decay curves of the Hahn-echo contrasts with and without exposure to imaging light for $\boson$. The solid lines are fitting curves by $f(T) = A\exp(-(T/T_2)^n)$. The shaded areas are the $1\sigma$-confidence intervals. The fitting results are summarized in Table \ref{tab:echo}. Although the coherence under the \qty{399}{nm} imaging decreases faster than that of without imaging, the contrast of the Hahn-echo signal is almost the same for our typical exposure time of \qty{20}{ms}, represented as an orange dashed line.}
    \label{fig:echo}
\end{figure}

To quantify this, we characterize the coherence property of the nuclear spin qubit using the Ramsey and Hahn-echo sequences under irradiation of imaging light for $\boson$, where we consider two imaging methods: ``\qty{399}{nm} imaging'' with a \qty{399}{nm} probe beam and \qty{556}{nm} cooling beams, and ``\qty{556}{nm} imaging'' with a single \qty{556}{nm} probe beam.
We first prepare the dual-Yb atom array and initialize the nuclear spin qubit via optical pumping by the $\ket{1}\rightarrow\ket{\tPo,F=3/2,m_F=1/2}$ transition. We then irradiate a circularly-polarized single \qty{556}{nm} beam in the horizontal plane to rotate the qubits by single-beam Raman transitions \cite{Jenkins2022-lt}. The qubit control beam has an intensity of \num{2.4e5}\,$I_\mathrm{s,556}$ and a detuning of \qty{-1.0}{GHz} (\qty{-3.1}{GHz}) from the resonance of $\ket{\tPo,F=1/2}$ of $\fermi$ ($\ket{\tPo}$ of $\boson$), yielding the Rabi frequency of $2\pi\times\qty{208}{kHz}$. After a $\pi/2$-pulse and a fixed waiting time of $T+\Delta T$, another $\pi/2$-pulse is applied to map the relative phase of the qubit states to their relative populations. Here, the additional waiting time of $\Delta T$ is used for scanning the relative phase between the qubit states. In the Hahn-echo sequence, we insert an additional $\pi$-pulse after a waiting time of $T/2$ from the first $\pi/2$-pulse. Finally, we blow away the atoms in the $\ket{0}$ state via the $\ket{0}\leftrightarrow\ket{\tPo,F=3/2,m_F=3/2}$ transition and image the remaining atoms. By scanning $\Delta T$, we obtain a Ramsey (Hahn-echo) signal and compare its contrast with and without irradiation of imaging light for $\boson$ during holding time $T$ to evaluate its influence.

We first present the results of the Hahn-echo experiment using our current \qty{399}{nm} imaging method. During the holding time $T$, with a trap depth of \qty{0.9}{mK}, we apply a resonant \qty{399}{nm} probe beam with a radius of \qty{1.3}{mm} at an intensity of $\num{1e-3}\,I_{\mathrm{s,399}}$, as well as \qty{556}{nm} cooling beams with radii of \qty{4}{mm} (vertical) and \qty{12}{mm} (horizontal) at a total intensity of $10\,I_{\mathrm{s,556}}$, which are detuned by $-5.5\Gamma$ from the resonance. Figure\,\ref{fig:echo}(a) shows the Hahn-echo signals for holding time of $T=\qty{20}{ms}$, which is typical in our current setup, with and without \qty{399}{nm} imaging. The oscillation frequency is determined by the Zeeman splitting of the nuclear spin states with a magnetic field of \qty{2.5}{G} applied during the holding time. The contrast ratio between the two conditions is 0.991(18), indicating that the coherence is maintained within the uncertainty. The histogram of the number of detected photons during imaging is shown in Fig.\,\ref{fig:echo}(b). The detection fidelity of $\ket{\sSz}=\ket{0}$ of $\boson$ in this case is 0.9992 with a survival probability after imaging of 0.988, which is estimated by employing the model-free method \cite{Manetsch2024-sp} (see Appendix \ref{sec:methods}).

To assess the trend of the influence on the coherence in the longer exposure time region, we extend the holding time up to \qty{1500}{ms}. Figure\,\ref{fig:echo}(c) illustrates the decay curve of the Hahn-echo contrast as a function of the exposure time. We estimate the coherence time $T_2$ in each case by fitting the data using the function $f(T) = A\exp(-(T/T_2)^n)$, where $A$, $T_2$, and $n$ are fitting parameters. The fitting results are summarized in Table \ref{tab:echo}. In the \qty{399}{nm} imaging, the coherence time is \qty{0.67(5)}{s}, which is shortened due to the scattering of the probe and cooling beams. However, as shown Fig.\,\ref{fig:echo}(a), we can still achieve high-fidelity imaging while maintaining the coherence. Moreover, considering the NA of our objective lens, there is still potential to improve the photon collection efficiency by a factor of three. If such an optimal imaging system is realized, the exposure time could be reduced by one-third, leading to a data qubit coherence retention of 99.7\%. We note that the coherence without imaging is limited by the scattering from the tweezer with a trapping depth of \qty{0.9}{mK}. This issue can be addressed by making the trap shallower at the data qubit sites while deepening it at the ancilla qubit sites.

\begin{table}[tb]
\caption{Fitting results of the coherence curves in the Hahn-echo sequence.}
\centering
\begin{tabular}{lccc}
\hline
                                             & $A$     & $T_2$ (s) & $n$    \\ \hline
without imaging                              & 0.97(2) & 4.3(2.3)  & 1.0(3) \\
\qty{399}{nm} imaging                        & 0.97(2) & 0.67(5)   & 1.2(2) \\
\hline
\end{tabular}\label{tab:echo}
\end{table}

\begin{figure}[tb]
    \centering
    \includegraphics[width=\linewidth]{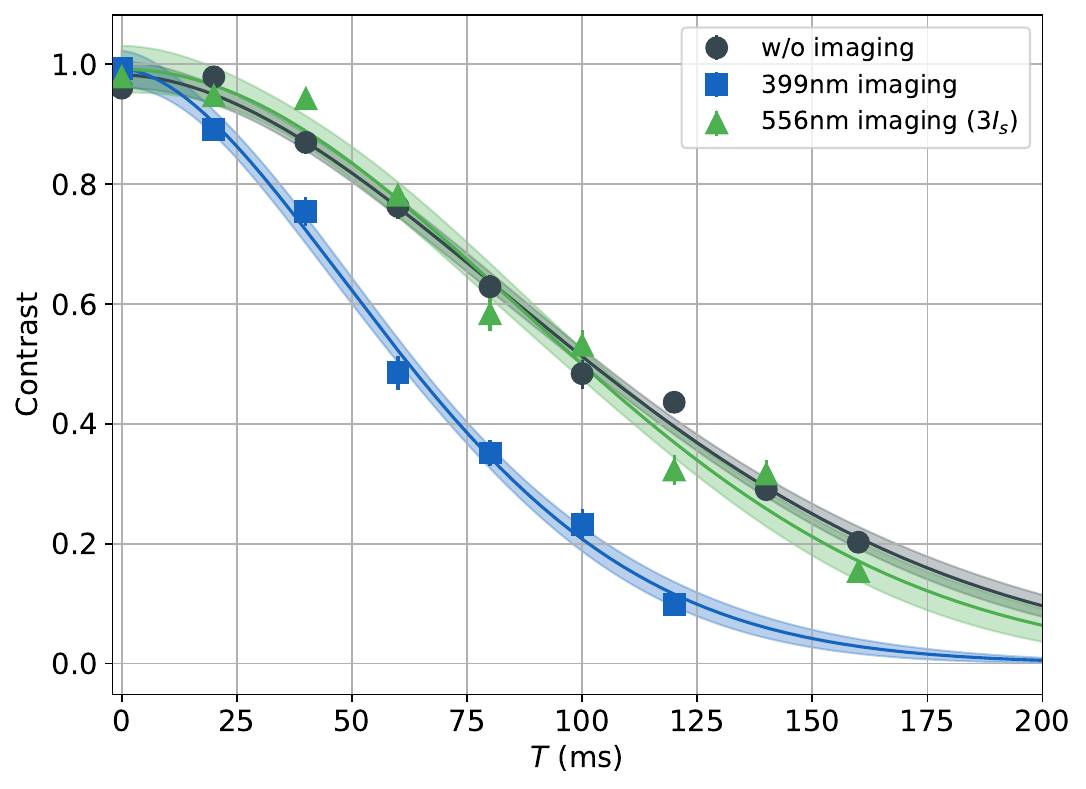}
    \caption{Influence on the coherence of the nuclear spin qubit from the imaging light for $\boson$ in the Ramsey sequence. Decay curves of the Ramsey contrast are shown as a function of exposure time. The solid lines are the fitting curves by $f(T)=A\exp(-(T/T_2^*)^n)$, where $A$, $T_2^*$, and $n$ are fitting parameters. The shaded areas are the $1\sigma$-confidence intervals. While the Ramsey contrast under \qty{399}{nm} imaging decays faster than that without imaging, the contrast with \qty{556}{nm} probe light with an intensity of $3\,I_{\mathrm{s,556}}$ (green triangle) remains the same as that of without irradiating a probe beam, suggesting that \qty{556}{nm} imaging is a possible future improvement of low cross-talk measurements.}
    \label{fig:Ramsey}
\end{figure}
\begin{table}[bt]
\caption{Fitting results of the coherence curves in the Ramsey sequence.}
\centering
\begin{tabular}{lccc}
\hline
                                             & $A$     & $T_2^*$ (ms) & $n$    \\ \hline
without imaging                              & 0.98(2) & 130(3)  & 2.0(2) \\
\qty{556}{nm} imaging                        & 0.98(2) & 122(5)  & 2.3(3) \\
\qty{399}{nm} imaging                        & 0.99(2) & 77(2)   & 1.8(1) \\
\hline
\end{tabular}\label{tab:Ramsey}
\end{table}
As a potential future improvement, we evaluate the influence on the coherence using the Ramsey sequence with the \qty{556}{nm} imaging, which has proven highly effective for $\boson$ trapped in \qty{532}{nm} tweezers \cite{Saskin2019-xk}. The trap depth of \qty{0.9}{mK} is the same as the Hahn-echo sequence. We apply a resonant single \qty{556}{nm} beam for $\boson$ with a radius of \qty{1.4}{mm} at an intensity of $3\,I_{\mathrm{s,556}}$, typical in the previous research \cite{Saskin2019-xk}, during the holding time for the \qty{556}{nm} imaging. The probe beam diagonally propagates from the horizontal plane by 16 degrees and is horizontally polarized. Figure\,\ref{fig:Ramsey} presents the trend of the Ramsey contrast with a holding time up to \qty{160}{ms}. While the contrast in the case of the \qty{399}{nm} imaging rapidly decreases, the coherence under the \qty{556}{nm} imaging remains almost the same as that without imaging. We again note that the observed coherence is reduced due to the tweezer potential of \qty{0.9}{mK}, and does not exhibit a fundamental limitation. The fitting results, using the same function for the Hahn-echo experiment, are summarized in Table.\ref{tab:Ramsey}. This result highlights the potential of the \qty{556}{nm} imaging for low cross-talk measurements. Based on the experimental result in Ref. \cite{Saskin2019-xk}, the imaging fidelity is expected to exceed 0.9999 with a \qty{99}{\%} survival probability for a \qty{30}{ms} exposure time, with negligible impact on the data qubit coherence. Since our current optical setup does not support \qty{556}{nm} imaging, we plan to verify its actual performance in the dual-Yb system in future work, which will be useful in fully evaluating its potential for improving low cross-talk measurements and enhancing overall system performance.

\section{Discussion}
The cross-talk measurement with our current \qty{399}{nm} imaging, shown in Fig.\,\ref{fig:echo}, has revealed that dual-Yb has the capability of low cross-talk measurement with high-fidelity detection of the ancilla qubits, which is already within reach of the requirement for quantum error correction \cite{Fowler2009-jf}. The coherence loss during imaging is less than \qty{1}{\%}, which is similarly qualified or even superior to the previous research using the omg-architecture \cite{Lis2023-cb}, energy isolation by local light shift \cite{Norcia2023-dt}, and shelving technique \cite{Graham2023-wl}. The one important key aspect of the dual-Yb system lies in the simplicity of the operation for mid-circuit measurements while the other approaches including the spatial isolation methods \cite{Bluvstein2024-yx, Deist2022-io} require technically demanding local operations. 

As is shown in Fig.\,\ref{fig:echo}(a), the crosstalk induced by the \qty{399}{nm} imaging with our current parameters is only \qty{0.9}{\%}. However, when comparing the dual-Yb system with the dual-alkali system \cite{Singh2023-ox}, the behavior of the contrast decay curves with and without imaging in the long exposure time region differs from that observed in the dual-alkali system. This is due to the limited frequency isolation, which is characterized by the ratio between the isotope shift $|\omega-\omega'|$ and the natural linewidth $\Gamma$ of the transition, with $|\omega_{\fermi}-\omega_{\boson}|/\Gamma_{399}\sim30$ for the $\sSz\text{--}\sPo$ transition of ytterbium, much smaller than $|\omega_{\mathrm{Rb}}-\omega_{\mathrm{Cs}}|/\Gamma\sim10^6$ for the dual-alkali system. However, as shown in Fig.\,5, \qty{556}{nm} imaging holds the potential to achieve comparably low cross-talk measurements as the dual-alkali system, owing to its large frequency isolation of $|\omega_{\fermi}-\omega_{\boson}|/\Gamma_{556}\sim10^4$.

Here, we point out that the important difference in the alkali and Yb systems arises from the types of qubits. In the alkali system, the hyperfine state qubits with millisecond-order coherence, with no dynamical decoupling, are used for both data and ancilla qubits. Typically, the qubit detection in this system relies on the state-selective loss before imaging, which requires more frequent atom replenishment. Although the non-destructive imaging has been demonstrated in alkali systems using techniques such as shelving \cite{Graham2023-wl} and state transfer to a stretched state \cite{Nikolov2023-fv, Radnaev2024-ac}, those operations may carry the risk of finite infidelity. In contrast, the dual-Yb system can offer a data qubit (nuclear spin qubit) with inherently long second-order coherence time, and an ancilla qubit (optical qubit) that is intrinsically capable of non-destructive readout without state-selective loss. These advantages make the dual-Yb system preferable in implementing mid-circuit measurements for quantum error correction.

\section{Conclusion}
In summary, we have reported the realization of a dual-Yb atom array consisting of fermionic $\fermi$ and bosonic $\boson$. This hybrid system offers a nuclear spin qubit as a data qubit with a long coherence time and an optical clock qubit as an ancilla qubit, which provides the ability to read out the qubit states non-destructively. We found the tunability of the dual occupancy rate from almost zero to several tens of percent. Thus, this remarkable versatility could open the avenue to the experiment with heteronuclear molecules trapped in optical tweezers \cite{Holland2023-uz, Ruttley2024-vx, Ruttley2024-iw, Picard2024-kr,Holland2024-wn}. We evaluated the coherence of the nuclear spin qubits under exposure to the imaging beams for $\boson$. In the case of \qty{399}{nm} imaging with an exposure time of \qty{20}{ms}, the coherence is maintained at 99.1(1.8)\,\% level with an imaging fidelity of 99.92\,\% and a survival probability after imaging of \qty{98.8}{\%} for $\boson$. This performance is comparably good as the previous work \cite{Lis2023-cb,Norcia2023-dt, Graham2023-wl}, even without technical local operations. Moreover, we confirm that the \qty{556}{nm} imaging with $3\,I_{\mathrm{s,556}}$, typical in previous study \cite{Saskin2019-xk}, causes no detrimental effect on the coherence and could enable high-fidelity and low atom loss imaging, making it a promising candidate for mid-circuit measurements.

As an outlook, we plan to characterize the properties of the optical clock qubit. Leveraging the insensitivity to magnetic field fluctuations \cite{Dzuba2018-tz} and the magic-wavelength tweezer that compensates the tensor light shifts, the $\tPt$ and $4f^{13}5d6s^2~(J=2)$ states could be good candidates for the optical qubit, as well as the $\tPz$ state. The single qubit operations on the optical qubits will utilize $\pi$-pulse for the X gates and phase shift of the excitation laser for the Z gate \cite{Cao2024-hp}. The fine-structure qubit that was recently demonstrated in strontium systems \cite{Unnikrishnan2024-ec, Pucher2024-pf} is also an interesting direction due to its fast controllability.
Regarding the generation of entanglement between nuclear spin qubits and clock transition qubits, the time-optimal two-qubit gates mediating the Rydberg interaction \cite{Jandura2022-ei} can be used. For this, it is necessary to investigate the Rydberg interaction between the isotopes. There may be F\"{o}rster resonances, which can enhance the gate speed and fidelity, as found in the dual-species system of rubidium and cesium \cite{Ireland2024-wa, Anand2024-xl}. Those studies combining our dual-Yb system with some additional improvement and development will pave the way to the ancilla-based QEC protocols in the neutral atom architecture.

\begin{acknowledgments}
We acknowledge K. Enomoto and H. Hara for the discussion about heteronuclear photoassociation resonances. We also thank H. Loh and S. de L\'{e}s\'{e}luec for the helpful advice on the atom rearrangement. This work was supported by the Grant-in-Aid for Scientific Research of JSPS (No.\ JP17H06138, No.\ JP18H05405, No.\ JP18H05228, No.\ JP21H01014, No.\ JP21K03384, No.\ JP22K20356), JST PRESTO (No.\ JPMJPR23F5), JST CREST (Nos.\ JPMJCR1673 and JPMJCR23I3), MEXT Quantum Leap Flagship Program (MEXT Q-
LEAP) Grant No.\ JPMXS0118069021, and JST Moon-shot R\&D Grant No.\ JPMJMS2269.
Y.N.\ acknowledges support from the JSPS (KAKENHI Grant No.\ 22KJ1949). 
T.K.\ acknowledges the support from the establishment of university fellowships towards the creation of science technology innovation, Grant No.\ JPMJFS2123.
\end{acknowledgments}

\appendix

\section{Methods\label{sec:methods}}
\subsection{Preparation of dual-Yb atom array}
\begin{figure}[t]
    \centering
    \includegraphics[width=\linewidth]{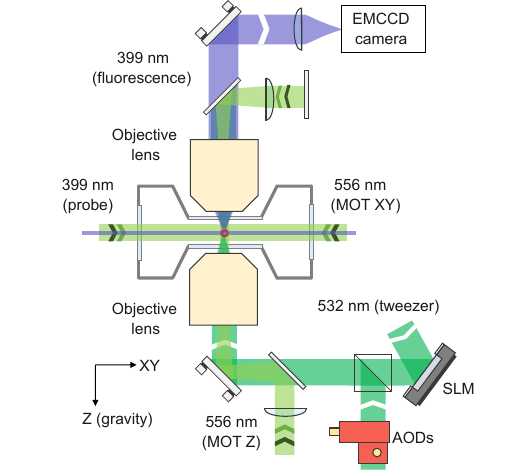}  
    \caption{Experimental setup. The optical tweezer array is generated by a spatial light modulator and an objective lens of NA = 0.6 with \qty{532}{nm} light. A pair of AODs is used to generate moving tweezers for atom rearrangement. The MOT beams for $\fermi$ and $\boson$ share the optical path in each axis. The MOT beam along the Z axis is combined with the tweezer beam by a dichroic mirror and is focused on the back-focal plane of the objective for collimating at the atom position. Atom fluorescence induced by \qty{399}{nm} probe beam in the horizontal plane is collected by another objective and imaged by an EMCCD camera.}
    \label{fig:setup}
\end{figure}

The experimental setup is shown in Fig.\,\ref{fig:setup}. The sequence starts with a dual MOT of $\fermi$ and $\boson$ on the $\sSz\text{--}\tPo$ transition, achieved by sequentially adjusting the frequency of the Zeeman slowing beam on the $\sSz\text{--}\sPo$ transition from $\fermi$ to $\boson$. The MOT beams for each isotope share the optical path. After MOT loading with a total duration of \qty{1}{s}, the MOTs are compressed by increasing the magnetic field gradient from \qty{10.4}{G/cm} to \qty{17.4}{G/cm} to load atoms into the optical tweezer array, which is generated by a spatial-light-modulator (Hamamatsu, X15213-L16) with \qty{532}{nm} light \cite{Nogrette2014-zl}. We then apply red-detuned LAC beams for \qty{200}{ms} to ensure the single atom loading. The single atoms trapped in the tweezers are imaged on the $\sSz\text{--}\sPo$ transition ($\lambda=\qty{399}{nm}$) with simultaneous cooling by the MOT beams. The emitted photons are collected onto an EMCCD camera (Andor, iXon-897) by an objective lens with a numerical aperture (NA) of 0.6. The imaging time is \qty{60}{ms}. The survival probability of the atoms after imaging is \qty{95}{\%} for both isotopes, which are limited by low photon collection efficiency. Since the isotope shifts of the relevant transitions are much larger than their natural linewidth, as shown in Fig.\,\ref{fig:overview}(c) of the main text, the probe and cooling beams for each isotope do not cause detrimental effects on the survival probability of the other isotope.

\subsection{Initialization and readout of the nuclear spin qubits}
The nuclear spin qubit is encoded in the ground state of $\fermi$ as $\ket{0} = \ket{\sSz,m_F=1/2}$ and $\ket{1} = \ket{\sSz,m_F=-1/2}$. The qubit initialization is done via optical pumping for \qty{1}{ms} by the $\ket{1}\rightarrow\ket{\tPo,F=3/2,m_F=1/2}$ transition. During the pumping, we apply a magnetic field of \qty{48}{G} to suppress the unwanted transition of $\ket{0}\rightarrow\ket{\tPo,F=3/2,m_F=3/2}$. The qubit state is read out through state-selective loss before imaging. To achieve this, we increase the magnetic field to \qty{48}{G} again, then ramp down the tweezer depth to \qty{35}{\mu K}, and apply a \qty{556}{nm} beam for \qty{5}{ms} to blow away the atoms in $\ket{0}$ state via $\ket{0}\rightarrow\ket{\tPo,F=3/2,m_F=3/2}$ transition. After that, we ramp up the tweezer potential to the original depth of \qty{0.9}{mK} and reduce the magnetic field to \qty{2.5}{G} and image the remaining atoms.

\subsection{Estimating the imaging fidelity}
The imaging fidelity is an important factor in assessing the performance of the low cross-talk measurements. To precisely evaluate the fidelity without any bias, we adopted a model-free method, as detailed in \cite{Manetsch2024-sp}. In this approach, we successively take three images under identical conditions during the cross-talk measurement with \qty{399}{nm} imaging as described in the main text. The model-free method yielded an imaging fidelity of 0.9992 with a survival probability of 0.988 under a \qty{20}{ms} exposure time.

\begin{figure}[tb]
    \centering
    \includegraphics[width=\linewidth]{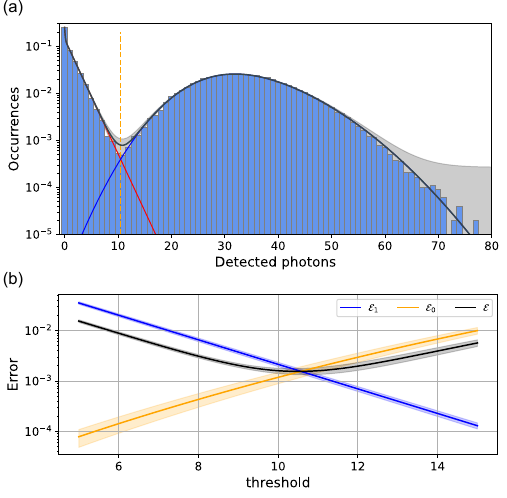}
    \caption{Fitting to the histogram of detected photons. (a) The histogram of the number of detected photons by \qty{399}{nm} imaging with an exposure time of \qty{20}{ms}. The black, red, and blue curves represent the fitting function (Eq.\,\ref{eq:fit_func}), $P_D(x)$, and $P_B(x)$, respectively. The dashed orange is the threshold that maximizes the imaging fidelity at 0.9985(2). The shaded areas are the 1$\sigma$-confidence intervals. (b) Imaging infidelities as a function of the threshold.}
    \label{fig:fit_histogram}
\end{figure}

As a double-check, we also employed the fitting method. To fit the histogram of the detected photons during imaging, we chose fitting functions as
\begin{equation*}
    P(x) = (1-F)P_D(x)+FP_B(x),
\end{equation*}
where $F$ is the loading probability of atoms, and $P_D(x)$ and $P_B(x)$ are the distributions for dark and bright peaks of the histogram as
\begin{equation*}
\begin{split}
P_D(x) &= \frac{1}{\sqrt{2\pi}\sigma_D}\exp\left(-\frac{(x-\mu_D)^2}{2\sigma_D^2}\right)+a\exp(-bx)\\
P_B(x) &= \frac{1}{\sqrt{2\pi}\sigma_B}\mathrm{exp}\left(-\frac{(x-\mu_B)^2}{2\sigma_B^2}\right)\left(1+\mathrm{erf}\left(c\frac{x-\mu_B}{\sqrt{2}\sigma_B}\right)\right)
\end{split}
\end{equation*}
with the fitting parameters $\sigma_D,~\mu_D~a,~b,~\sigma_B,~\mu_B,~c$. The $P_D(x)$ is composed of the Gaussian and exponential distributions, which account for the readout and amplification noise from the EMCCD camera. We note that strictly speaking, the amplification process of the camera count is explained by the Erlang distribution $P(x)=x^{n-1}e^{-x/g}/(g^n(n-1)!)$ with a photo-electron number $n$ and a gain $g$ \cite{Bergschneider2018-ki}. However, for the dark distribution, where the number of photo-electrons is almost zero ($n\le1$), the Erlang distribution can be reduced to an exponential distribution. For the fitting to the bright peak, we empirically chose a skewed Gaussian distribution. Figure\,\ref{fig:fit_histogram} shows the fitting result and corresponding infidelities calculated as
\begin{equation}\label{eq:fit_func}
    \begin{split}
        \mathcal{E} &= (1-F)\mathcal{E}_0+F\mathcal{E}_1 \\
        &= (1-F)\int_{x_{\mathrm{th}}}^\infty P_D(x)dx + F\int_0^{x_{\mathrm{th}}} P_B(x)dx
    \end{split}
\end{equation}
with the threshold for determining the presence of atoms and $\mathcal{E}_0$ and $\mathcal{E}_1$ are the false-negative and false-positive errors. The best fidelity is found to be 0.9985(2), which is in reasonable agreement with the model-free method. However, we have chosen to present the fidelity from the model-free method in the main text, as the fitting did not perfectly match the histogram data.

\section{Calculation of heteronuclear PA resonances\label{sec:PA_calc}}

\begin{figure}[tb]
    \centering
    \includegraphics[width=\linewidth]{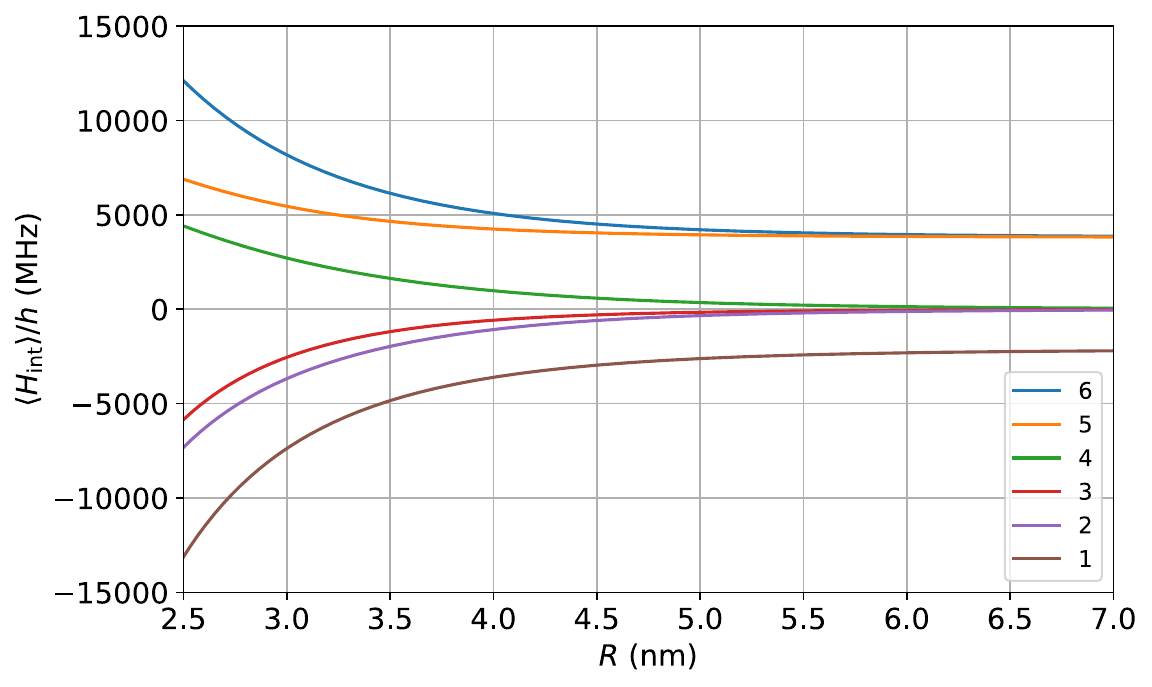}
    \caption{Interaction potentials between $\fermi$ and $\boson$ obtained from Eq.\,\ref{eq:int_hamiltonian}. The potentials are plotted with respect to the $\tPo$ state of $\boson$. Potential No.\,2 and 3 asymptotically connect to the $\ket{\sSz}_{171}\ket{\tPo}_{174}$ atomic state in the dissociation limit.}
    \label{fig:potential}
\end{figure}

To calculate the PA resonances of the excited $\fermi\boson$ molecule, which asymptotically connects to the $\sSz\text{--}\tPo$ state, we start with the interaction Hamiltonian under a Born-Oppenheimer approximation \cite{Jones2006-oh}, given as
\begin{equation}
    \mathcal{H}_{\mathrm{int}} = \frac{\vec{d}_{171}\cdot\vec{d}_{174}-3d_{z,171}d_{z,174}}{4\pi\epsilon_0 R^3}+A h\vec{I}_{171}\cdot\vec{J}_{171}.
    \label{eq:int_hamiltonian}
\end{equation}
Here, $\vec{d}$ and $d_z$ are the dipole operator and its projection onto the interatomic axis for each atom, $\epsilon_0$ is the permittivity of vacuum, $R$ is the interatomic distance, $h$ is the Planck constant, $A$ is the hyperfine coupling constant of $\fermi$, and $\vec{I}$ and $\vec{J}$ are the nuclear spin and electronic angular momentum operators of $\fermi$, respectively. We take the basis of $\ket{G_i}_{171}\otimes\ket{E_j}_{174}$ and $\ket{E_i}_{171}\otimes\ket{G}_{174}$ for the diagonalization, where $\ket{G_i}_{171}~(i=1,2)$ and $\ket{E_i}_{171}~(i=1\text{--}6)$ are the ground states and excited $\tPo$ states of $\fermi$, and $\ket{G}_{174}$ and $\ket{E_j}_{174}~(j=1\text{--}3)$ are the ground state and excited $\tPo$ states of $\boson$, respectively. By diagonalization of the Hamiltonian (Eq. \ref{eq:int_hamiltonian}), we obtain six interaction potentials, as shown in Fig.\,\ref{fig:potential}. Note that the retardation effect on the resonant dipole interaction \cite{Weiner1999-tv} is involved in this calculation. The potential No.\,1--3 are attractive, thus they can support weakly-bound heteronuclear molecules. Then, we add a centrifugal potential, van der Waals potential, and a short-range repulsive potential to the obtained potential $\langle\mathcal{H}_{\mathrm{int}}\rangle$ and get a model potential \cite{Enomoto2008-au} as
\begin{equation}
    V(R) = \langle\mathcal{H}_{\mathrm{int}}\rangle + \hbar^2\frac{T_e(T_e+1)+\langle\vec{F}^2\rangle-2\Omega^2}{2\mu R^2}-\frac{C_6}{R^6}+\frac{C_{12}}{R^{12}},
    \label{eq:model_potential}
\end{equation}
where $\hbar = h/(2\pi)$ is reduced Planck constant, $\mu$ is the reduced mass of $\fermi$ and $\boson$, $T_e$ and $\Omega$ are the total electronic angular momentum and its projection onto the molecular axis, $C_6$ and $C_{12}$ are the coefficients for the van der Waals potential and the short-range 
repulsive potential, respectively. $\vec{F} = \vec{I}_{171}+\vec{J}_{171}+\vec{J}_{174}$ is the sum of electronic angular momenta of $\fermi$ and $\boson$ and nuclear spin of $\fermi$. Finally, we solve the Schr\"{o}dinger equation with the model potential (Eq.\,\ref{eq:model_potential}) by Numerov method \cite{BLATT1967382} and obtain the eigenenergies. The parameters for this calculation are summarized in Table\,\ref{tab:PA_calc}.

Table\, \ref{tab:PA_lines} shows the resonance frequencies of the PA below \qty{1}{GHz} for the potential No.\,2, in which the molecular states have $T_e=1/2,~3/2$ and $|\Omega|=1/2$. The calculated PA resonance frequencies agree with the experimental values with a few MHz precision. 

\begin{table}[]
\caption{Parameters for PA resonance calculation. $E_h$ is Hartree energy and $a_0$ is Bohr radius.}
\begin{tabular}{lr}
\hline
$A$                      & \qty{3957.781}{MHz} \cite{Pandey2009-fp}           \\
$d^2/(4\pi\epsilon_0)$   & $0.08744313\,E_h a_0^3$ \cite{Borkowski2011-tq} \\
$C_6$                    & $2405.364747\,E_h a_0^6$ \cite{Borkowski2011-tq}     \\
$C_{12}$                 & $9.318\times10^8\,E_h a_0^{12}$ (fitting) \\ \hline
\end{tabular}\label{tab:PA_calc}
\end{table}

\begin{table}[]
\centering
\caption{PA resonances below \qty{1}{GHz} for the potential No.\,2.}
\begin{tabular}{lrr}\hline
$T_e$ & Calculation (MHz) & Experiment (MHz) \\ \hline
$1/2$ & \num{-0.05}           & \num{-0.50(2)}   \\
$3/2$ & \num{-55.3}           & ---              \\
$1/2$ & \num{-62.9}           & ---              \\
$3/2$ & \num{-270.5}          & \num{-268(2)}         \\
$1/2$ & \num{-283.6}          & \num{-281(1)}         \\
$3/2$ & \num{-699.4}          & \num{-698.8(8)}       \\
$1/2$ & \num{-717.8}          & \num{-716.4(5)}       \\ \hline
\end{tabular}
\label{tab:PA_lines}
\end{table}

\section{The difference of dual occupancy probability in dual-alkali and dual-Yb systems\label{sec:DO_diff}}
We have presented the tunability of the dual occupancy in the dual-Yb array while this behavior is not observed in the dual-isotope alkali system \cite{Sheng2022-az}. The difference between the dual-isotope ytterbium and alkali systems arises from the difference in the natural linewidth $\Gamma$ of the cooling transition.

The interatomic potential of the heteronuclear molecule bound near the dissociation limit is governed by the van der Waals interaction as $V(R)\propto -C_6/R^6$. From the LeRoy-Bernstein formula \cite{LeRoy1970-lb}, the derivative of the binding energy $E(v)$ to the vibrational level $v$ for the heteronuclear molecule is analytically calculated as
\begin{equation*}
    \frac{dE(v)}{dv} \propto C_6^{-1/6}(D-E(v))^{2/3},
\end{equation*}
where $D$ is the dissociation limit energy. Since the detuning of the cooling beams, which induces light-assisted collisions or photo-association, is on the order of several $\Gamma$, $D-E(v)$ in the above equation can be replaced with $\Gamma$. Furthermore, considering that the $C_6$ coefficient is derived from the second-order perturbation of the dipole-dipole interaction with a strength $d^2$ and thus, $C_6\propto|d^2|^2 = \Gamma^2$, we get the scaling of the binding energy as
\begin{equation*}
    \frac{dE(v)}{dv} \propto (\Gamma^2)^{-1/6}\Gamma^{2/3} = \Gamma^{1/3}.
\end{equation*}
From the above equation, we can estimate the number of the weakly-bound states per $\Gamma$ as $\Gamma/(dE/dv) = \Gamma^{2/3}$. 

Since the ratio of the linewidth between the ${}^1\mathrm{S}_0\text{--}{}^3\mathrm{P}_1$ transition of ytterbium and the $D_2$ transition of rubidium is around $1:30$, the number of the bound states per $\Gamma$ around the cooling detuning for alkali atoms is approximately 10 times as large as ytterbium. As a result, the cooling beams of alkali atoms are likely to overlap the photo-association resonance and naturally induce heteronuclear light-assisted collisions while the photo-association resonances of ytterbium are resolved compared to the linewidth of ${}^1\mathrm{S}_0\text{--}{}^3\mathrm{P}_1$ transition as shown in Table \ref{tab:PA_lines}. 

We note that dual occupancy is naturally avoided in the dual-species alkali system by working with the two-color tweezer system \cite{Brooks2021-ec, Singh2022-zg} .

\section{Basic properties of the nuclear spin qubit\label{sec:ns_qubit_property}}
\begin{figure}[tb]
    \centering
    \includegraphics[width=\linewidth]{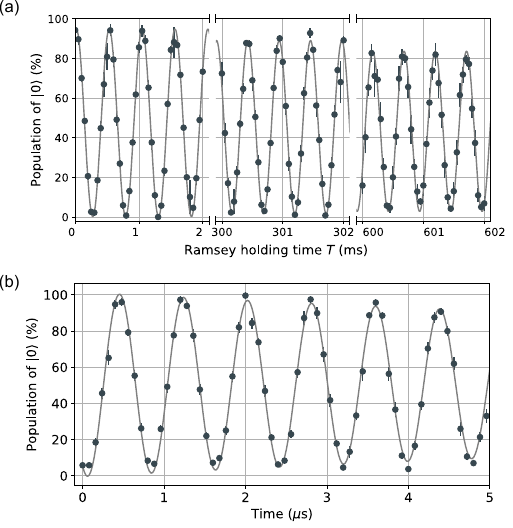}
    \caption{Coherence measurement and fast control of the nuclear spin qubit. (a) Ramsey signal in tweezers of $\qty{35}{\mu K}$ depth. The coherence time is $T_2^* = \qty{3.8(4)}{s}$. The solid line is a fitting curve of damped sinusoidal function. (b) Fast qubit control by a fictitious magnetic field. We obtain Rabi frequency up to \qty{1.3}{MHz} with $\pi$-pulse fidelity of $\sim0.995$. The solid line is a fitting curve of a damped sinusoidal function.}
    \label{fig:qubit_property}
\end{figure}

Since the nuclear spin in the ground state of $\fermi$ is decoupled to the electronic angular momentum, it serves as an excellent qubit with long coherence. We measure the coherence time $T_2^*$ by the Ramsey sequence in a shallow trap with a depth of \qty{35}{\mu K}, as shown in Fig.\,\ref{fig:qubit_property}(a). The qubits exhibit a second-order coherence time of $T_2^*=\qty{3.8(4)}{s}$ in our setup, which is similar to the result in the previous research \cite{Jenkins2022-lt,Ma2022-gy}.

The qubit control is done by a fictitious magnetic field induced by a circularly-polarized \qty{556}{nm} single beam \cite{Jenkins2022-lt}. We set the detuning of the control beam to \qty{-1.0}{GHz} from the $\ket{\tPo,F=1/2}$ resonance and observe the fast Rabi oscillation up to \qty{1.3}{MHz} (Fig.\,\ref{fig:qubit_property}(b)). The fidelity of $\pi$-pulse is found to be $\sim0.995$ by comparing the survival probabilities with and without blasting out the atoms in $\ket{0}$ after applying a $\pi$-pulse. The fidelity can be further improved by optimizing the detuning of the laser detuning and the angle between the quantization and laser propagation axis \cite{Lis2023-cb}. After these improvements and incorporating the recent development of high-fidelity two-qubit gates \cite{Evered2023-nv, Scholl2023-jb, Peper2024-wc, Tsai2024-tu, Radnaev2024-ac}, we believe that the dual-Yb system is a good candidate for implementing mid-circuit measurements and facilitating quantum error correction.

\clearpage


\begin{thebibliography}{98}%
  \makeatletter
  \providecommand \@ifxundefined [1]{%
   \@ifx{#1\undefined}
  }%
  \providecommand \@ifnum [1]{%
   \ifnum #1\expandafter \@firstoftwo
   \else \expandafter \@secondoftwo
   \fi
  }%
  \providecommand \@ifx [1]{%
   \ifx #1\expandafter \@firstoftwo
   \else \expandafter \@secondoftwo
   \fi
  }%
  \providecommand \natexlab [1]{#1}%
  \providecommand \enquote  [1]{``#1''}%
  \providecommand \bibnamefont  [1]{#1}%
  \providecommand \bibfnamefont [1]{#1}%
  \providecommand \citenamefont [1]{#1}%
  \providecommand \href@noop [0]{\@secondoftwo}%
  \providecommand \href [0]{\begingroup \@sanitize@url \@href}%
  \providecommand \@href[1]{\@@startlink{#1}\@@href}%
  \providecommand \@@href[1]{\endgroup#1\@@endlink}%
  \providecommand \@sanitize@url [0]{\catcode `\\12\catcode `\$12\catcode `\&12\catcode `\#12\catcode `\^12\catcode `\_12\catcode `\%12\relax}%
  \providecommand \@@startlink[1]{}%
  \providecommand \@@endlink[0]{}%
  \providecommand \url  [0]{\begingroup\@sanitize@url \@url }%
  \providecommand \@url [1]{\endgroup\@href {#1}{\urlprefix }}%
  \providecommand \urlprefix  [0]{URL }%
  \providecommand \Eprint [0]{\href }%
  \providecommand \doibase [0]{https://doi.org/}%
  \providecommand \selectlanguage [0]{\@gobble}%
  \providecommand \bibinfo  [0]{\@secondoftwo}%
  \providecommand \bibfield  [0]{\@secondoftwo}%
  \providecommand \translation [1]{[#1]}%
  \providecommand \BibitemOpen [0]{}%
  \providecommand \bibitemStop [0]{}%
  \providecommand \bibitemNoStop [0]{.\EOS\space}%
  \providecommand \EOS [0]{\spacefactor3000\relax}%
  \providecommand \BibitemShut  [1]{\csname bibitem#1\endcsname}%
  \let\auto@bib@innerbib\@empty
  \bibitem [{\citenamefont {Gottesman}(1997)}]{Gottesman1997-zd}%
    \BibitemOpen
    \bibfield  {author} {\bibinfo {author} {\bibfnamefont {D.}~\bibnamefont {Gottesman}},\ }\emph {\bibinfo {title} {{Stabilizer Codes and Quantum Error Correction}}},\ \href@noop {} {Ph.D. thesis},\ \bibinfo  {school} {California Institute of Technology} (\bibinfo {year} {1997})\BibitemShut {NoStop}%
  \bibitem [{\citenamefont {{Google Quantum AI}}(2021)}]{Google_Quantum_AI2021-td}%
    \BibitemOpen
    \bibfield  {author} {\bibinfo {author} {\bibnamefont {{Google Quantum AI}}},\ }\bibfield  {title} {\bibinfo {title} {Exponential suppression of bit or phase errors with cyclic error correction},\ }\href {https://doi.org/10.1038/s41586-021-03588-y} {\bibfield  {journal} {\bibinfo  {journal} {Nature (London)}\ }\textbf {\bibinfo {volume} {595}},\ \bibinfo {pages} {383} (\bibinfo {year} {2021})}\BibitemShut {NoStop}%
  \bibitem [{\citenamefont {Zhao}\ \emph {et~al.}(2022)\citenamefont {Zhao}, \citenamefont {Ye}, \citenamefont {Huang}, \citenamefont {Zhang}, \citenamefont {Wu}, \citenamefont {Guan}, \citenamefont {Zhu}, \citenamefont {Wei}, \citenamefont {He}, \citenamefont {Cao}, \citenamefont {Chen}, \citenamefont {Chung}, \citenamefont {Deng}, \citenamefont {Fan}, \citenamefont {Gong}, \citenamefont {Guo}, \citenamefont {Guo}, \citenamefont {Han}, \citenamefont {Li}, \citenamefont {Li}, \citenamefont {Li}, \citenamefont {Liang}, \citenamefont {Lin}, \citenamefont {Qian}, \citenamefont {Rong}, \citenamefont {Su}, \citenamefont {Sun}, \citenamefont {Wang}, \citenamefont {Wu}, \citenamefont {Xu}, \citenamefont {Ying}, \citenamefont {Yu}, \citenamefont {Zha}, \citenamefont {Zhang}, \citenamefont {Huo}, \citenamefont {Lu}, \citenamefont {Peng}, \citenamefont {Zhu},\ and\ \citenamefont {Pan}}]{Zhao2022-qf}%
    \BibitemOpen
    \bibfield  {author} {\bibinfo {author} {\bibfnamefont {Y.}~\bibnamefont {Zhao}}, \bibinfo {author} {\bibfnamefont {Y.}~\bibnamefont {Ye}}, \bibinfo {author} {\bibfnamefont {H.-L.}\ \bibnamefont {Huang}}, \bibinfo {author} {\bibfnamefont {Y.}~\bibnamefont {Zhang}}, \bibinfo {author} {\bibfnamefont {D.}~\bibnamefont {Wu}}, \bibinfo {author} {\bibfnamefont {H.}~\bibnamefont {Guan}}, \bibinfo {author} {\bibfnamefont {Q.}~\bibnamefont {Zhu}}, \bibinfo {author} {\bibfnamefont {Z.}~\bibnamefont {Wei}}, \bibinfo {author} {\bibfnamefont {T.}~\bibnamefont {He}}, \bibinfo {author} {\bibfnamefont {S.}~\bibnamefont {Cao}}, \bibinfo {author} {\bibfnamefont {F.}~\bibnamefont {Chen}}, \bibinfo {author} {\bibfnamefont {T.-H.}\ \bibnamefont {Chung}}, \bibinfo {author} {\bibfnamefont {H.}~\bibnamefont {Deng}}, \bibinfo {author} {\bibfnamefont {D.}~\bibnamefont {Fan}}, \bibinfo {author} {\bibfnamefont {M.}~\bibnamefont {Gong}}, \bibinfo {author} {\bibfnamefont {C.}~\bibnamefont {Guo}}, \bibinfo {author} {\bibfnamefont
    {S.}~\bibnamefont {Guo}}, \bibinfo {author} {\bibfnamefont {L.}~\bibnamefont {Han}}, \bibinfo {author} {\bibfnamefont {N.}~\bibnamefont {Li}}, \bibinfo {author} {\bibfnamefont {S.}~\bibnamefont {Li}}, \bibinfo {author} {\bibfnamefont {Y.}~\bibnamefont {Li}}, \bibinfo {author} {\bibfnamefont {F.}~\bibnamefont {Liang}}, \bibinfo {author} {\bibfnamefont {J.}~\bibnamefont {Lin}}, \bibinfo {author} {\bibfnamefont {H.}~\bibnamefont {Qian}}, \bibinfo {author} {\bibfnamefont {H.}~\bibnamefont {Rong}}, \bibinfo {author} {\bibfnamefont {H.}~\bibnamefont {Su}}, \bibinfo {author} {\bibfnamefont {L.}~\bibnamefont {Sun}}, \bibinfo {author} {\bibfnamefont {S.}~\bibnamefont {Wang}}, \bibinfo {author} {\bibfnamefont {Y.}~\bibnamefont {Wu}}, \bibinfo {author} {\bibfnamefont {Y.}~\bibnamefont {Xu}}, \bibinfo {author} {\bibfnamefont {C.}~\bibnamefont {Ying}}, \bibinfo {author} {\bibfnamefont {J.}~\bibnamefont {Yu}}, \bibinfo {author} {\bibfnamefont {C.}~\bibnamefont {Zha}}, \bibinfo {author} {\bibfnamefont {K.}~\bibnamefont
    {Zhang}}, \bibinfo {author} {\bibfnamefont {Y.-H.}\ \bibnamefont {Huo}}, \bibinfo {author} {\bibfnamefont {C.-Y.}\ \bibnamefont {Lu}}, \bibinfo {author} {\bibfnamefont {C.-Z.}\ \bibnamefont {Peng}}, \bibinfo {author} {\bibfnamefont {X.}~\bibnamefont {Zhu}},\ and\ \bibinfo {author} {\bibfnamefont {J.-W.}\ \bibnamefont {Pan}},\ }\bibfield  {title} {\bibinfo {title} {{Realization of an Error-Correcting Surface Code with Superconducting Qubits}},\ }\href {https://doi.org/10.1103/PhysRevLett.129.030501} {\bibfield  {journal} {\bibinfo  {journal} {Phys. Rev. Lett.}\ }\textbf {\bibinfo {volume} {129}},\ \bibinfo {pages} {030501} (\bibinfo {year} {2022})}\BibitemShut {NoStop}%
  \bibitem [{\citenamefont {Krinner}\ \emph {et~al.}(2022)\citenamefont {Krinner}, \citenamefont {Lacroix}, \citenamefont {Remm}, \citenamefont {Di~Paolo}, \citenamefont {Genois}, \citenamefont {Leroux}, \citenamefont {Hellings}, \citenamefont {Lazar}, \citenamefont {Swiadek}, \citenamefont {Herrmann}, \citenamefont {Norris}, \citenamefont {Andersen}, \citenamefont {M{\"u}ller}, \citenamefont {Blais}, \citenamefont {Eichler},\ and\ \citenamefont {Wallraff}}]{Krinner2022-td}%
    \BibitemOpen
    \bibfield  {author} {\bibinfo {author} {\bibfnamefont {S.}~\bibnamefont {Krinner}}, \bibinfo {author} {\bibfnamefont {N.}~\bibnamefont {Lacroix}}, \bibinfo {author} {\bibfnamefont {A.}~\bibnamefont {Remm}}, \bibinfo {author} {\bibfnamefont {A.}~\bibnamefont {Di~Paolo}}, \bibinfo {author} {\bibfnamefont {E.}~\bibnamefont {Genois}}, \bibinfo {author} {\bibfnamefont {C.}~\bibnamefont {Leroux}}, \bibinfo {author} {\bibfnamefont {C.}~\bibnamefont {Hellings}}, \bibinfo {author} {\bibfnamefont {S.}~\bibnamefont {Lazar}}, \bibinfo {author} {\bibfnamefont {F.}~\bibnamefont {Swiadek}}, \bibinfo {author} {\bibfnamefont {J.}~\bibnamefont {Herrmann}}, \bibinfo {author} {\bibfnamefont {G.~J.}\ \bibnamefont {Norris}}, \bibinfo {author} {\bibfnamefont {C.~K.}\ \bibnamefont {Andersen}}, \bibinfo {author} {\bibfnamefont {M.}~\bibnamefont {M{\"u}ller}}, \bibinfo {author} {\bibfnamefont {A.}~\bibnamefont {Blais}}, \bibinfo {author} {\bibfnamefont {C.}~\bibnamefont {Eichler}},\ and\ \bibinfo {author} {\bibfnamefont
    {A.}~\bibnamefont {Wallraff}},\ }\bibfield  {title} {\bibinfo {title} {Realizing repeated quantum error correction in a distance-three surface code},\ }\href {https://doi.org/10.1038/s41586-022-04566-8} {\bibfield  {journal} {\bibinfo  {journal} {Nature (London)}\ }\textbf {\bibinfo {volume} {605}},\ \bibinfo {pages} {669} (\bibinfo {year} {2022})}\BibitemShut {NoStop}%
  \bibitem [{\citenamefont {Egan}\ \emph {et~al.}(2021)\citenamefont {Egan}, \citenamefont {Debroy}, \citenamefont {Noel}, \citenamefont {Risinger}, \citenamefont {Zhu}, \citenamefont {Biswas}, \citenamefont {Newman}, \citenamefont {Li}, \citenamefont {Brown}, \citenamefont {Cetina},\ and\ \citenamefont {Monroe}}]{Egan2021-ai}%
    \BibitemOpen
    \bibfield  {author} {\bibinfo {author} {\bibfnamefont {L.}~\bibnamefont {Egan}}, \bibinfo {author} {\bibfnamefont {D.~M.}\ \bibnamefont {Debroy}}, \bibinfo {author} {\bibfnamefont {C.}~\bibnamefont {Noel}}, \bibinfo {author} {\bibfnamefont {A.}~\bibnamefont {Risinger}}, \bibinfo {author} {\bibfnamefont {D.}~\bibnamefont {Zhu}}, \bibinfo {author} {\bibfnamefont {D.}~\bibnamefont {Biswas}}, \bibinfo {author} {\bibfnamefont {M.}~\bibnamefont {Newman}}, \bibinfo {author} {\bibfnamefont {M.}~\bibnamefont {Li}}, \bibinfo {author} {\bibfnamefont {K.~R.}\ \bibnamefont {Brown}}, \bibinfo {author} {\bibfnamefont {M.}~\bibnamefont {Cetina}},\ and\ \bibinfo {author} {\bibfnamefont {C.}~\bibnamefont {Monroe}},\ }\bibfield  {title} {\bibinfo {title} {Fault-tolerant control of an error-corrected qubit},\ }\href {https://doi.org/10.1038/s41586-021-03928-y} {\bibfield  {journal} {\bibinfo  {journal} {Nature (London)}\ }\textbf {\bibinfo {volume} {598}},\ \bibinfo {pages} {281} (\bibinfo {year} {2021})}\BibitemShut {NoStop}%
  \bibitem [{\citenamefont {da~Silva}\ \emph {et~al.}(2024)\citenamefont {da~Silva}, \citenamefont {Ryan-Anderson}, \citenamefont {Bello-Rivas}, \citenamefont {Chernoguzov}, \citenamefont {Dreiling}, \citenamefont {Foltz}, \citenamefont {Frachon}, \citenamefont {Gaebler}, \citenamefont {Gatterman}, \citenamefont {Grans-Samuelsson}, \citenamefont {Hayes}, \citenamefont {Hewitt}, \citenamefont {Johansen}, \citenamefont {Lucchetti}, \citenamefont {Mills}, \citenamefont {Moses}, \citenamefont {Neyenhuis}, \citenamefont {Paz}, \citenamefont {Pino}, \citenamefont {Siegfried}, \citenamefont {Strabley}, \citenamefont {Sundaram}, \citenamefont {Tom}, \citenamefont {Wernli}, \citenamefont {Zanner}, \citenamefont {Stutz},\ and\ \citenamefont {Svore}}]{Da_Silva2024-ek}%
    \BibitemOpen
    \bibfield  {author} {\bibinfo {author} {\bibfnamefont {M.~P.}\ \bibnamefont {da~Silva}}, \bibinfo {author} {\bibfnamefont {C.}~\bibnamefont {Ryan-Anderson}}, \bibinfo {author} {\bibfnamefont {J.~M.}\ \bibnamefont {Bello-Rivas}}, \bibinfo {author} {\bibfnamefont {A.}~\bibnamefont {Chernoguzov}}, \bibinfo {author} {\bibfnamefont {J.~M.}\ \bibnamefont {Dreiling}}, \bibinfo {author} {\bibfnamefont {C.}~\bibnamefont {Foltz}}, \bibinfo {author} {\bibfnamefont {F.}~\bibnamefont {Frachon}}, \bibinfo {author} {\bibfnamefont {J.~P.}\ \bibnamefont {Gaebler}}, \bibinfo {author} {\bibfnamefont {T.~M.}\ \bibnamefont {Gatterman}}, \bibinfo {author} {\bibfnamefont {L.}~\bibnamefont {Grans-Samuelsson}}, \bibinfo {author} {\bibfnamefont {D.}~\bibnamefont {Hayes}}, \bibinfo {author} {\bibfnamefont {N.}~\bibnamefont {Hewitt}}, \bibinfo {author} {\bibfnamefont {J.}~\bibnamefont {Johansen}}, \bibinfo {author} {\bibfnamefont {D.}~\bibnamefont {Lucchetti}}, \bibinfo {author} {\bibfnamefont {M.}~\bibnamefont {Mills}}, \bibinfo
    {author} {\bibfnamefont {S.~A.}\ \bibnamefont {Moses}}, \bibinfo {author} {\bibfnamefont {B.}~\bibnamefont {Neyenhuis}}, \bibinfo {author} {\bibfnamefont {A.}~\bibnamefont {Paz}}, \bibinfo {author} {\bibfnamefont {J.}~\bibnamefont {Pino}}, \bibinfo {author} {\bibfnamefont {P.}~\bibnamefont {Siegfried}}, \bibinfo {author} {\bibfnamefont {J.}~\bibnamefont {Strabley}}, \bibinfo {author} {\bibfnamefont {A.}~\bibnamefont {Sundaram}}, \bibinfo {author} {\bibfnamefont {D.}~\bibnamefont {Tom}}, \bibinfo {author} {\bibfnamefont {S.~J.}\ \bibnamefont {Wernli}}, \bibinfo {author} {\bibfnamefont {M.}~\bibnamefont {Zanner}}, \bibinfo {author} {\bibfnamefont {R.~P.}\ \bibnamefont {Stutz}},\ and\ \bibinfo {author} {\bibfnamefont {K.~M.}\ \bibnamefont {Svore}},\ }\bibfield  {title} {\bibinfo {title} {Demonstration of logical qubits and repeated error correction with better-than-physical error rates},\ }\Eprint {https://arxiv.org/abs/2404.02280} {arXiv:2404.02280}  (\bibinfo {year} {2024})\BibitemShut {NoStop}%
  \bibitem [{\citenamefont {Huft}\ \emph {et~al.}(2022)\citenamefont {Huft}, \citenamefont {Song}, \citenamefont {Graham}, \citenamefont {Jooya}, \citenamefont {Deshpande}, \citenamefont {Fang}, \citenamefont {Kats},\ and\ \citenamefont {Saffman}}]{Huft2022-ke}%
    \BibitemOpen
    \bibfield  {author} {\bibinfo {author} {\bibfnamefont {P.}~\bibnamefont {Huft}}, \bibinfo {author} {\bibfnamefont {Y.}~\bibnamefont {Song}}, \bibinfo {author} {\bibfnamefont {T.~M.}\ \bibnamefont {Graham}}, \bibinfo {author} {\bibfnamefont {K.}~\bibnamefont {Jooya}}, \bibinfo {author} {\bibfnamefont {S.}~\bibnamefont {Deshpande}}, \bibinfo {author} {\bibfnamefont {C.}~\bibnamefont {Fang}}, \bibinfo {author} {\bibfnamefont {M.}~\bibnamefont {Kats}},\ and\ \bibinfo {author} {\bibfnamefont {M.}~\bibnamefont {Saffman}},\ }\bibfield  {title} {\bibinfo {title} {Simple, passive design for large optical trap arrays for single atoms},\ }\href {https://doi.org/10.1103/PhysRevA.105.063111} {\bibfield  {journal} {\bibinfo  {journal} {Phys. Rev. A}\ }\textbf {\bibinfo {volume} {105}},\ \bibinfo {pages} {063111} (\bibinfo {year} {2022})}\BibitemShut {NoStop}%
  \bibitem [{\citenamefont {Tao}\ \emph {et~al.}(2023)\citenamefont {Tao}, \citenamefont {Ammenwerth}, \citenamefont {Gyger}, \citenamefont {Bloch},\ and\ \citenamefont {Zeiher}}]{Tao2023-ri}%
    \BibitemOpen
    \bibfield  {author} {\bibinfo {author} {\bibfnamefont {R.}~\bibnamefont {Tao}}, \bibinfo {author} {\bibfnamefont {M.}~\bibnamefont {Ammenwerth}}, \bibinfo {author} {\bibfnamefont {F.}~\bibnamefont {Gyger}}, \bibinfo {author} {\bibfnamefont {I.}~\bibnamefont {Bloch}},\ and\ \bibinfo {author} {\bibfnamefont {J.}~\bibnamefont {Zeiher}},\ }\bibfield  {title} {\bibinfo {title} {High-fidelity detection of large-scale atom arrays in an optical lattice},\ }\Eprint {https://arxiv.org/abs/2309.04717} {arXiv:2309.04717}  (\bibinfo {year} {2023})\BibitemShut {NoStop}%
  \bibitem [{\citenamefont {Pause}\ \emph {et~al.}(2024)\citenamefont {Pause}, \citenamefont {Sturm}, \citenamefont {Mittenb{\"u}hler}, \citenamefont {Amann}, \citenamefont {Preuschoff}, \citenamefont {Sch{\"a}ffner}, \citenamefont {Schlosser},\ and\ \citenamefont {Birkl}}]{Pause2024-wk}%
    \BibitemOpen
    \bibfield  {author} {\bibinfo {author} {\bibfnamefont {L.}~\bibnamefont {Pause}}, \bibinfo {author} {\bibfnamefont {L.}~\bibnamefont {Sturm}}, \bibinfo {author} {\bibfnamefont {M.}~\bibnamefont {Mittenb{\"u}hler}}, \bibinfo {author} {\bibfnamefont {S.}~\bibnamefont {Amann}}, \bibinfo {author} {\bibfnamefont {T.}~\bibnamefont {Preuschoff}}, \bibinfo {author} {\bibfnamefont {D.}~\bibnamefont {Sch{\"a}ffner}}, \bibinfo {author} {\bibfnamefont {M.}~\bibnamefont {Schlosser}},\ and\ \bibinfo {author} {\bibfnamefont {G.}~\bibnamefont {Birkl}},\ }\bibfield  {title} {\bibinfo {title} {Supercharged two-dimensional tweezer array with more than 1000 atomic qubits},\ }\href {https://doi.org/10.1364/OPTICA.513551} {\bibfield  {journal} {\bibinfo  {journal} {Optica}\ }\textbf {\bibinfo {volume} {11}},\ \bibinfo {pages} {222} (\bibinfo {year} {2024})}\BibitemShut {NoStop}%
  \bibitem [{\citenamefont {Norcia}\ \emph {et~al.}(2024)\citenamefont {Norcia}, \citenamefont {Kim}, \citenamefont {Cairncross}, \citenamefont {Stone}, \citenamefont {Ryou}, \citenamefont {Jaffe}, \citenamefont {Brown}, \citenamefont {Barnes}, \citenamefont {Battaglino}, \citenamefont {Brown}, \citenamefont {Cassella}, \citenamefont {Chen}, \citenamefont {Coxe}, \citenamefont {Crow}, \citenamefont {Epstein}, \citenamefont {Griger}, \citenamefont {Halperin}, \citenamefont {Hummel}, \citenamefont {Jones}, \citenamefont {Kindem}, \citenamefont {King}, \citenamefont {Kotru}, \citenamefont {Lauigan}, \citenamefont {Li}, \citenamefont {Lu}, \citenamefont {Megidish}, \citenamefont {Marjanovic}, \citenamefont {McDonald}, \citenamefont {Mittiga}, \citenamefont {Muniz}, \citenamefont {Narayanaswami}, \citenamefont {Nishiguchi}, \citenamefont {Paule}, \citenamefont {Pawlak}, \citenamefont {Peng}, \citenamefont {Pudenz}, \citenamefont {Smull}, \citenamefont {Stack}, \citenamefont {Urbanek}, \citenamefont {van~de
    Veerdonk}, \citenamefont {Vendeiro}, \citenamefont {Wadleigh}, \citenamefont {Wilkason}, \citenamefont {Wu}, \citenamefont {Xie}, \citenamefont {Zalys-Geller}, \citenamefont {Zhang},\ and\ \citenamefont {Bloom}}]{Norcia2024-zh}%
    \BibitemOpen
    \bibfield  {author} {\bibinfo {author} {\bibfnamefont {M.~A.}\ \bibnamefont {Norcia}}, \bibinfo {author} {\bibfnamefont {H.}~\bibnamefont {Kim}}, \bibinfo {author} {\bibfnamefont {W.~B.}\ \bibnamefont {Cairncross}}, \bibinfo {author} {\bibfnamefont {M.}~\bibnamefont {Stone}}, \bibinfo {author} {\bibfnamefont {A.}~\bibnamefont {Ryou}}, \bibinfo {author} {\bibfnamefont {M.}~\bibnamefont {Jaffe}}, \bibinfo {author} {\bibfnamefont {M.~O.}\ \bibnamefont {Brown}}, \bibinfo {author} {\bibfnamefont {K.}~\bibnamefont {Barnes}}, \bibinfo {author} {\bibfnamefont {P.}~\bibnamefont {Battaglino}}, \bibinfo {author} {\bibfnamefont {A.}~\bibnamefont {Brown}}, \bibinfo {author} {\bibfnamefont {K.}~\bibnamefont {Cassella}}, \bibinfo {author} {\bibfnamefont {C.-A.}\ \bibnamefont {Chen}}, \bibinfo {author} {\bibfnamefont {R.}~\bibnamefont {Coxe}}, \bibinfo {author} {\bibfnamefont {D.}~\bibnamefont {Crow}}, \bibinfo {author} {\bibfnamefont {J.}~\bibnamefont {Epstein}}, \bibinfo {author} {\bibfnamefont {C.}~\bibnamefont
    {Griger}}, \bibinfo {author} {\bibfnamefont {E.}~\bibnamefont {Halperin}}, \bibinfo {author} {\bibfnamefont {F.}~\bibnamefont {Hummel}}, \bibinfo {author} {\bibfnamefont {A.~M.~W.}\ \bibnamefont {Jones}}, \bibinfo {author} {\bibfnamefont {J.~M.}\ \bibnamefont {Kindem}}, \bibinfo {author} {\bibfnamefont {J.}~\bibnamefont {King}}, \bibinfo {author} {\bibfnamefont {K.}~\bibnamefont {Kotru}}, \bibinfo {author} {\bibfnamefont {J.}~\bibnamefont {Lauigan}}, \bibinfo {author} {\bibfnamefont {M.}~\bibnamefont {Li}}, \bibinfo {author} {\bibfnamefont {M.}~\bibnamefont {Lu}}, \bibinfo {author} {\bibfnamefont {E.}~\bibnamefont {Megidish}}, \bibinfo {author} {\bibfnamefont {J.}~\bibnamefont {Marjanovic}}, \bibinfo {author} {\bibfnamefont {M.}~\bibnamefont {McDonald}}, \bibinfo {author} {\bibfnamefont {T.}~\bibnamefont {Mittiga}}, \bibinfo {author} {\bibfnamefont {J.~A.}\ \bibnamefont {Muniz}}, \bibinfo {author} {\bibfnamefont {S.}~\bibnamefont {Narayanaswami}}, \bibinfo {author} {\bibfnamefont {C.}~\bibnamefont
    {Nishiguchi}}, \bibinfo {author} {\bibfnamefont {T.}~\bibnamefont {Paule}}, \bibinfo {author} {\bibfnamefont {K.~A.}\ \bibnamefont {Pawlak}}, \bibinfo {author} {\bibfnamefont {L.~S.}\ \bibnamefont {Peng}}, \bibinfo {author} {\bibfnamefont {K.~L.}\ \bibnamefont {Pudenz}}, \bibinfo {author} {\bibfnamefont {A.}~\bibnamefont {Smull}}, \bibinfo {author} {\bibfnamefont {D.}~\bibnamefont {Stack}}, \bibinfo {author} {\bibfnamefont {M.}~\bibnamefont {Urbanek}}, \bibinfo {author} {\bibfnamefont {R.~J.~M.}\ \bibnamefont {van~de Veerdonk}}, \bibinfo {author} {\bibfnamefont {Z.}~\bibnamefont {Vendeiro}}, \bibinfo {author} {\bibfnamefont {L.}~\bibnamefont {Wadleigh}}, \bibinfo {author} {\bibfnamefont {T.}~\bibnamefont {Wilkason}}, \bibinfo {author} {\bibfnamefont {T.-Y.}\ \bibnamefont {Wu}}, \bibinfo {author} {\bibfnamefont {X.}~\bibnamefont {Xie}}, \bibinfo {author} {\bibfnamefont {E.}~\bibnamefont {Zalys-Geller}}, \bibinfo {author} {\bibfnamefont {X.}~\bibnamefont {Zhang}},\ and\ \bibinfo {author} {\bibfnamefont
    {B.~J.}\ \bibnamefont {Bloom}},\ }\bibfield  {title} {\bibinfo {title} {Iterative assembly of {$^{171}$Yb} atom arrays in cavity-enhanced optical lattices},\ }\Eprint {https://arxiv.org/abs/2401.16177} {arXiv:2401.16177}  (\bibinfo {year} {2024})\BibitemShut {NoStop}%
  \bibitem [{\citenamefont {Manetsch}\ \emph {et~al.}(2024)\citenamefont {Manetsch}, \citenamefont {Nomura}, \citenamefont {Bataille}, \citenamefont {Leung}, \citenamefont {Lv},\ and\ \citenamefont {Endres}}]{Manetsch2024-sp}%
    \BibitemOpen
    \bibfield  {author} {\bibinfo {author} {\bibfnamefont {H.~J.}\ \bibnamefont {Manetsch}}, \bibinfo {author} {\bibfnamefont {G.}~\bibnamefont {Nomura}}, \bibinfo {author} {\bibfnamefont {E.}~\bibnamefont {Bataille}}, \bibinfo {author} {\bibfnamefont {K.~H.}\ \bibnamefont {Leung}}, \bibinfo {author} {\bibfnamefont {X.}~\bibnamefont {Lv}},\ and\ \bibinfo {author} {\bibfnamefont {M.}~\bibnamefont {Endres}},\ }\bibfield  {title} {\bibinfo {title} {A tweezer array with 6100 highly coherent atomic qubits},\ }\Eprint {https://arxiv.org/abs/2403.12021} {arXiv:2403.12021}  (\bibinfo {year} {2024})\BibitemShut {NoStop}%
  \bibitem [{\citenamefont {Pichard}\ \emph {et~al.}(2024)\citenamefont {Pichard}, \citenamefont {Lim}, \citenamefont {Bloch}, \citenamefont {Vaneecloo}, \citenamefont {Bourachot}, \citenamefont {Both}, \citenamefont {M\'{e}riaux}, \citenamefont {Dutartre}, \citenamefont {Hostein}, \citenamefont {Paris}, \citenamefont {Ximenez}, \citenamefont {Signoles}, \citenamefont {Browaeys}, \citenamefont {Lahaye},\ and\ \citenamefont {Dreon}}]{Pichard2024-ba}%
    \BibitemOpen
    \bibfield  {author} {\bibinfo {author} {\bibfnamefont {G.}~\bibnamefont {Pichard}}, \bibinfo {author} {\bibfnamefont {D.}~\bibnamefont {Lim}}, \bibinfo {author} {\bibfnamefont {E.}~\bibnamefont {Bloch}}, \bibinfo {author} {\bibfnamefont {J.}~\bibnamefont {Vaneecloo}}, \bibinfo {author} {\bibfnamefont {L.}~\bibnamefont {Bourachot}}, \bibinfo {author} {\bibfnamefont {G.-J.}\ \bibnamefont {Both}}, \bibinfo {author} {\bibfnamefont {G.}~\bibnamefont {M\'{e}riaux}}, \bibinfo {author} {\bibfnamefont {S.}~\bibnamefont {Dutartre}}, \bibinfo {author} {\bibfnamefont {R.}~\bibnamefont {Hostein}}, \bibinfo {author} {\bibfnamefont {J.}~\bibnamefont {Paris}}, \bibinfo {author} {\bibfnamefont {B.}~\bibnamefont {Ximenez}}, \bibinfo {author} {\bibfnamefont {A.}~\bibnamefont {Signoles}}, \bibinfo {author} {\bibfnamefont {A.}~\bibnamefont {Browaeys}}, \bibinfo {author} {\bibfnamefont {T.}~\bibnamefont {Lahaye}},\ and\ \bibinfo {author} {\bibfnamefont {D.}~\bibnamefont {Dreon}},\ }\bibfield  {title} {\bibinfo {title}
    {Rearrangement of individual atoms in a 2000-site optical-tweezer array at cryogenic temperatures},\ }\href {https://doi.org/10.1103/physrevapplied.22.024073} {\bibfield  {journal} {\bibinfo  {journal} {Phys. Rev. Appl.}\ }\textbf {\bibinfo {volume} {22}},\ \bibinfo {pages} {024073} (\bibinfo {year} {2024})}\BibitemShut {NoStop}%
  \bibitem [{\citenamefont {Evered}\ \emph {et~al.}(2023)\citenamefont {Evered}, \citenamefont {Bluvstein}, \citenamefont {Kalinowski}, \citenamefont {Ebadi}, \citenamefont {Manovitz}, \citenamefont {Zhou}, \citenamefont {Li}, \citenamefont {Geim}, \citenamefont {Wang}, \citenamefont {Maskara}, \citenamefont {Levine}, \citenamefont {Semeghini}, \citenamefont {Greiner}, \citenamefont {Vuleti{\'c}},\ and\ \citenamefont {Lukin}}]{Evered2023-nv}%
    \BibitemOpen
    \bibfield  {author} {\bibinfo {author} {\bibfnamefont {S.~J.}\ \bibnamefont {Evered}}, \bibinfo {author} {\bibfnamefont {D.}~\bibnamefont {Bluvstein}}, \bibinfo {author} {\bibfnamefont {M.}~\bibnamefont {Kalinowski}}, \bibinfo {author} {\bibfnamefont {S.}~\bibnamefont {Ebadi}}, \bibinfo {author} {\bibfnamefont {T.}~\bibnamefont {Manovitz}}, \bibinfo {author} {\bibfnamefont {H.}~\bibnamefont {Zhou}}, \bibinfo {author} {\bibfnamefont {S.~H.}\ \bibnamefont {Li}}, \bibinfo {author} {\bibfnamefont {A.~A.}\ \bibnamefont {Geim}}, \bibinfo {author} {\bibfnamefont {T.~T.}\ \bibnamefont {Wang}}, \bibinfo {author} {\bibfnamefont {N.}~\bibnamefont {Maskara}}, \bibinfo {author} {\bibfnamefont {H.}~\bibnamefont {Levine}}, \bibinfo {author} {\bibfnamefont {G.}~\bibnamefont {Semeghini}}, \bibinfo {author} {\bibfnamefont {M.}~\bibnamefont {Greiner}}, \bibinfo {author} {\bibfnamefont {V.}~\bibnamefont {Vuleti{\'c}}},\ and\ \bibinfo {author} {\bibfnamefont {M.~D.}\ \bibnamefont {Lukin}},\ }\bibfield  {title} {\bibinfo {title}
    {High-fidelity parallel entangling gates on a neutral-atom quantum computer},\ }\href {https://doi.org/10.1038/s41586-023-06481-y} {\bibfield  {journal} {\bibinfo  {journal} {Nature (London)}\ }\textbf {\bibinfo {volume} {622}},\ \bibinfo {pages} {268} (\bibinfo {year} {2023})}\BibitemShut {NoStop}%
  \bibitem [{\citenamefont {Scholl}\ \emph {et~al.}(2023)\citenamefont {Scholl}, \citenamefont {Shaw}, \citenamefont {Tsai}, \citenamefont {Finkelstein}, \citenamefont {Choi},\ and\ \citenamefont {Endres}}]{Scholl2023-jb}%
    \BibitemOpen
    \bibfield  {author} {\bibinfo {author} {\bibfnamefont {P.}~\bibnamefont {Scholl}}, \bibinfo {author} {\bibfnamefont {A.~L.}\ \bibnamefont {Shaw}}, \bibinfo {author} {\bibfnamefont {R.~B.-S.}\ \bibnamefont {Tsai}}, \bibinfo {author} {\bibfnamefont {R.}~\bibnamefont {Finkelstein}}, \bibinfo {author} {\bibfnamefont {J.}~\bibnamefont {Choi}},\ and\ \bibinfo {author} {\bibfnamefont {M.}~\bibnamefont {Endres}},\ }\bibfield  {title} {\bibinfo {title} {{Erasure conversion in a high-fidelity Rydberg quantum simulator}},\ }\href {https://doi.org/10.1038/s41586-023-06516-4} {\bibfield  {journal} {\bibinfo  {journal} {Nature (London)}\ }\textbf {\bibinfo {volume} {622}},\ \bibinfo {pages} {273} (\bibinfo {year} {2023})}\BibitemShut {NoStop}%
  \bibitem [{\citenamefont {Peper}\ \emph {et~al.}(2024)\citenamefont {Peper}, \citenamefont {Li}, \citenamefont {Knapp}, \citenamefont {Bileska}, \citenamefont {Ma}, \citenamefont {Liu}, \citenamefont {Peng}, \citenamefont {Zhang}, \citenamefont {Horvath}, \citenamefont {Burgers},\ and\ \citenamefont {Thompson}}]{Peper2024-wc}%
    \BibitemOpen
    \bibfield  {author} {\bibinfo {author} {\bibfnamefont {M.}~\bibnamefont {Peper}}, \bibinfo {author} {\bibfnamefont {Y.}~\bibnamefont {Li}}, \bibinfo {author} {\bibfnamefont {D.~Y.}\ \bibnamefont {Knapp}}, \bibinfo {author} {\bibfnamefont {M.}~\bibnamefont {Bileska}}, \bibinfo {author} {\bibfnamefont {S.}~\bibnamefont {Ma}}, \bibinfo {author} {\bibfnamefont {G.}~\bibnamefont {Liu}}, \bibinfo {author} {\bibfnamefont {P.}~\bibnamefont {Peng}}, \bibinfo {author} {\bibfnamefont {B.}~\bibnamefont {Zhang}}, \bibinfo {author} {\bibfnamefont {S.~P.}\ \bibnamefont {Horvath}}, \bibinfo {author} {\bibfnamefont {A.~P.}\ \bibnamefont {Burgers}},\ and\ \bibinfo {author} {\bibfnamefont {J.~D.}\ \bibnamefont {Thompson}},\ }\bibfield  {title} {\bibinfo {title} {{Spectroscopy and modeling of ${}^{171}\mathrm{Yb}$ Rydberg states for high-fidelity two-qubit gates}},\ }\Eprint {https://arxiv.org/abs/2406.01482} {arXiv:2406.01482}  (\bibinfo {year} {2024})\BibitemShut {NoStop}%
  \bibitem [{\citenamefont {Tsai}\ \emph {et~al.}(2024)\citenamefont {Tsai}, \citenamefont {Sun}, \citenamefont {Shaw}, \citenamefont {Finkelstein},\ and\ \citenamefont {Endres}}]{Tsai2024-tu}%
    \BibitemOpen
    \bibfield  {author} {\bibinfo {author} {\bibfnamefont {R.~B.-S.}\ \bibnamefont {Tsai}}, \bibinfo {author} {\bibfnamefont {X.}~\bibnamefont {Sun}}, \bibinfo {author} {\bibfnamefont {A.~L.}\ \bibnamefont {Shaw}}, \bibinfo {author} {\bibfnamefont {R.}~\bibnamefont {Finkelstein}},\ and\ \bibinfo {author} {\bibfnamefont {M.}~\bibnamefont {Endres}},\ }\bibfield  {title} {\bibinfo {title} {{Benchmarking and linear response modeling of high-fidelity Rydberg gates}},\ }\Eprint {https://arxiv.org/abs/2407.20184} {arXiv:2407.20184}  (\bibinfo {year} {2024})\BibitemShut {NoStop}%
  \bibitem [{\citenamefont {Radnaev}\ \emph {et~al.}(2024)\citenamefont {Radnaev}, \citenamefont {Chung}, \citenamefont {Cole}, \citenamefont {Mason}, \citenamefont {Ballance}, \citenamefont {Bedalov}, \citenamefont {Belknap}, \citenamefont {Berman}, \citenamefont {Blakely}, \citenamefont {Bloomfield}, \citenamefont {Buttler}, \citenamefont {Campbell}, \citenamefont {Chopinaud}, \citenamefont {Copenhaver}, \citenamefont {Dawes}, \citenamefont {Eubanks}, \citenamefont {Friss}, \citenamefont {Garcia}, \citenamefont {Gilbert}, \citenamefont {Gillette}, \citenamefont {Goiporia}, \citenamefont {Gokhale}, \citenamefont {Goldwin}, \citenamefont {Goodwin}, \citenamefont {Graham}, \citenamefont {Guttormsson}, \citenamefont {Hickman}, \citenamefont {Hurtley}, \citenamefont {Iliev}, \citenamefont {Jones}, \citenamefont {Jones}, \citenamefont {Kuper}, \citenamefont {Lewis}, \citenamefont {Lichtman}, \citenamefont {Majdeteimouri}, \citenamefont {Mason}, \citenamefont {McMaster}, \citenamefont {Miles}, \citenamefont
    {Mitchell}, \citenamefont {Murphree}, \citenamefont {Neff-Mallon}, \citenamefont {Oh}, \citenamefont {Omole}, \citenamefont {Simon}, \citenamefont {Pederson}, \citenamefont {Perlin}, \citenamefont {Reiter}, \citenamefont {Rines}, \citenamefont {Romlow}, \citenamefont {Scott}, \citenamefont {Stiefvater}, \citenamefont {Tanner}, \citenamefont {Tucker}, \citenamefont {Vinogradov}, \citenamefont {Warter}, \citenamefont {Yeo}, \citenamefont {Saffman},\ and\ \citenamefont {Noel}}]{Radnaev2024-ac}%
    \BibitemOpen
    \bibfield  {author} {\bibinfo {author} {\bibfnamefont {A.~G.}\ \bibnamefont {Radnaev}}, \bibinfo {author} {\bibfnamefont {W.~C.}\ \bibnamefont {Chung}}, \bibinfo {author} {\bibfnamefont {D.~C.}\ \bibnamefont {Cole}}, \bibinfo {author} {\bibfnamefont {D.}~\bibnamefont {Mason}}, \bibinfo {author} {\bibfnamefont {T.~G.}\ \bibnamefont {Ballance}}, \bibinfo {author} {\bibfnamefont {M.~J.}\ \bibnamefont {Bedalov}}, \bibinfo {author} {\bibfnamefont {D.~A.}\ \bibnamefont {Belknap}}, \bibinfo {author} {\bibfnamefont {M.~R.}\ \bibnamefont {Berman}}, \bibinfo {author} {\bibfnamefont {M.}~\bibnamefont {Blakely}}, \bibinfo {author} {\bibfnamefont {I.~L.}\ \bibnamefont {Bloomfield}}, \bibinfo {author} {\bibfnamefont {P.~D.}\ \bibnamefont {Buttler}}, \bibinfo {author} {\bibfnamefont {C.}~\bibnamefont {Campbell}}, \bibinfo {author} {\bibfnamefont {A.}~\bibnamefont {Chopinaud}}, \bibinfo {author} {\bibfnamefont {E.}~\bibnamefont {Copenhaver}}, \bibinfo {author} {\bibfnamefont {M.~K.}\ \bibnamefont {Dawes}}, \bibinfo
    {author} {\bibfnamefont {S.~Y.}\ \bibnamefont {Eubanks}}, \bibinfo {author} {\bibfnamefont {A.~J.}\ \bibnamefont {Friss}}, \bibinfo {author} {\bibfnamefont {D.~M.}\ \bibnamefont {Garcia}}, \bibinfo {author} {\bibfnamefont {J.}~\bibnamefont {Gilbert}}, \bibinfo {author} {\bibfnamefont {M.}~\bibnamefont {Gillette}}, \bibinfo {author} {\bibfnamefont {P.}~\bibnamefont {Goiporia}}, \bibinfo {author} {\bibfnamefont {P.}~\bibnamefont {Gokhale}}, \bibinfo {author} {\bibfnamefont {J.}~\bibnamefont {Goldwin}}, \bibinfo {author} {\bibfnamefont {D.}~\bibnamefont {Goodwin}}, \bibinfo {author} {\bibfnamefont {T.~M.}\ \bibnamefont {Graham}}, \bibinfo {author} {\bibfnamefont {C.~J.}\ \bibnamefont {Guttormsson}}, \bibinfo {author} {\bibfnamefont {G.~T.}\ \bibnamefont {Hickman}}, \bibinfo {author} {\bibfnamefont {L.}~\bibnamefont {Hurtley}}, \bibinfo {author} {\bibfnamefont {M.}~\bibnamefont {Iliev}}, \bibinfo {author} {\bibfnamefont {E.~B.}\ \bibnamefont {Jones}}, \bibinfo {author} {\bibfnamefont {R.~A.}\ \bibnamefont
    {Jones}}, \bibinfo {author} {\bibfnamefont {K.~W.}\ \bibnamefont {Kuper}}, \bibinfo {author} {\bibfnamefont {T.~B.}\ \bibnamefont {Lewis}}, \bibinfo {author} {\bibfnamefont {M.~T.}\ \bibnamefont {Lichtman}}, \bibinfo {author} {\bibfnamefont {F.}~\bibnamefont {Majdeteimouri}}, \bibinfo {author} {\bibfnamefont {J.~J.}\ \bibnamefont {Mason}}, \bibinfo {author} {\bibfnamefont {J.~K.}\ \bibnamefont {McMaster}}, \bibinfo {author} {\bibfnamefont {J.~A.}\ \bibnamefont {Miles}}, \bibinfo {author} {\bibfnamefont {P.~T.}\ \bibnamefont {Mitchell}}, \bibinfo {author} {\bibfnamefont {J.~D.}\ \bibnamefont {Murphree}}, \bibinfo {author} {\bibfnamefont {N.~A.}\ \bibnamefont {Neff-Mallon}}, \bibinfo {author} {\bibfnamefont {T.}~\bibnamefont {Oh}}, \bibinfo {author} {\bibfnamefont {V.}~\bibnamefont {Omole}}, \bibinfo {author} {\bibfnamefont {C.~P.}\ \bibnamefont {Simon}}, \bibinfo {author} {\bibfnamefont {N.}~\bibnamefont {Pederson}}, \bibinfo {author} {\bibfnamefont {M.~A.}\ \bibnamefont {Perlin}}, \bibinfo {author}
    {\bibfnamefont {A.}~\bibnamefont {Reiter}}, \bibinfo {author} {\bibfnamefont {R.}~\bibnamefont {Rines}}, \bibinfo {author} {\bibfnamefont {P.}~\bibnamefont {Romlow}}, \bibinfo {author} {\bibfnamefont {A.~M.}\ \bibnamefont {Scott}}, \bibinfo {author} {\bibfnamefont {D.}~\bibnamefont {Stiefvater}}, \bibinfo {author} {\bibfnamefont {J.~R.}\ \bibnamefont {Tanner}}, \bibinfo {author} {\bibfnamefont {A.~K.}\ \bibnamefont {Tucker}}, \bibinfo {author} {\bibfnamefont {I.~V.}\ \bibnamefont {Vinogradov}}, \bibinfo {author} {\bibfnamefont {M.~L.}\ \bibnamefont {Warter}}, \bibinfo {author} {\bibfnamefont {M.}~\bibnamefont {Yeo}}, \bibinfo {author} {\bibfnamefont {M.}~\bibnamefont {Saffman}},\ and\ \bibinfo {author} {\bibfnamefont {T.~W.}\ \bibnamefont {Noel}},\ }\bibfield  {title} {\bibinfo {title} {A universal neutral-atom quantum computer with individual optical addressing and non-destructive readout},\ }\Eprint {https://arxiv.org/abs/2408.08288} {arXiv:2408.08288}  (\bibinfo {year} {2024})\BibitemShut {NoStop}%
  \bibitem [{\citenamefont {Bluvstein}\ \emph {et~al.}(2022)\citenamefont {Bluvstein}, \citenamefont {Levine}, \citenamefont {Semeghini}, \citenamefont {Wang}, \citenamefont {Ebadi}, \citenamefont {Kalinowski}, \citenamefont {Keesling}, \citenamefont {Maskara}, \citenamefont {Pichler}, \citenamefont {Greiner}, \citenamefont {Vuleti{\'c}},\ and\ \citenamefont {Lukin}}]{Bluvstein2022-tf}%
    \BibitemOpen
    \bibfield  {author} {\bibinfo {author} {\bibfnamefont {D.}~\bibnamefont {Bluvstein}}, \bibinfo {author} {\bibfnamefont {H.}~\bibnamefont {Levine}}, \bibinfo {author} {\bibfnamefont {G.}~\bibnamefont {Semeghini}}, \bibinfo {author} {\bibfnamefont {T.~T.}\ \bibnamefont {Wang}}, \bibinfo {author} {\bibfnamefont {S.}~\bibnamefont {Ebadi}}, \bibinfo {author} {\bibfnamefont {M.}~\bibnamefont {Kalinowski}}, \bibinfo {author} {\bibfnamefont {A.}~\bibnamefont {Keesling}}, \bibinfo {author} {\bibfnamefont {N.}~\bibnamefont {Maskara}}, \bibinfo {author} {\bibfnamefont {H.}~\bibnamefont {Pichler}}, \bibinfo {author} {\bibfnamefont {M.}~\bibnamefont {Greiner}}, \bibinfo {author} {\bibfnamefont {V.}~\bibnamefont {Vuleti{\'c}}},\ and\ \bibinfo {author} {\bibfnamefont {M.~D.}\ \bibnamefont {Lukin}},\ }\bibfield  {title} {\bibinfo {title} {A quantum processor based on coherent transport of entangled atom arrays},\ }\href {https://doi.org/10.1038/s41586-022-04592-6} {\bibfield  {journal} {\bibinfo  {journal} {Nature
    (London)}\ }\textbf {\bibinfo {volume} {604}},\ \bibinfo {pages} {451} (\bibinfo {year} {2022})}\BibitemShut {NoStop}%
  \bibitem [{\citenamefont {Bluvstein}\ \emph {et~al.}(2024)\citenamefont {Bluvstein}, \citenamefont {Evered}, \citenamefont {Geim}, \citenamefont {Li}, \citenamefont {Zhou}, \citenamefont {Manovitz}, \citenamefont {Ebadi}, \citenamefont {Cain}, \citenamefont {Kalinowski}, \citenamefont {Hangleiter}, \citenamefont {Bonilla~Ataides}, \citenamefont {Maskara}, \citenamefont {Cong}, \citenamefont {Gao}, \citenamefont {Sales~Rodriguez}, \citenamefont {Karolyshyn}, \citenamefont {Semeghini}, \citenamefont {Gullans}, \citenamefont {Greiner}, \citenamefont {Vuleti{\'c}},\ and\ \citenamefont {Lukin}}]{Bluvstein2024-yx}%
    \BibitemOpen
    \bibfield  {author} {\bibinfo {author} {\bibfnamefont {D.}~\bibnamefont {Bluvstein}}, \bibinfo {author} {\bibfnamefont {S.~J.}\ \bibnamefont {Evered}}, \bibinfo {author} {\bibfnamefont {A.~A.}\ \bibnamefont {Geim}}, \bibinfo {author} {\bibfnamefont {S.~H.}\ \bibnamefont {Li}}, \bibinfo {author} {\bibfnamefont {H.}~\bibnamefont {Zhou}}, \bibinfo {author} {\bibfnamefont {T.}~\bibnamefont {Manovitz}}, \bibinfo {author} {\bibfnamefont {S.}~\bibnamefont {Ebadi}}, \bibinfo {author} {\bibfnamefont {M.}~\bibnamefont {Cain}}, \bibinfo {author} {\bibfnamefont {M.}~\bibnamefont {Kalinowski}}, \bibinfo {author} {\bibfnamefont {D.}~\bibnamefont {Hangleiter}}, \bibinfo {author} {\bibfnamefont {J.~P.}\ \bibnamefont {Bonilla~Ataides}}, \bibinfo {author} {\bibfnamefont {N.}~\bibnamefont {Maskara}}, \bibinfo {author} {\bibfnamefont {I.}~\bibnamefont {Cong}}, \bibinfo {author} {\bibfnamefont {X.}~\bibnamefont {Gao}}, \bibinfo {author} {\bibfnamefont {P.}~\bibnamefont {Sales~Rodriguez}}, \bibinfo {author} {\bibfnamefont
    {T.}~\bibnamefont {Karolyshyn}}, \bibinfo {author} {\bibfnamefont {G.}~\bibnamefont {Semeghini}}, \bibinfo {author} {\bibfnamefont {M.~J.}\ \bibnamefont {Gullans}}, \bibinfo {author} {\bibfnamefont {M.}~\bibnamefont {Greiner}}, \bibinfo {author} {\bibfnamefont {V.}~\bibnamefont {Vuleti{\'c}}},\ and\ \bibinfo {author} {\bibfnamefont {M.~D.}\ \bibnamefont {Lukin}},\ }\bibfield  {title} {\bibinfo {title} {Logical quantum processor based on reconfigurable atom arrays},\ }\href {https://doi.org/10.1038/s41586-023-06927-3} {\bibfield  {journal} {\bibinfo  {journal} {Nature (London)}\ }\textbf {\bibinfo {volume} {626}},\ \bibinfo {pages} {58} (\bibinfo {year} {2024})}\BibitemShut {NoStop}%
  \bibitem [{\citenamefont {Ma}\ \emph {et~al.}(2023)\citenamefont {Ma}, \citenamefont {Liu}, \citenamefont {Peng}, \citenamefont {Zhang}, \citenamefont {Jandura}, \citenamefont {Claes}, \citenamefont {Burgers}, \citenamefont {Pupillo}, \citenamefont {Puri},\ and\ \citenamefont {Thompson}}]{Ma2023-mo}%
    \BibitemOpen
    \bibfield  {author} {\bibinfo {author} {\bibfnamefont {S.}~\bibnamefont {Ma}}, \bibinfo {author} {\bibfnamefont {G.}~\bibnamefont {Liu}}, \bibinfo {author} {\bibfnamefont {P.}~\bibnamefont {Peng}}, \bibinfo {author} {\bibfnamefont {B.}~\bibnamefont {Zhang}}, \bibinfo {author} {\bibfnamefont {S.}~\bibnamefont {Jandura}}, \bibinfo {author} {\bibfnamefont {J.}~\bibnamefont {Claes}}, \bibinfo {author} {\bibfnamefont {A.~P.}\ \bibnamefont {Burgers}}, \bibinfo {author} {\bibfnamefont {G.}~\bibnamefont {Pupillo}}, \bibinfo {author} {\bibfnamefont {S.}~\bibnamefont {Puri}},\ and\ \bibinfo {author} {\bibfnamefont {J.~D.}\ \bibnamefont {Thompson}},\ }\bibfield  {title} {\bibinfo {title} {High-fidelity gates and mid-circuit erasure conversion in an atomic qubit},\ }\href {https://doi.org/10.1038/s41586-023-06438-1} {\bibfield  {journal} {\bibinfo  {journal} {Nature (London)}\ }\textbf {\bibinfo {volume} {622}},\ \bibinfo {pages} {279} (\bibinfo {year} {2023})}\BibitemShut {NoStop}%
  \bibitem [{\citenamefont {Wu}\ \emph {et~al.}(2022)\citenamefont {Wu}, \citenamefont {Kolkowitz}, \citenamefont {Puri},\ and\ \citenamefont {Thompson}}]{Wu2022-ze}%
    \BibitemOpen
    \bibfield  {author} {\bibinfo {author} {\bibfnamefont {Y.}~\bibnamefont {Wu}}, \bibinfo {author} {\bibfnamefont {S.}~\bibnamefont {Kolkowitz}}, \bibinfo {author} {\bibfnamefont {S.}~\bibnamefont {Puri}},\ and\ \bibinfo {author} {\bibfnamefont {J.~D.}\ \bibnamefont {Thompson}},\ }\bibfield  {title} {\bibinfo {title} {{Erasure conversion for fault-tolerant quantum computing in alkaline earth Rydberg atom arrays}},\ }\href {https://doi.org/10.1038/s41467-022-32094-6} {\bibfield  {journal} {\bibinfo  {journal} {Nat. Commun.}\ }\textbf {\bibinfo {volume} {13}},\ \bibinfo {pages} {4657} (\bibinfo {year} {2022})}\BibitemShut {NoStop}%
  \bibitem [{\citenamefont {Cong}\ \emph {et~al.}(2022)\citenamefont {Cong}, \citenamefont {Levine}, \citenamefont {Keesling}, \citenamefont {Bluvstein}, \citenamefont {Wang},\ and\ \citenamefont {Lukin}}]{Cong2022-bm}%
    \BibitemOpen
    \bibfield  {author} {\bibinfo {author} {\bibfnamefont {I.}~\bibnamefont {Cong}}, \bibinfo {author} {\bibfnamefont {H.}~\bibnamefont {Levine}}, \bibinfo {author} {\bibfnamefont {A.}~\bibnamefont {Keesling}}, \bibinfo {author} {\bibfnamefont {D.}~\bibnamefont {Bluvstein}}, \bibinfo {author} {\bibfnamefont {S.-T.}\ \bibnamefont {Wang}},\ and\ \bibinfo {author} {\bibfnamefont {M.~D.}\ \bibnamefont {Lukin}},\ }\bibfield  {title} {\bibinfo {title} {{Hardware-Efficient, Fault-Tolerant Quantum Computation with Rydberg Atoms}},\ }\href {https://doi.org/10.1103/PhysRevX.12.021049} {\bibfield  {journal} {\bibinfo  {journal} {Phys. Rev. X}\ }\textbf {\bibinfo {volume} {12}},\ \bibinfo {pages} {021049} (\bibinfo {year} {2022})}\BibitemShut {NoStop}%
  \bibitem [{\citenamefont {Xu}\ \emph {et~al.}(2023)\citenamefont {Xu}, \citenamefont {Pablo Bonilla~Ataides}, \citenamefont {Pattison}, \citenamefont {Raveendran}, \citenamefont {Bluvstein}, \citenamefont {Wurtz}, \citenamefont {Vasic}, \citenamefont {Lukin}, \citenamefont {Jiang},\ and\ \citenamefont {Zhou}}]{Xu2023-bt}%
    \BibitemOpen
    \bibfield  {author} {\bibinfo {author} {\bibfnamefont {Q.}~\bibnamefont {Xu}}, \bibinfo {author} {\bibfnamefont {J.}~\bibnamefont {Pablo Bonilla~Ataides}}, \bibinfo {author} {\bibfnamefont {C.~A.}\ \bibnamefont {Pattison}}, \bibinfo {author} {\bibfnamefont {N.}~\bibnamefont {Raveendran}}, \bibinfo {author} {\bibfnamefont {D.}~\bibnamefont {Bluvstein}}, \bibinfo {author} {\bibfnamefont {J.}~\bibnamefont {Wurtz}}, \bibinfo {author} {\bibfnamefont {B.}~\bibnamefont {Vasic}}, \bibinfo {author} {\bibfnamefont {M.~D.}\ \bibnamefont {Lukin}}, \bibinfo {author} {\bibfnamefont {L.}~\bibnamefont {Jiang}},\ and\ \bibinfo {author} {\bibfnamefont {H.}~\bibnamefont {Zhou}},\ }\bibfield  {title} {\bibinfo {title} {{Constant-Overhead Fault-Tolerant Quantum Computation with Reconfigurable Atom Arrays}},\ }\Eprint {https://arxiv.org/abs/2308.08648} {arXiv:2308.08648}  (\bibinfo {year} {2023})\BibitemShut {NoStop}%
  \bibitem [{\citenamefont {Sahay}\ \emph {et~al.}(2023)\citenamefont {Sahay}, \citenamefont {Jin}, \citenamefont {Claes}, \citenamefont {Thompson},\ and\ \citenamefont {Puri}}]{Sahay2023-eb}%
    \BibitemOpen
    \bibfield  {author} {\bibinfo {author} {\bibfnamefont {K.}~\bibnamefont {Sahay}}, \bibinfo {author} {\bibfnamefont {J.}~\bibnamefont {Jin}}, \bibinfo {author} {\bibfnamefont {J.}~\bibnamefont {Claes}}, \bibinfo {author} {\bibfnamefont {J.~D.}\ \bibnamefont {Thompson}},\ and\ \bibinfo {author} {\bibfnamefont {S.}~\bibnamefont {Puri}},\ }\bibfield  {title} {\bibinfo {title} {{High-Threshold Codes for Neutral-Atom Qubits with Biased Erasure Errors}},\ }\href {https://doi.org/10.1103/PhysRevX.13.041013} {\bibfield  {journal} {\bibinfo  {journal} {Phys. Rev. X}\ }\textbf {\bibinfo {volume} {13}},\ \bibinfo {pages} {041013} (\bibinfo {year} {2023})}\BibitemShut {NoStop}%
  \bibitem [{\citenamefont {Jia}\ \emph {et~al.}(2024)\citenamefont {Jia}, \citenamefont {Huie}, \citenamefont {Li}, \citenamefont {Sun}, \citenamefont {Hu}, \citenamefont {{Aakash}}, \citenamefont {Kogan}, \citenamefont {Karve}, \citenamefont {Lee},\ and\ \citenamefont {Covey}}]{Jia2024-ox}%
    \BibitemOpen
    \bibfield  {author} {\bibinfo {author} {\bibfnamefont {Z.}~\bibnamefont {Jia}}, \bibinfo {author} {\bibfnamefont {W.}~\bibnamefont {Huie}}, \bibinfo {author} {\bibfnamefont {L.}~\bibnamefont {Li}}, \bibinfo {author} {\bibfnamefont {W.~K.~C.}\ \bibnamefont {Sun}}, \bibinfo {author} {\bibfnamefont {X.}~\bibnamefont {Hu}}, \bibinfo {author} {\bibnamefont {{Aakash}}}, \bibinfo {author} {\bibfnamefont {H.}~\bibnamefont {Kogan}}, \bibinfo {author} {\bibfnamefont {A.}~\bibnamefont {Karve}}, \bibinfo {author} {\bibfnamefont {J.~Y.}\ \bibnamefont {Lee}},\ and\ \bibinfo {author} {\bibfnamefont {J.~P.}\ \bibnamefont {Covey}},\ }\bibfield  {title} {\bibinfo {title} {An architecture for two-qubit encoding in neutral ytterbium-171 atoms},\ }\Eprint {https://arxiv.org/abs/2402.13134} {arXiv:2402.13134}  (\bibinfo {year} {2024})\BibitemShut {NoStop}%
  \bibitem [{\citenamefont {Saffman}(2016)}]{Saffman2016-jh}%
    \BibitemOpen
    \bibfield  {author} {\bibinfo {author} {\bibfnamefont {M.}~\bibnamefont {Saffman}},\ }\bibfield  {title} {\bibinfo {title} {{Quantum computing with atomic qubits and Rydberg interactions: progress and challenges}},\ }\href {https://doi.org/10.1088/0953-4075/49/20/202001} {\bibfield  {journal} {\bibinfo  {journal} {J. Phys. B: At. Mol. Opt. Phys.}\ }\textbf {\bibinfo {volume} {49}},\ \bibinfo {pages} {202001} (\bibinfo {year} {2016})}\BibitemShut {NoStop}%
  \bibitem [{\citenamefont {Deist}\ \emph {et~al.}(2022)\citenamefont {Deist}, \citenamefont {Lu}, \citenamefont {Ho}, \citenamefont {Pasha}, \citenamefont {Zeiher}, \citenamefont {Yan},\ and\ \citenamefont {Stamper-Kurn}}]{Deist2022-io}%
    \BibitemOpen
    \bibfield  {author} {\bibinfo {author} {\bibfnamefont {E.}~\bibnamefont {Deist}}, \bibinfo {author} {\bibfnamefont {Y.-H.}\ \bibnamefont {Lu}}, \bibinfo {author} {\bibfnamefont {J.}~\bibnamefont {Ho}}, \bibinfo {author} {\bibfnamefont {M.~K.}\ \bibnamefont {Pasha}}, \bibinfo {author} {\bibfnamefont {J.}~\bibnamefont {Zeiher}}, \bibinfo {author} {\bibfnamefont {Z.}~\bibnamefont {Yan}},\ and\ \bibinfo {author} {\bibfnamefont {D.~M.}\ \bibnamefont {Stamper-Kurn}},\ }\bibfield  {title} {\bibinfo {title} {{Mid-Circuit Cavity Measurement in a Neutral Atom Array}},\ }\href {https://doi.org/10.1103/PhysRevLett.129.203602} {\bibfield  {journal} {\bibinfo  {journal} {Phys. Rev. Lett.}\ }\textbf {\bibinfo {volume} {129}},\ \bibinfo {pages} {203602} (\bibinfo {year} {2022})}\BibitemShut {NoStop}%
  \bibitem [{\citenamefont {Graham}\ \emph {et~al.}(2023)\citenamefont {Graham}, \citenamefont {Phuttitarn}, \citenamefont {Chinnarasu}, \citenamefont {Song}, \citenamefont {Poole}, \citenamefont {Jooya}, \citenamefont {Scott}, \citenamefont {Scott}, \citenamefont {Eichler},\ and\ \citenamefont {Saffman}}]{Graham2023-wl}%
    \BibitemOpen
    \bibfield  {author} {\bibinfo {author} {\bibfnamefont {T.~M.}\ \bibnamefont {Graham}}, \bibinfo {author} {\bibfnamefont {L.}~\bibnamefont {Phuttitarn}}, \bibinfo {author} {\bibfnamefont {R.}~\bibnamefont {Chinnarasu}}, \bibinfo {author} {\bibfnamefont {Y.}~\bibnamefont {Song}}, \bibinfo {author} {\bibfnamefont {C.}~\bibnamefont {Poole}}, \bibinfo {author} {\bibfnamefont {K.}~\bibnamefont {Jooya}}, \bibinfo {author} {\bibfnamefont {J.}~\bibnamefont {Scott}}, \bibinfo {author} {\bibfnamefont {A.}~\bibnamefont {Scott}}, \bibinfo {author} {\bibfnamefont {P.}~\bibnamefont {Eichler}},\ and\ \bibinfo {author} {\bibfnamefont {M.}~\bibnamefont {Saffman}},\ }\bibfield  {title} {\bibinfo {title} {{Midcircuit Measurements on a Single-Species Neutral Alkali Atom Quantum Processor}},\ }\href {https://doi.org/10.1103/PhysRevX.13.041051} {\bibfield  {journal} {\bibinfo  {journal} {Phys. Rev. X}\ }\textbf {\bibinfo {volume} {13}},\ \bibinfo {pages} {041051} (\bibinfo {year} {2023})}\BibitemShut {NoStop}%
  \bibitem [{\citenamefont {Norcia}\ \emph {et~al.}(2023)\citenamefont {Norcia}, \citenamefont {Cairncross}, \citenamefont {Barnes}, \citenamefont {Battaglino}, \citenamefont {Brown}, \citenamefont {Brown}, \citenamefont {Cassella}, \citenamefont {Chen}, \citenamefont {Coxe}, \citenamefont {Crow}, \citenamefont {Epstein}, \citenamefont {Griger}, \citenamefont {Jones}, \citenamefont {Kim}, \citenamefont {Kindem}, \citenamefont {King}, \citenamefont {Kondov}, \citenamefont {Kotru}, \citenamefont {Lauigan}, \citenamefont {Li}, \citenamefont {Lu}, \citenamefont {Megidish}, \citenamefont {Marjanovic}, \citenamefont {McDonald}, \citenamefont {Mittiga}, \citenamefont {Muniz}, \citenamefont {Narayanaswami}, \citenamefont {Nishiguchi}, \citenamefont {Notermans}, \citenamefont {Paule}, \citenamefont {Pawlak}, \citenamefont {Peng}, \citenamefont {Ryou}, \citenamefont {Smull}, \citenamefont {Stack}, \citenamefont {Stone}, \citenamefont {Sucich}, \citenamefont {Urbanek}, \citenamefont {van~de Veerdonk}, \citenamefont
    {Vendeiro}, \citenamefont {Wilkason}, \citenamefont {Wu}, \citenamefont {Xie}, \citenamefont {Zhang},\ and\ \citenamefont {Bloom}}]{Norcia2023-dt}%
    \BibitemOpen
    \bibfield  {author} {\bibinfo {author} {\bibfnamefont {M.~A.}\ \bibnamefont {Norcia}}, \bibinfo {author} {\bibfnamefont {W.~B.}\ \bibnamefont {Cairncross}}, \bibinfo {author} {\bibfnamefont {K.}~\bibnamefont {Barnes}}, \bibinfo {author} {\bibfnamefont {P.}~\bibnamefont {Battaglino}}, \bibinfo {author} {\bibfnamefont {A.}~\bibnamefont {Brown}}, \bibinfo {author} {\bibfnamefont {M.~O.}\ \bibnamefont {Brown}}, \bibinfo {author} {\bibfnamefont {K.}~\bibnamefont {Cassella}}, \bibinfo {author} {\bibfnamefont {C.-A.}\ \bibnamefont {Chen}}, \bibinfo {author} {\bibfnamefont {R.}~\bibnamefont {Coxe}}, \bibinfo {author} {\bibfnamefont {D.}~\bibnamefont {Crow}}, \bibinfo {author} {\bibfnamefont {J.}~\bibnamefont {Epstein}}, \bibinfo {author} {\bibfnamefont {C.}~\bibnamefont {Griger}}, \bibinfo {author} {\bibfnamefont {A.~M.~W.}\ \bibnamefont {Jones}}, \bibinfo {author} {\bibfnamefont {H.}~\bibnamefont {Kim}}, \bibinfo {author} {\bibfnamefont {J.~M.}\ \bibnamefont {Kindem}}, \bibinfo {author} {\bibfnamefont
    {J.}~\bibnamefont {King}}, \bibinfo {author} {\bibfnamefont {S.~S.}\ \bibnamefont {Kondov}}, \bibinfo {author} {\bibfnamefont {K.}~\bibnamefont {Kotru}}, \bibinfo {author} {\bibfnamefont {J.}~\bibnamefont {Lauigan}}, \bibinfo {author} {\bibfnamefont {M.}~\bibnamefont {Li}}, \bibinfo {author} {\bibfnamefont {M.}~\bibnamefont {Lu}}, \bibinfo {author} {\bibfnamefont {E.}~\bibnamefont {Megidish}}, \bibinfo {author} {\bibfnamefont {J.}~\bibnamefont {Marjanovic}}, \bibinfo {author} {\bibfnamefont {M.}~\bibnamefont {McDonald}}, \bibinfo {author} {\bibfnamefont {T.}~\bibnamefont {Mittiga}}, \bibinfo {author} {\bibfnamefont {J.~A.}\ \bibnamefont {Muniz}}, \bibinfo {author} {\bibfnamefont {S.}~\bibnamefont {Narayanaswami}}, \bibinfo {author} {\bibfnamefont {C.}~\bibnamefont {Nishiguchi}}, \bibinfo {author} {\bibfnamefont {R.}~\bibnamefont {Notermans}}, \bibinfo {author} {\bibfnamefont {T.}~\bibnamefont {Paule}}, \bibinfo {author} {\bibfnamefont {K.~A.}\ \bibnamefont {Pawlak}}, \bibinfo {author} {\bibfnamefont
    {L.~S.}\ \bibnamefont {Peng}}, \bibinfo {author} {\bibfnamefont {A.}~\bibnamefont {Ryou}}, \bibinfo {author} {\bibfnamefont {A.}~\bibnamefont {Smull}}, \bibinfo {author} {\bibfnamefont {D.}~\bibnamefont {Stack}}, \bibinfo {author} {\bibfnamefont {M.}~\bibnamefont {Stone}}, \bibinfo {author} {\bibfnamefont {A.}~\bibnamefont {Sucich}}, \bibinfo {author} {\bibfnamefont {M.}~\bibnamefont {Urbanek}}, \bibinfo {author} {\bibfnamefont {R.~J.~M.}\ \bibnamefont {van~de Veerdonk}}, \bibinfo {author} {\bibfnamefont {Z.}~\bibnamefont {Vendeiro}}, \bibinfo {author} {\bibfnamefont {T.}~\bibnamefont {Wilkason}}, \bibinfo {author} {\bibfnamefont {T.-Y.}\ \bibnamefont {Wu}}, \bibinfo {author} {\bibfnamefont {X.}~\bibnamefont {Xie}}, \bibinfo {author} {\bibfnamefont {X.}~\bibnamefont {Zhang}},\ and\ \bibinfo {author} {\bibfnamefont {B.~J.}\ \bibnamefont {Bloom}},\ }\bibfield  {title} {\bibinfo {title} {{Midcircuit Qubit Measurement and Rearrangement in a {$^{171}$Yb} Atomic Array}},\ }\href
    {https://doi.org/10.1103/PhysRevX.13.041034} {\bibfield  {journal} {\bibinfo  {journal} {Phys. Rev. X}\ }\textbf {\bibinfo {volume} {13}},\ \bibinfo {pages} {041034} (\bibinfo {year} {2023})}\BibitemShut {NoStop}%
  \bibitem [{\citenamefont {Lis}\ \emph {et~al.}(2023)\citenamefont {Lis}, \citenamefont {Senoo}, \citenamefont {McGrew}, \citenamefont {R{\"o}nchen}, \citenamefont {Jenkins},\ and\ \citenamefont {Kaufman}}]{Lis2023-cb}%
    \BibitemOpen
    \bibfield  {author} {\bibinfo {author} {\bibfnamefont {J.~W.}\ \bibnamefont {Lis}}, \bibinfo {author} {\bibfnamefont {A.}~\bibnamefont {Senoo}}, \bibinfo {author} {\bibfnamefont {W.~F.}\ \bibnamefont {McGrew}}, \bibinfo {author} {\bibfnamefont {F.}~\bibnamefont {R{\"o}nchen}}, \bibinfo {author} {\bibfnamefont {A.}~\bibnamefont {Jenkins}},\ and\ \bibinfo {author} {\bibfnamefont {A.~M.}\ \bibnamefont {Kaufman}},\ }\bibfield  {title} {\bibinfo {title} {{Midcircuit Operations Using the omg Architecture in Neutral Atom Arrays}},\ }\href {https://doi.org/10.1103/PhysRevX.13.041035} {\bibfield  {journal} {\bibinfo  {journal} {Phys. Rev. X}\ }\textbf {\bibinfo {volume} {13}},\ \bibinfo {pages} {041035} (\bibinfo {year} {2023})}\BibitemShut {NoStop}%
  \bibitem [{\citenamefont {Singh}\ \emph {et~al.}(2022)\citenamefont {Singh}, \citenamefont {Anand}, \citenamefont {Pocklington}, \citenamefont {Kemp},\ and\ \citenamefont {Bernien}}]{Singh2022-zg}%
    \BibitemOpen
    \bibfield  {author} {\bibinfo {author} {\bibfnamefont {K.}~\bibnamefont {Singh}}, \bibinfo {author} {\bibfnamefont {S.}~\bibnamefont {Anand}}, \bibinfo {author} {\bibfnamefont {A.}~\bibnamefont {Pocklington}}, \bibinfo {author} {\bibfnamefont {J.~T.}\ \bibnamefont {Kemp}},\ and\ \bibinfo {author} {\bibfnamefont {H.}~\bibnamefont {Bernien}},\ }\bibfield  {title} {\bibinfo {title} {{Dual-Element Two-Dimensional Atom Array with Continuous-Mode Operation}},\ }\href {https://doi.org/10.1103/PhysRevX.12.011040} {\bibfield  {journal} {\bibinfo  {journal} {Phys. Rev. X}\ }\textbf {\bibinfo {volume} {12}},\ \bibinfo {pages} {011040} (\bibinfo {year} {2022})}\BibitemShut {NoStop}%
  \bibitem [{\citenamefont {Singh}\ \emph {et~al.}(2023)\citenamefont {Singh}, \citenamefont {Bradley}, \citenamefont {Anand}, \citenamefont {Ramesh}, \citenamefont {White},\ and\ \citenamefont {Bernien}}]{Singh2023-ox}%
    \BibitemOpen
    \bibfield  {author} {\bibinfo {author} {\bibfnamefont {K.}~\bibnamefont {Singh}}, \bibinfo {author} {\bibfnamefont {C.~E.}\ \bibnamefont {Bradley}}, \bibinfo {author} {\bibfnamefont {S.}~\bibnamefont {Anand}}, \bibinfo {author} {\bibfnamefont {V.}~\bibnamefont {Ramesh}}, \bibinfo {author} {\bibfnamefont {R.}~\bibnamefont {White}},\ and\ \bibinfo {author} {\bibfnamefont {H.}~\bibnamefont {Bernien}},\ }\bibfield  {title} {\bibinfo {title} {Mid-circuit correction of correlated phase errors using an array of spectator qubits},\ }\href {https://doi.org/10.1126/science.ade5337} {\bibfield  {journal} {\bibinfo  {journal} {Science}\ }\textbf {\bibinfo {volume} {380}},\ \bibinfo {pages} {1265} (\bibinfo {year} {2023})}\BibitemShut {NoStop}%
  \bibitem [{\citenamefont {Zeng}\ \emph {et~al.}(2017)\citenamefont {Zeng}, \citenamefont {Xu}, \citenamefont {He}, \citenamefont {Liu}, \citenamefont {Liu}, \citenamefont {Wang}, \citenamefont {Papoular}, \citenamefont {Shlyapnikov},\ and\ \citenamefont {Zhan}}]{Zeng2017-mi}%
    \BibitemOpen
    \bibfield  {author} {\bibinfo {author} {\bibfnamefont {Y.}~\bibnamefont {Zeng}}, \bibinfo {author} {\bibfnamefont {P.}~\bibnamefont {Xu}}, \bibinfo {author} {\bibfnamefont {X.}~\bibnamefont {He}}, \bibinfo {author} {\bibfnamefont {Y.}~\bibnamefont {Liu}}, \bibinfo {author} {\bibfnamefont {M.}~\bibnamefont {Liu}}, \bibinfo {author} {\bibfnamefont {J.}~\bibnamefont {Wang}}, \bibinfo {author} {\bibfnamefont {D.~J.}\ \bibnamefont {Papoular}}, \bibinfo {author} {\bibfnamefont {G.~V.}\ \bibnamefont {Shlyapnikov}},\ and\ \bibinfo {author} {\bibfnamefont {M.}~\bibnamefont {Zhan}},\ }\bibfield  {title} {\bibinfo {title} {{Entangling Two Individual Atoms of Different Isotopes via Rydberg Blockade}},\ }\href {https://doi.org/10.1103/PhysRevLett.119.160502} {\bibfield  {journal} {\bibinfo  {journal} {Phys. Rev. Lett.}\ }\textbf {\bibinfo {volume} {119}},\ \bibinfo {pages} {160502} (\bibinfo {year} {2017})}\BibitemShut {NoStop}%
  \bibitem [{\citenamefont {Das}\ \emph {et~al.}(2005)\citenamefont {Das}, \citenamefont {Barthwal}, \citenamefont {Banerjee},\ and\ \citenamefont {Natarajan}}]{Das2005-fx}%
    \BibitemOpen
    \bibfield  {author} {\bibinfo {author} {\bibfnamefont {D.}~\bibnamefont {Das}}, \bibinfo {author} {\bibfnamefont {S.}~\bibnamefont {Barthwal}}, \bibinfo {author} {\bibfnamefont {A.}~\bibnamefont {Banerjee}},\ and\ \bibinfo {author} {\bibfnamefont {V.}~\bibnamefont {Natarajan}},\ }\bibfield  {title} {\bibinfo {title} {{Absolute frequency measurements in Yb with $0.08\phantom{\rule{0.3em}{0ex}}\mathrm{ppb}$ uncertainty: Isotope shifts and hyperfine structure in the $399\text{\ensuremath{-}}\mathrm{nm}$ ${^{1}S_{0}}\ensuremath{\rightarrow}{^{1}P_{1}}$ line}},\ }\href {https://doi.org/10.1103/PhysRevA.72.032506} {\bibfield  {journal} {\bibinfo  {journal} {Phys. Rev. A}\ }\textbf {\bibinfo {volume} {72}},\ \bibinfo {pages} {032506} (\bibinfo {year} {2005})}\BibitemShut {NoStop}%
  \bibitem [{\citenamefont {Pandey}\ \emph {et~al.}(2009)\citenamefont {Pandey}, \citenamefont {Singh}, \citenamefont {Kumar}, \citenamefont {Suryanarayana},\ and\ \citenamefont {Natarajan}}]{Pandey2009-fp}%
    \BibitemOpen
    \bibfield  {author} {\bibinfo {author} {\bibfnamefont {K.}~\bibnamefont {Pandey}}, \bibinfo {author} {\bibfnamefont {A.~K.}\ \bibnamefont {Singh}}, \bibinfo {author} {\bibfnamefont {P.~V.~K.}\ \bibnamefont {Kumar}}, \bibinfo {author} {\bibfnamefont {M.~V.}\ \bibnamefont {Suryanarayana}},\ and\ \bibinfo {author} {\bibfnamefont {V.}~\bibnamefont {Natarajan}},\ }\bibfield  {title} {\bibinfo {title} {{Isotope shifts and hyperfine structure in the 555.8-nm ${^{1}S}_{0}\ensuremath{\rightarrow}{^{3}P}_{1}$ line of Yb}},\ }\href {https://doi.org/10.1103/PhysRevA.80.022518} {\bibfield  {journal} {\bibinfo  {journal} {Phys. Rev. A}\ }\textbf {\bibinfo {volume} {80}},\ \bibinfo {pages} {022518} (\bibinfo {year} {2009})}\BibitemShut {NoStop}%
  \bibitem [{\citenamefont {Fowler}\ \emph {et~al.}(2012)\citenamefont {Fowler}, \citenamefont {Mariantoni}, \citenamefont {Martinis},\ and\ \citenamefont {Cleland}}]{Fowler2012-sb}%
    \BibitemOpen
    \bibfield  {author} {\bibinfo {author} {\bibfnamefont {A.~G.}\ \bibnamefont {Fowler}}, \bibinfo {author} {\bibfnamefont {M.}~\bibnamefont {Mariantoni}}, \bibinfo {author} {\bibfnamefont {J.~M.}\ \bibnamefont {Martinis}},\ and\ \bibinfo {author} {\bibfnamefont {A.~N.}\ \bibnamefont {Cleland}},\ }\bibfield  {title} {\bibinfo {title} {{Surface codes: Towards practical large-scale quantum computation}},\ }\href {https://doi.org/10.1103/PhysRevA.86.032324} {\bibfield  {journal} {\bibinfo  {journal} {Phys. Rev. A}\ }\textbf {\bibinfo {volume} {86}},\ \bibinfo {pages} {032324} (\bibinfo {year} {2012})}\BibitemShut {NoStop}%
  \bibitem [{\citenamefont {Jenkins}\ \emph {et~al.}(2022)\citenamefont {Jenkins}, \citenamefont {Lis}, \citenamefont {Senoo}, \citenamefont {McGrew},\ and\ \citenamefont {Kaufman}}]{Jenkins2022-lt}%
    \BibitemOpen
    \bibfield  {author} {\bibinfo {author} {\bibfnamefont {A.}~\bibnamefont {Jenkins}}, \bibinfo {author} {\bibfnamefont {J.~W.}\ \bibnamefont {Lis}}, \bibinfo {author} {\bibfnamefont {A.}~\bibnamefont {Senoo}}, \bibinfo {author} {\bibfnamefont {W.~F.}\ \bibnamefont {McGrew}},\ and\ \bibinfo {author} {\bibfnamefont {A.~M.}\ \bibnamefont {Kaufman}},\ }\bibfield  {title} {\bibinfo {title} {{Ytterbium Nuclear-Spin Qubits in an Optical Tweezer Array}},\ }\href {https://doi.org/10.1103/PhysRevX.12.021027} {\bibfield  {journal} {\bibinfo  {journal} {Phys. Rev. X}\ }\textbf {\bibinfo {volume} {12}},\ \bibinfo {pages} {021027} (\bibinfo {year} {2022})}\BibitemShut {NoStop}%
  \bibitem [{\citenamefont {Ma}\ \emph {et~al.}(2022)\citenamefont {Ma}, \citenamefont {Burgers}, \citenamefont {Liu}, \citenamefont {Wilson}, \citenamefont {Zhang},\ and\ \citenamefont {Thompson}}]{Ma2022-gy}%
    \BibitemOpen
    \bibfield  {author} {\bibinfo {author} {\bibfnamefont {S.}~\bibnamefont {Ma}}, \bibinfo {author} {\bibfnamefont {A.~P.}\ \bibnamefont {Burgers}}, \bibinfo {author} {\bibfnamefont {G.}~\bibnamefont {Liu}}, \bibinfo {author} {\bibfnamefont {J.}~\bibnamefont {Wilson}}, \bibinfo {author} {\bibfnamefont {B.}~\bibnamefont {Zhang}},\ and\ \bibinfo {author} {\bibfnamefont {J.~D.}\ \bibnamefont {Thompson}},\ }\bibfield  {title} {\bibinfo {title} {{Universal Gate Operations on Nuclear Spin Qubits in an Optical Tweezer Array of {$^{171}$Yb} Atoms}},\ }\href {https://doi.org/10.1103/PhysRevX.12.021028} {\bibfield  {journal} {\bibinfo  {journal} {Phys. Rev. X}\ }\textbf {\bibinfo {volume} {12}},\ \bibinfo {pages} {021028} (\bibinfo {year} {2022})}\BibitemShut {NoStop}%
  \bibitem [{\citenamefont {Schine}\ \emph {et~al.}(2022)\citenamefont {Schine}, \citenamefont {Young}, \citenamefont {Eckner}, \citenamefont {Martin},\ and\ \citenamefont {Kaufman}}]{Schine2022-sn}%
    \BibitemOpen
    \bibfield  {author} {\bibinfo {author} {\bibfnamefont {N.}~\bibnamefont {Schine}}, \bibinfo {author} {\bibfnamefont {A.~W.}\ \bibnamefont {Young}}, \bibinfo {author} {\bibfnamefont {W.~J.}\ \bibnamefont {Eckner}}, \bibinfo {author} {\bibfnamefont {M.~J.}\ \bibnamefont {Martin}},\ and\ \bibinfo {author} {\bibfnamefont {A.~M.}\ \bibnamefont {Kaufman}},\ }\bibfield  {title} {\bibinfo {title} {{Long-lived Bell states in an array of optical clock qubits}},\ }\href {https://doi.org/10.1038/s41567-022-01678-w} {\bibfield  {journal} {\bibinfo  {journal} {Nat. Phys.}\ }\textbf {\bibinfo {volume} {18}},\ \bibinfo {pages} {1067} (\bibinfo {year} {2022})}\BibitemShut {NoStop}%
  \bibitem [{\citenamefont {Yamamoto}\ \emph {et~al.}(2016)\citenamefont {Yamamoto}, \citenamefont {Kobayashi}, \citenamefont {Kuno}, \citenamefont {Kato},\ and\ \citenamefont {Takahashi}}]{Yamamoto2016-vk}%
    \BibitemOpen
    \bibfield  {author} {\bibinfo {author} {\bibfnamefont {R.}~\bibnamefont {Yamamoto}}, \bibinfo {author} {\bibfnamefont {J.}~\bibnamefont {Kobayashi}}, \bibinfo {author} {\bibfnamefont {T.}~\bibnamefont {Kuno}}, \bibinfo {author} {\bibfnamefont {K.}~\bibnamefont {Kato}},\ and\ \bibinfo {author} {\bibfnamefont {Y.}~\bibnamefont {Takahashi}},\ }\bibfield  {title} {\bibinfo {title} {An ytterbium quantum gas microscope with narrow-line laser cooling},\ }\href {https://doi.org/10.1088/1367-2630/18/2/023016} {\bibfield  {journal} {\bibinfo  {journal} {New J. Phys.}\ }\textbf {\bibinfo {volume} {18}},\ \bibinfo {pages} {023016} (\bibinfo {year} {2016})}\BibitemShut {NoStop}%
  \bibitem [{\citenamefont {Saskin}\ \emph {et~al.}(2019)\citenamefont {Saskin}, \citenamefont {Wilson}, \citenamefont {Grinkemeyer},\ and\ \citenamefont {Thompson}}]{Saskin2019-xk}%
    \BibitemOpen
    \bibfield  {author} {\bibinfo {author} {\bibfnamefont {S.}~\bibnamefont {Saskin}}, \bibinfo {author} {\bibfnamefont {J.~T.}\ \bibnamefont {Wilson}}, \bibinfo {author} {\bibfnamefont {B.}~\bibnamefont {Grinkemeyer}},\ and\ \bibinfo {author} {\bibfnamefont {J.~D.}\ \bibnamefont {Thompson}},\ }\bibfield  {title} {\bibinfo {title} {{Narrow-Line Cooling and Imaging of Ytterbium Atoms in an Optical Tweezer Array}},\ }\href {https://doi.org/10.1103/PhysRevLett.122.143002} {\bibfield  {journal} {\bibinfo  {journal} {Phys. Rev. Lett.}\ }\textbf {\bibinfo {volume} {122}},\ \bibinfo {pages} {143002} (\bibinfo {year} {2019})}\BibitemShut {NoStop}%
  \bibitem [{\citenamefont {Tomita}\ \emph {et~al.}(2019)\citenamefont {Tomita}, \citenamefont {Nakajima}, \citenamefont {Takasu},\ and\ \citenamefont {Takahashi}}]{Tomita2019-ie}%
    \BibitemOpen
    \bibfield  {author} {\bibinfo {author} {\bibfnamefont {T.}~\bibnamefont {Tomita}}, \bibinfo {author} {\bibfnamefont {S.}~\bibnamefont {Nakajima}}, \bibinfo {author} {\bibfnamefont {Y.}~\bibnamefont {Takasu}},\ and\ \bibinfo {author} {\bibfnamefont {Y.}~\bibnamefont {Takahashi}},\ }\bibfield  {title} {\bibinfo {title} {{Dissipative Bose-Hubbard system with intrinsic two-body loss}},\ }\href {https://doi.org/10.1103/PhysRevA.99.031601} {\bibfield  {journal} {\bibinfo  {journal} {Phys. Rev. A}\ }\textbf {\bibinfo {volume} {99}},\ \bibinfo {pages} {031601(R)} (\bibinfo {year} {2019})}\BibitemShut {NoStop}%
  \bibitem [{\citenamefont {Okuno}\ \emph {et~al.}(2022)\citenamefont {Okuno}, \citenamefont {Nakamura}, \citenamefont {Kusano}, \citenamefont {Takasu}, \citenamefont {Takei}, \citenamefont {Konishi},\ and\ \citenamefont {Takahashi}}]{Okuno2022-cd}%
    \BibitemOpen
    \bibfield  {author} {\bibinfo {author} {\bibfnamefont {D.}~\bibnamefont {Okuno}}, \bibinfo {author} {\bibfnamefont {Y.}~\bibnamefont {Nakamura}}, \bibinfo {author} {\bibfnamefont {T.}~\bibnamefont {Kusano}}, \bibinfo {author} {\bibfnamefont {Y.}~\bibnamefont {Takasu}}, \bibinfo {author} {\bibfnamefont {N.}~\bibnamefont {Takei}}, \bibinfo {author} {\bibfnamefont {H.}~\bibnamefont {Konishi}},\ and\ \bibinfo {author} {\bibfnamefont {Y.}~\bibnamefont {Takahashi}},\ }\bibfield  {title} {\bibinfo {title} {{High-resolution Spectroscopy and Single-photon Rydberg Excitation of Reconfigurable Ytterbium Atom Tweezer Arrays Utilizing a Metastable State}},\ }\href {https://doi.org/10.7566/JPSJ.91.084301} {\bibfield  {journal} {\bibinfo  {journal} {J. Phys. Soc. Jpn.}\ }\textbf {\bibinfo {volume} {91}},\ \bibinfo {pages} {084301} (\bibinfo {year} {2022})}\BibitemShut {NoStop}%
  \bibitem [{\citenamefont {Sheng}\ \emph {et~al.}(2022)\citenamefont {Sheng}, \citenamefont {Hou}, \citenamefont {He}, \citenamefont {Wang}, \citenamefont {Guo}, \citenamefont {Zhuang}, \citenamefont {Mamat}, \citenamefont {Xu}, \citenamefont {Liu}, \citenamefont {Wang},\ and\ \citenamefont {Zhan}}]{Sheng2022-az}%
    \BibitemOpen
    \bibfield  {author} {\bibinfo {author} {\bibfnamefont {C.}~\bibnamefont {Sheng}}, \bibinfo {author} {\bibfnamefont {J.}~\bibnamefont {Hou}}, \bibinfo {author} {\bibfnamefont {X.}~\bibnamefont {He}}, \bibinfo {author} {\bibfnamefont {K.}~\bibnamefont {Wang}}, \bibinfo {author} {\bibfnamefont {R.}~\bibnamefont {Guo}}, \bibinfo {author} {\bibfnamefont {J.}~\bibnamefont {Zhuang}}, \bibinfo {author} {\bibfnamefont {B.}~\bibnamefont {Mamat}}, \bibinfo {author} {\bibfnamefont {P.}~\bibnamefont {Xu}}, \bibinfo {author} {\bibfnamefont {M.}~\bibnamefont {Liu}}, \bibinfo {author} {\bibfnamefont {J.}~\bibnamefont {Wang}},\ and\ \bibinfo {author} {\bibfnamefont {M.}~\bibnamefont {Zhan}},\ }\bibfield  {title} {\bibinfo {title} {{Defect-Free Arbitrary-Geometry Assembly of Mixed-Species Atom Arrays}},\ }\href {https://doi.org/10.1103/PhysRevLett.128.083202} {\bibfield  {journal} {\bibinfo  {journal} {Phys. Rev. Lett.}\ }\textbf {\bibinfo {volume} {128}},\ \bibinfo {pages} {083202} (\bibinfo {year} {2022})}\BibitemShut
    {NoStop}%
  \bibitem [{\citenamefont {Tao}\ \emph {et~al.}(2022)\citenamefont {Tao}, \citenamefont {Yu}, \citenamefont {Xu}, \citenamefont {Hou}, \citenamefont {He},\ and\ \citenamefont {Zhan}}]{Tao2022-pr}%
    \BibitemOpen
    \bibfield  {author} {\bibinfo {author} {\bibfnamefont {Z.-J.}\ \bibnamefont {Tao}}, \bibinfo {author} {\bibfnamefont {L.-G.}\ \bibnamefont {Yu}}, \bibinfo {author} {\bibfnamefont {P.}~\bibnamefont {Xu}}, \bibinfo {author} {\bibfnamefont {J.-Y.}\ \bibnamefont {Hou}}, \bibinfo {author} {\bibfnamefont {X.-D.}\ \bibnamefont {He}},\ and\ \bibinfo {author} {\bibfnamefont {M.-S.}\ \bibnamefont {Zhan}},\ }\bibfield  {title} {\bibinfo {title} {Efficient two-dimensional defect-free dual-species atom arrays rearrangement algorithm with near-fewest atom moves},\ }\href {https://doi.org/10.1088/0256-307x/39/8/083701} {\bibfield  {journal} {\bibinfo  {journal} {Chin. Physics Lett.}\ }\textbf {\bibinfo {volume} {39}},\ \bibinfo {pages} {083701} (\bibinfo {year} {2022})}\BibitemShut {NoStop}%
  \bibitem [{\citenamefont {Tian}\ \emph {et~al.}(2023)\citenamefont {Tian}, \citenamefont {Wee}, \citenamefont {Qu}, \citenamefont {Lim}, \citenamefont {Datla}, \citenamefont {Koh},\ and\ \citenamefont {Loh}}]{Tian2023-xj}%
    \BibitemOpen
    \bibfield  {author} {\bibinfo {author} {\bibfnamefont {W.}~\bibnamefont {Tian}}, \bibinfo {author} {\bibfnamefont {W.~J.}\ \bibnamefont {Wee}}, \bibinfo {author} {\bibfnamefont {A.}~\bibnamefont {Qu}}, \bibinfo {author} {\bibfnamefont {B.~J.~M.}\ \bibnamefont {Lim}}, \bibinfo {author} {\bibfnamefont {P.~R.}\ \bibnamefont {Datla}}, \bibinfo {author} {\bibfnamefont {V.~P.~W.}\ \bibnamefont {Koh}},\ and\ \bibinfo {author} {\bibfnamefont {H.}~\bibnamefont {Loh}},\ }\bibfield  {title} {\bibinfo {title} {Parallel assembly of arbitrary defect-free atom arrays with a multitweezer algorithm},\ }\href {https://doi.org/10.1103/physrevapplied.19.034048} {\bibfield  {journal} {\bibinfo  {journal} {Phys. Rev. Appl.}\ }\textbf {\bibinfo {volume} {19}},\ \bibinfo {pages} {034048} (\bibinfo {year} {2023})}\BibitemShut {NoStop}%
  \bibitem [{\citenamefont {Shaw}\ \emph {et~al.}(2023)\citenamefont {Shaw}, \citenamefont {Scholl}, \citenamefont {Finklestein}, \citenamefont {Madjarov}, \citenamefont {Grinkemeyer},\ and\ \citenamefont {Endres}}]{Shaw2023-gg}%
    \BibitemOpen
    \bibfield  {author} {\bibinfo {author} {\bibfnamefont {A.~L.}\ \bibnamefont {Shaw}}, \bibinfo {author} {\bibfnamefont {P.}~\bibnamefont {Scholl}}, \bibinfo {author} {\bibfnamefont {R.}~\bibnamefont {Finklestein}}, \bibinfo {author} {\bibfnamefont {I.~S.}\ \bibnamefont {Madjarov}}, \bibinfo {author} {\bibfnamefont {B.}~\bibnamefont {Grinkemeyer}},\ and\ \bibinfo {author} {\bibfnamefont {M.}~\bibnamefont {Endres}},\ }\bibfield  {title} {\bibinfo {title} {{Dark-State Enhanced Loading of an Optical Tweezer Array}},\ }\href {https://doi.org/10.1103/PhysRevLett.130.193402} {\bibfield  {journal} {\bibinfo  {journal} {Phys. Rev. Lett.}\ }\textbf {\bibinfo {volume} {130}},\ \bibinfo {pages} {193402} (\bibinfo {year} {2023})}\BibitemShut {NoStop}%
  \bibitem [{\citenamefont {Grünzweig}\ \emph {et~al.}(2010)\citenamefont {Grünzweig}, \citenamefont {Hilliard}, \citenamefont {McGovern},\ and\ \citenamefont {Andersen}}]{Grunzweig2010-lo}%
    \BibitemOpen
    \bibfield  {author} {\bibinfo {author} {\bibfnamefont {T.}~\bibnamefont {Grünzweig}}, \bibinfo {author} {\bibfnamefont {A.}~\bibnamefont {Hilliard}}, \bibinfo {author} {\bibfnamefont {M.}~\bibnamefont {McGovern}},\ and\ \bibinfo {author} {\bibfnamefont {M.~F.}\ \bibnamefont {Andersen}},\ }\bibfield  {title} {\bibinfo {title} {Near-deterministic preparation of a single atom in an optical microtrap},\ }\href {https://doi.org/10.1038/nphys1778} {\bibfield  {journal} {\bibinfo  {journal} {Nat. Phys.}\ }\textbf {\bibinfo {volume} {6}},\ \bibinfo {pages} {951} (\bibinfo {year} {2010})}\BibitemShut {NoStop}%
  \bibitem [{\citenamefont {Brown}\ \emph {et~al.}(2019)\citenamefont {Brown}, \citenamefont {Thiele}, \citenamefont {Kiehl}, \citenamefont {Hsu},\ and\ \citenamefont {Regal}}]{Brown2019-rr}%
    \BibitemOpen
    \bibfield  {author} {\bibinfo {author} {\bibfnamefont {M.~O.}\ \bibnamefont {Brown}}, \bibinfo {author} {\bibfnamefont {T.}~\bibnamefont {Thiele}}, \bibinfo {author} {\bibfnamefont {C.}~\bibnamefont {Kiehl}}, \bibinfo {author} {\bibfnamefont {T.-W.}\ \bibnamefont {Hsu}},\ and\ \bibinfo {author} {\bibfnamefont {C.~A.}\ \bibnamefont {Regal}},\ }\bibfield  {title} {\bibinfo {title} {{Gray-Molasses Optical-Tweezer Loading: Controlling Collisions for Scaling Atom-Array Assembly}},\ }\href {https://doi.org/10.1103/PhysRevX.9.011057} {\bibfield  {journal} {\bibinfo  {journal} {Phys. Rev. X}\ }\textbf {\bibinfo {volume} {9}},\ \bibinfo {pages} {011057} (\bibinfo {year} {2019})}\BibitemShut {NoStop}%
  \bibitem [{\citenamefont {Ni}\ \emph {et~al.}(2008)\citenamefont {Ni}, \citenamefont {Ospelkaus}, \citenamefont {de~Miranda}, \citenamefont {Pe'er}, \citenamefont {Neyenhuis}, \citenamefont {Zirbel}, \citenamefont {Kotochigova}, \citenamefont {Julienne}, \citenamefont {Jin},\ and\ \citenamefont {Ye}}]{Ni2008-kk}%
    \BibitemOpen
    \bibfield  {author} {\bibinfo {author} {\bibfnamefont {K.-K.}\ \bibnamefont {Ni}}, \bibinfo {author} {\bibfnamefont {S.}~\bibnamefont {Ospelkaus}}, \bibinfo {author} {\bibfnamefont {M.~H.~G.}\ \bibnamefont {de~Miranda}}, \bibinfo {author} {\bibfnamefont {A.}~\bibnamefont {Pe'er}}, \bibinfo {author} {\bibfnamefont {B.}~\bibnamefont {Neyenhuis}}, \bibinfo {author} {\bibfnamefont {J.~J.}\ \bibnamefont {Zirbel}}, \bibinfo {author} {\bibfnamefont {S.}~\bibnamefont {Kotochigova}}, \bibinfo {author} {\bibfnamefont {P.~S.}\ \bibnamefont {Julienne}}, \bibinfo {author} {\bibfnamefont {D.~S.}\ \bibnamefont {Jin}},\ and\ \bibinfo {author} {\bibfnamefont {J.}~\bibnamefont {Ye}},\ }\bibfield  {title} {\bibinfo {title} {A high phase-space-density gas of polar molecules},\ }\href {https://doi.org/10.1126/science.1163861} {\bibfield  {journal} {\bibinfo  {journal} {Science}\ }\textbf {\bibinfo {volume} {322}},\ \bibinfo {pages} {231} (\bibinfo {year} {2008})}\BibitemShut {NoStop}%
  \bibitem [{\citenamefont {Chotia}\ \emph {et~al.}(2012)\citenamefont {Chotia}, \citenamefont {Neyenhuis}, \citenamefont {Moses}, \citenamefont {Yan}, \citenamefont {Covey}, \citenamefont {Foss-Feig}, \citenamefont {Rey}, \citenamefont {Jin},\ and\ \citenamefont {Ye}}]{Chotia2012-xv}%
    \BibitemOpen
    \bibfield  {author} {\bibinfo {author} {\bibfnamefont {A.}~\bibnamefont {Chotia}}, \bibinfo {author} {\bibfnamefont {B.}~\bibnamefont {Neyenhuis}}, \bibinfo {author} {\bibfnamefont {S.~A.}\ \bibnamefont {Moses}}, \bibinfo {author} {\bibfnamefont {B.}~\bibnamefont {Yan}}, \bibinfo {author} {\bibfnamefont {J.~P.}\ \bibnamefont {Covey}}, \bibinfo {author} {\bibfnamefont {M.}~\bibnamefont {Foss-Feig}}, \bibinfo {author} {\bibfnamefont {A.~M.}\ \bibnamefont {Rey}}, \bibinfo {author} {\bibfnamefont {D.~S.}\ \bibnamefont {Jin}},\ and\ \bibinfo {author} {\bibfnamefont {J.}~\bibnamefont {Ye}},\ }\bibfield  {title} {\bibinfo {title} {{Long-lived dipolar molecules and Feshbach molecules in a {3D} optical lattice}},\ }\href {https://doi.org/10.1103/PhysRevLett.108.080405} {\bibfield  {journal} {\bibinfo  {journal} {Phys. Rev. Lett.}\ }\textbf {\bibinfo {volume} {108}},\ \bibinfo {pages} {080405} (\bibinfo {year} {2012})}\BibitemShut {NoStop}%
  \bibitem [{\citenamefont {Kitagawa}\ \emph {et~al.}(2008)\citenamefont {Kitagawa}, \citenamefont {Enomoto}, \citenamefont {Kasa}, \citenamefont {Takahashi}, \citenamefont {Ciury{\l}o}, \citenamefont {Naidon},\ and\ \citenamefont {Julienne}}]{Kitagawa2008-fj}%
    \BibitemOpen
    \bibfield  {author} {\bibinfo {author} {\bibfnamefont {M.}~\bibnamefont {Kitagawa}}, \bibinfo {author} {\bibfnamefont {K.}~\bibnamefont {Enomoto}}, \bibinfo {author} {\bibfnamefont {K.}~\bibnamefont {Kasa}}, \bibinfo {author} {\bibfnamefont {Y.}~\bibnamefont {Takahashi}}, \bibinfo {author} {\bibfnamefont {R.}~\bibnamefont {Ciury{\l}o}}, \bibinfo {author} {\bibfnamefont {P.}~\bibnamefont {Naidon}},\ and\ \bibinfo {author} {\bibfnamefont {P.~S.}\ \bibnamefont {Julienne}},\ }\bibfield  {title} {\bibinfo {title} {Two-color photoassociation spectroscopy of ytterbium atoms and the precise determinations of $s$-wave scattering lengths},\ }\href {https://doi.org/10.1103/PhysRevA.77.012719} {\bibfield  {journal} {\bibinfo  {journal} {Phys. Rev. A}\ }\textbf {\bibinfo {volume} {77}},\ \bibinfo {pages} {012719} (\bibinfo {year} {2008})}\BibitemShut {NoStop}%
  \bibitem [{\citenamefont {Yu}\ \emph {et~al.}(2021)\citenamefont {Yu}, \citenamefont {Wang}, \citenamefont {Hood}, \citenamefont {Picard}, \citenamefont {Zhang}, \citenamefont {Cairncross}, \citenamefont {Hutson}, \citenamefont {Gonzalez-Ferez}, \citenamefont {Rosenband},\ and\ \citenamefont {Ni}}]{Yu2021-wk}%
    \BibitemOpen
    \bibfield  {author} {\bibinfo {author} {\bibfnamefont {Y.}~\bibnamefont {Yu}}, \bibinfo {author} {\bibfnamefont {K.}~\bibnamefont {Wang}}, \bibinfo {author} {\bibfnamefont {J.~D.}\ \bibnamefont {Hood}}, \bibinfo {author} {\bibfnamefont {L.~R.~B.}\ \bibnamefont {Picard}}, \bibinfo {author} {\bibfnamefont {J.~T.}\ \bibnamefont {Zhang}}, \bibinfo {author} {\bibfnamefont {W.~B.}\ \bibnamefont {Cairncross}}, \bibinfo {author} {\bibfnamefont {J.~M.}\ \bibnamefont {Hutson}}, \bibinfo {author} {\bibfnamefont {R.}~\bibnamefont {Gonzalez-Ferez}}, \bibinfo {author} {\bibfnamefont {T.}~\bibnamefont {Rosenband}},\ and\ \bibinfo {author} {\bibfnamefont {K.-K.}\ \bibnamefont {Ni}},\ }\bibfield  {title} {\bibinfo {title} {{Coherent Optical Creation of a Single Molecule}},\ }\href {https://doi.org/10.1103/PhysRevX.11.031061} {\bibfield  {journal} {\bibinfo  {journal} {Phys. Rev. X}\ }\textbf {\bibinfo {volume} {11}},\ \bibinfo {pages} {031061} (\bibinfo {year} {2021})}\BibitemShut {NoStop}%
  \bibitem [{\citenamefont {Liu}\ \emph {et~al.}(2018)\citenamefont {Liu}, \citenamefont {Hood}, \citenamefont {Yu}, \citenamefont {Zhang}, \citenamefont {Hutzler}, \citenamefont {Rosenband},\ and\ \citenamefont {Ni}}]{Liu2018-ec}%
    \BibitemOpen
    \bibfield  {author} {\bibinfo {author} {\bibfnamefont {L.~R.}\ \bibnamefont {Liu}}, \bibinfo {author} {\bibfnamefont {J.~D.}\ \bibnamefont {Hood}}, \bibinfo {author} {\bibfnamefont {Y.}~\bibnamefont {Yu}}, \bibinfo {author} {\bibfnamefont {J.~T.}\ \bibnamefont {Zhang}}, \bibinfo {author} {\bibfnamefont {N.~R.}\ \bibnamefont {Hutzler}}, \bibinfo {author} {\bibfnamefont {T.}~\bibnamefont {Rosenband}},\ and\ \bibinfo {author} {\bibfnamefont {K.-K.}\ \bibnamefont {Ni}},\ }\bibfield  {title} {\bibinfo {title} {Building one molecule from a reservoir of two atoms},\ }\href {https://doi.org/10.1126/science.aar7797} {\bibfield  {journal} {\bibinfo  {journal} {Science}\ }\textbf {\bibinfo {volume} {360}},\ \bibinfo {pages} {900} (\bibinfo {year} {2018})}\BibitemShut {NoStop}%
  \bibitem [{\citenamefont {Liu}\ \emph {et~al.}(2019)\citenamefont {Liu}, \citenamefont {Hood}, \citenamefont {Yu}, \citenamefont {Zhang}, \citenamefont {Wang}, \citenamefont {Lin}, \citenamefont {Rosenband},\ and\ \citenamefont {Ni}}]{Liu2019-ho}%
    \BibitemOpen
    \bibfield  {author} {\bibinfo {author} {\bibfnamefont {L.~R.}\ \bibnamefont {Liu}}, \bibinfo {author} {\bibfnamefont {J.~D.}\ \bibnamefont {Hood}}, \bibinfo {author} {\bibfnamefont {Y.}~\bibnamefont {Yu}}, \bibinfo {author} {\bibfnamefont {J.~T.}\ \bibnamefont {Zhang}}, \bibinfo {author} {\bibfnamefont {K.}~\bibnamefont {Wang}}, \bibinfo {author} {\bibfnamefont {Y.-W.}\ \bibnamefont {Lin}}, \bibinfo {author} {\bibfnamefont {T.}~\bibnamefont {Rosenband}},\ and\ \bibinfo {author} {\bibfnamefont {K.-K.}\ \bibnamefont {Ni}},\ }\bibfield  {title} {\bibinfo {title} {{Molecular Assembly of Ground-State Cooled Single Atoms}},\ }\href {https://doi.org/10.1103/PhysRevX.9.021039} {\bibfield  {journal} {\bibinfo  {journal} {Phys. Rev. X}\ }\textbf {\bibinfo {volume} {9}},\ \bibinfo {pages} {021039} (\bibinfo {year} {2019})}\BibitemShut {NoStop}%
  \bibitem [{\citenamefont {Zhang}\ \emph {et~al.}(2020)\citenamefont {Zhang}, \citenamefont {Yu}, \citenamefont {Cairncross}, \citenamefont {Wang}, \citenamefont {Picard}, \citenamefont {Hood}, \citenamefont {Lin}, \citenamefont {Hutson},\ and\ \citenamefont {Ni}}]{Zhang2020-up}%
    \BibitemOpen
    \bibfield  {author} {\bibinfo {author} {\bibfnamefont {J.~T.}\ \bibnamefont {Zhang}}, \bibinfo {author} {\bibfnamefont {Y.}~\bibnamefont {Yu}}, \bibinfo {author} {\bibfnamefont {W.~B.}\ \bibnamefont {Cairncross}}, \bibinfo {author} {\bibfnamefont {K.}~\bibnamefont {Wang}}, \bibinfo {author} {\bibfnamefont {L.~R.~B.}\ \bibnamefont {Picard}}, \bibinfo {author} {\bibfnamefont {J.~D.}\ \bibnamefont {Hood}}, \bibinfo {author} {\bibfnamefont {Y.-W.}\ \bibnamefont {Lin}}, \bibinfo {author} {\bibfnamefont {J.~M.}\ \bibnamefont {Hutson}},\ and\ \bibinfo {author} {\bibfnamefont {K.-K.}\ \bibnamefont {Ni}},\ }\bibfield  {title} {\bibinfo {title} {{Forming a Single Molecule by Magnetoassociation in an Optical Tweezer}},\ }\href {https://doi.org/10.1103/PhysRevLett.124.253401} {\bibfield  {journal} {\bibinfo  {journal} {Phys. Rev. Lett.}\ }\textbf {\bibinfo {volume} {124}},\ \bibinfo {pages} {253401} (\bibinfo {year} {2020})}\BibitemShut {NoStop}%
  \bibitem [{\citenamefont {Anderegg}\ \emph {et~al.}(2019)\citenamefont {Anderegg}, \citenamefont {Cheuk}, \citenamefont {Bao}, \citenamefont {Burchesky}, \citenamefont {Ketterle}, \citenamefont {Ni},\ and\ \citenamefont {Doyle}}]{Anderegg2019-kk}%
    \BibitemOpen
    \bibfield  {author} {\bibinfo {author} {\bibfnamefont {L.}~\bibnamefont {Anderegg}}, \bibinfo {author} {\bibfnamefont {L.~W.}\ \bibnamefont {Cheuk}}, \bibinfo {author} {\bibfnamefont {Y.}~\bibnamefont {Bao}}, \bibinfo {author} {\bibfnamefont {S.}~\bibnamefont {Burchesky}}, \bibinfo {author} {\bibfnamefont {W.}~\bibnamefont {Ketterle}}, \bibinfo {author} {\bibfnamefont {K.-K.}\ \bibnamefont {Ni}},\ and\ \bibinfo {author} {\bibfnamefont {J.~M.}\ \bibnamefont {Doyle}},\ }\bibfield  {title} {\bibinfo {title} {An optical tweezer array of ultracold molecules},\ }\href {https://doi.org/10.1126/science.aax1265} {\bibfield  {journal} {\bibinfo  {journal} {Science}\ }\textbf {\bibinfo {volume} {365}},\ \bibinfo {pages} {1156} (\bibinfo {year} {2019})}\BibitemShut {NoStop}%
  \bibitem [{\citenamefont {He}\ \emph {et~al.}(2020)\citenamefont {He}, \citenamefont {Wang}, \citenamefont {Zhuang}, \citenamefont {Xu}, \citenamefont {Gao}, \citenamefont {Guo}, \citenamefont {Sheng}, \citenamefont {Liu}, \citenamefont {Wang}, \citenamefont {Li}, \citenamefont {Shlyapnikov},\ and\ \citenamefont {Zhan}}]{He2020-wd}%
    \BibitemOpen
    \bibfield  {author} {\bibinfo {author} {\bibfnamefont {X.}~\bibnamefont {He}}, \bibinfo {author} {\bibfnamefont {K.}~\bibnamefont {Wang}}, \bibinfo {author} {\bibfnamefont {J.}~\bibnamefont {Zhuang}}, \bibinfo {author} {\bibfnamefont {P.}~\bibnamefont {Xu}}, \bibinfo {author} {\bibfnamefont {X.}~\bibnamefont {Gao}}, \bibinfo {author} {\bibfnamefont {R.}~\bibnamefont {Guo}}, \bibinfo {author} {\bibfnamefont {C.}~\bibnamefont {Sheng}}, \bibinfo {author} {\bibfnamefont {M.}~\bibnamefont {Liu}}, \bibinfo {author} {\bibfnamefont {J.}~\bibnamefont {Wang}}, \bibinfo {author} {\bibfnamefont {J.}~\bibnamefont {Li}}, \bibinfo {author} {\bibfnamefont {G.~V.}\ \bibnamefont {Shlyapnikov}},\ and\ \bibinfo {author} {\bibfnamefont {M.}~\bibnamefont {Zhan}},\ }\bibfield  {title} {\bibinfo {title} {Coherently forming a single molecule in an optical trap},\ }\href {https://doi.org/10.1126/science.aba7468} {\bibfield  {journal} {\bibinfo  {journal} {Science}\ }\textbf {\bibinfo {volume} {370}},\ \bibinfo {pages} {331}
    (\bibinfo {year} {2020})}\BibitemShut {NoStop}%
  \bibitem [{\citenamefont {Ruttley}\ \emph {et~al.}(2023)\citenamefont {Ruttley}, \citenamefont {Guttridge}, \citenamefont {Spence}, \citenamefont {Bird}, \citenamefont {Le~Sueur}, \citenamefont {Hutson},\ and\ \citenamefont {Cornish}}]{Ruttley2023-nl}%
    \BibitemOpen
    \bibfield  {author} {\bibinfo {author} {\bibfnamefont {D.~K.}\ \bibnamefont {Ruttley}}, \bibinfo {author} {\bibfnamefont {A.}~\bibnamefont {Guttridge}}, \bibinfo {author} {\bibfnamefont {S.}~\bibnamefont {Spence}}, \bibinfo {author} {\bibfnamefont {R.~C.}\ \bibnamefont {Bird}}, \bibinfo {author} {\bibfnamefont {C.~R.}\ \bibnamefont {Le~Sueur}}, \bibinfo {author} {\bibfnamefont {J.~M.}\ \bibnamefont {Hutson}},\ and\ \bibinfo {author} {\bibfnamefont {S.~L.}\ \bibnamefont {Cornish}},\ }\bibfield  {title} {\bibinfo {title} {{Formation of Ultracold Molecules by Merging Optical Tweezers}},\ }\href {https://doi.org/10.1103/PhysRevLett.130.223401} {\bibfield  {journal} {\bibinfo  {journal} {Phys. Rev. Lett.}\ }\textbf {\bibinfo {volume} {130}},\ \bibinfo {pages} {223401} (\bibinfo {year} {2023})}\BibitemShut {NoStop}%
  \bibitem [{\citenamefont {Park}\ \emph {et~al.}(2017)\citenamefont {Park}, \citenamefont {Yan}, \citenamefont {Loh}, \citenamefont {Will},\ and\ \citenamefont {Zwierlein}}]{Park2017-qq}%
    \BibitemOpen
    \bibfield  {author} {\bibinfo {author} {\bibfnamefont {J.~W.}\ \bibnamefont {Park}}, \bibinfo {author} {\bibfnamefont {Z.~Z.}\ \bibnamefont {Yan}}, \bibinfo {author} {\bibfnamefont {H.}~\bibnamefont {Loh}}, \bibinfo {author} {\bibfnamefont {S.~A.}\ \bibnamefont {Will}},\ and\ \bibinfo {author} {\bibfnamefont {M.~W.}\ \bibnamefont {Zwierlein}},\ }\bibfield  {title} {\bibinfo {title} {Second-scale nuclear spin coherence time of ultracold {${}^{23}\mathrm{Na}{}^{40}\mathrm{K}$} molecules},\ }\href {https://doi.org/10.1126/science.aal5066} {\bibfield  {journal} {\bibinfo  {journal} {Science}\ }\textbf {\bibinfo {volume} {357}},\ \bibinfo {pages} {372} (\bibinfo {year} {2017})}\BibitemShut {NoStop}%
  \bibitem [{\citenamefont {Gregory}\ \emph {et~al.}(2021)\citenamefont {Gregory}, \citenamefont {Blackmore}, \citenamefont {Bromley}, \citenamefont {Hutson},\ and\ \citenamefont {Cornish}}]{Gregory2021-yx}%
    \BibitemOpen
    \bibfield  {author} {\bibinfo {author} {\bibfnamefont {P.~D.}\ \bibnamefont {Gregory}}, \bibinfo {author} {\bibfnamefont {J.~A.}\ \bibnamefont {Blackmore}}, \bibinfo {author} {\bibfnamefont {S.~L.}\ \bibnamefont {Bromley}}, \bibinfo {author} {\bibfnamefont {J.~M.}\ \bibnamefont {Hutson}},\ and\ \bibinfo {author} {\bibfnamefont {S.~L.}\ \bibnamefont {Cornish}},\ }\bibfield  {title} {\bibinfo {title} {Robust storage qubits in ultracold polar molecules},\ }\href {https://doi.org/10.1038/s41567-021-01328-7} {\bibfield  {journal} {\bibinfo  {journal} {Nat. Phys.}\ }\textbf {\bibinfo {volume} {17}},\ \bibinfo {pages} {1149} (\bibinfo {year} {2021})}\BibitemShut {NoStop}%
  \bibitem [{\citenamefont {Sawant}\ \emph {et~al.}(2020)\citenamefont {Sawant}, \citenamefont {Blackmore}, \citenamefont {Gregory}, \citenamefont {Mur-Petit}, \citenamefont {Jaksch}, \citenamefont {Aldegunde}, \citenamefont {Hutson}, \citenamefont {Tarbutt},\ and\ \citenamefont {Cornish}}]{Sawant2020-qj}%
    \BibitemOpen
    \bibfield  {author} {\bibinfo {author} {\bibfnamefont {R.}~\bibnamefont {Sawant}}, \bibinfo {author} {\bibfnamefont {J.~A.}\ \bibnamefont {Blackmore}}, \bibinfo {author} {\bibfnamefont {P.~D.}\ \bibnamefont {Gregory}}, \bibinfo {author} {\bibfnamefont {J.}~\bibnamefont {Mur-Petit}}, \bibinfo {author} {\bibfnamefont {D.}~\bibnamefont {Jaksch}}, \bibinfo {author} {\bibfnamefont {J.}~\bibnamefont {Aldegunde}}, \bibinfo {author} {\bibfnamefont {J.~M.}\ \bibnamefont {Hutson}}, \bibinfo {author} {\bibfnamefont {M.~R.}\ \bibnamefont {Tarbutt}},\ and\ \bibinfo {author} {\bibfnamefont {S.~L.}\ \bibnamefont {Cornish}},\ }\bibfield  {title} {\bibinfo {title} {Ultracold polar molecules as qudits},\ }\href {https://doi.org/10.1088/1367-2630/ab60f4} {\bibfield  {journal} {\bibinfo  {journal} {New J. Phys.}\ }\textbf {\bibinfo {volume} {22}},\ \bibinfo {pages} {013027} (\bibinfo {year} {2020})}\BibitemShut {NoStop}%
  \bibitem [{\citenamefont {Albert}\ \emph {et~al.}(2020)\citenamefont {Albert}, \citenamefont {Covey},\ and\ \citenamefont {Preskill}}]{Albert2020-bl}%
    \BibitemOpen
    \bibfield  {author} {\bibinfo {author} {\bibfnamefont {V.~V.}\ \bibnamefont {Albert}}, \bibinfo {author} {\bibfnamefont {J.~P.}\ \bibnamefont {Covey}},\ and\ \bibinfo {author} {\bibfnamefont {J.}~\bibnamefont {Preskill}},\ }\bibfield  {title} {\bibinfo {title} {Robust encoding of a qubit in a molecule},\ }\href {https://doi.org/10.1103/PhysRevX.10.031050} {\bibfield  {journal} {\bibinfo  {journal} {Phys. Rev. X}\ }\textbf {\bibinfo {volume} {10}},\ \bibinfo {pages} {031050} (\bibinfo {year} {2020})}\BibitemShut {NoStop}%
  \bibitem [{\citenamefont {Furey}\ \emph {et~al.}(2024)\citenamefont {Furey}, \citenamefont {Wu}, \citenamefont {Isaza-Monsalve}, \citenamefont {Walser}, \citenamefont {Mattivi}, \citenamefont {Nardi},\ and\ \citenamefont {Schindler}}]{Furey2024-bk}%
    \BibitemOpen
    \bibfield  {author} {\bibinfo {author} {\bibfnamefont {B.~J.}\ \bibnamefont {Furey}}, \bibinfo {author} {\bibfnamefont {Z.}~\bibnamefont {Wu}}, \bibinfo {author} {\bibfnamefont {M.}~\bibnamefont {Isaza-Monsalve}}, \bibinfo {author} {\bibfnamefont {S.}~\bibnamefont {Walser}}, \bibinfo {author} {\bibfnamefont {E.}~\bibnamefont {Mattivi}}, \bibinfo {author} {\bibfnamefont {R.}~\bibnamefont {Nardi}},\ and\ \bibinfo {author} {\bibfnamefont {P.}~\bibnamefont {Schindler}},\ }\bibfield  {title} {\bibinfo {title} {Strategies for implementing quantum error correction in molecular rotation},\ }\href {http://arxiv.org/abs/2405.02236} {\bibfield  {journal} {\bibinfo  {journal} {arXiv}\ } (\bibinfo {year} {2024})},\ \Eprint {https://arxiv.org/abs/2405.02236} {arXiv:2405.02236 [quant-ph]} \BibitemShut {NoStop}%
  \bibitem [{\citenamefont {Cairncross}\ \emph {et~al.}(2021)\citenamefont {Cairncross}, \citenamefont {Zhang}, \citenamefont {Picard}, \citenamefont {Yu}, \citenamefont {Wang},\ and\ \citenamefont {Ni}}]{Cairncross2021-gw}%
    \BibitemOpen
    \bibfield  {author} {\bibinfo {author} {\bibfnamefont {W.~B.}\ \bibnamefont {Cairncross}}, \bibinfo {author} {\bibfnamefont {J.~T.}\ \bibnamefont {Zhang}}, \bibinfo {author} {\bibfnamefont {L.~R.~B.}\ \bibnamefont {Picard}}, \bibinfo {author} {\bibfnamefont {Y.}~\bibnamefont {Yu}}, \bibinfo {author} {\bibfnamefont {K.}~\bibnamefont {Wang}},\ and\ \bibinfo {author} {\bibfnamefont {K.-K.}\ \bibnamefont {Ni}},\ }\bibfield  {title} {\bibinfo {title} {{Assembly of a Rovibrational Ground State Molecule in an Optical Tweezer}},\ }\href {https://doi.org/10.1103/PhysRevLett.126.123402} {\bibfield  {journal} {\bibinfo  {journal} {Phys. Rev. Lett.}\ }\textbf {\bibinfo {volume} {126}},\ \bibinfo {pages} {123402} (\bibinfo {year} {2021})}\BibitemShut {NoStop}%
  \bibitem [{\citenamefont {Reinaudi}\ \emph {et~al.}(2012)\citenamefont {Reinaudi}, \citenamefont {Osborn}, \citenamefont {McDonald}, \citenamefont {Kotochigova},\ and\ \citenamefont {Zelevinsky}}]{Reinaudi2012-il}%
    \BibitemOpen
    \bibfield  {author} {\bibinfo {author} {\bibfnamefont {G.}~\bibnamefont {Reinaudi}}, \bibinfo {author} {\bibfnamefont {C.~B.}\ \bibnamefont {Osborn}}, \bibinfo {author} {\bibfnamefont {M.}~\bibnamefont {McDonald}}, \bibinfo {author} {\bibfnamefont {S.}~\bibnamefont {Kotochigova}},\ and\ \bibinfo {author} {\bibfnamefont {T.}~\bibnamefont {Zelevinsky}},\ }\bibfield  {title} {\bibinfo {title} {{Optical production of stable ultracold ${}^{88}\mathrm{Sr}_2$ molecules}},\ }\href {https://doi.org/10.1103/PhysRevLett.109.115303} {\bibfield  {journal} {\bibinfo  {journal} {Phys. Rev. Lett.}\ }\textbf {\bibinfo {volume} {109}},\ \bibinfo {pages} {115303} (\bibinfo {year} {2012})}\BibitemShut {NoStop}%
  \bibitem [{\citenamefont {Stellmer}\ \emph {et~al.}(2012)\citenamefont {Stellmer}, \citenamefont {Pasquiou}, \citenamefont {Grimm},\ and\ \citenamefont {Schreck}}]{Stellmer2012-pg}%
    \BibitemOpen
    \bibfield  {author} {\bibinfo {author} {\bibfnamefont {S.}~\bibnamefont {Stellmer}}, \bibinfo {author} {\bibfnamefont {B.}~\bibnamefont {Pasquiou}}, \bibinfo {author} {\bibfnamefont {R.}~\bibnamefont {Grimm}},\ and\ \bibinfo {author} {\bibfnamefont {F.}~\bibnamefont {Schreck}},\ }\bibfield  {title} {\bibinfo {title} {{Creation of ultracold $\mathrm{Sr}_2$ molecules in the electronic ground state}},\ }\href {https://doi.org/10.1103/PhysRevLett.109.115302} {\bibfield  {journal} {\bibinfo  {journal} {Phys. Rev. Lett.}\ }\textbf {\bibinfo {volume} {109}},\ \bibinfo {pages} {115302} (\bibinfo {year} {2012})}\BibitemShut {NoStop}%
  \bibitem [{\citenamefont {Kato}\ \emph {et~al.}(2012)\citenamefont {Kato}, \citenamefont {Yamazaki}, \citenamefont {Shibata}, \citenamefont {Yamamoto}, \citenamefont {Yamada},\ and\ \citenamefont {Takahashi}}]{Kato2012-dw}%
    \BibitemOpen
    \bibfield  {author} {\bibinfo {author} {\bibfnamefont {S.}~\bibnamefont {Kato}}, \bibinfo {author} {\bibfnamefont {R.}~\bibnamefont {Yamazaki}}, \bibinfo {author} {\bibfnamefont {K.}~\bibnamefont {Shibata}}, \bibinfo {author} {\bibfnamefont {R.}~\bibnamefont {Yamamoto}}, \bibinfo {author} {\bibfnamefont {H.}~\bibnamefont {Yamada}},\ and\ \bibinfo {author} {\bibfnamefont {Y.}~\bibnamefont {Takahashi}},\ }\bibfield  {title} {\bibinfo {title} {Observation of long-lived van der waals molecules in an optical lattice},\ }\href {https://doi.org/10.1103/physreva.86.043411} {\bibfield  {journal} {\bibinfo  {journal} {Phys. Rev. A}\ }\textbf {\bibinfo {volume} {86}},\ \bibinfo {pages} {043411} (\bibinfo {year} {2012})}\BibitemShut {NoStop}%
  \bibitem [{\citenamefont {Thompson}\ \emph {et~al.}(2013)\citenamefont {Thompson}, \citenamefont {Tiecke}, \citenamefont {Zibrov}, \citenamefont {Vuletić},\ and\ \citenamefont {Lukin}}]{Thompson2013-jp}%
    \BibitemOpen
    \bibfield  {author} {\bibinfo {author} {\bibfnamefont {J.~D.}\ \bibnamefont {Thompson}}, \bibinfo {author} {\bibfnamefont {T.~G.}\ \bibnamefont {Tiecke}}, \bibinfo {author} {\bibfnamefont {A.~S.}\ \bibnamefont {Zibrov}}, \bibinfo {author} {\bibfnamefont {V.}~\bibnamefont {Vuletić}},\ and\ \bibinfo {author} {\bibfnamefont {M.~D.}\ \bibnamefont {Lukin}},\ }\bibfield  {title} {\bibinfo {title} {{Coherence and Raman sideband cooling of a single atom in an optical tweezer}},\ }\href {https://doi.org/10.1103/PhysRevLett.110.133001} {\bibfield  {journal} {\bibinfo  {journal} {Phys. Rev. Lett.}\ }\textbf {\bibinfo {volume} {110}},\ \bibinfo {pages} {133001} (\bibinfo {year} {2013})}\BibitemShut {NoStop}%
  \bibitem [{\citenamefont {Zhang}\ \emph {et~al.}(2022)\citenamefont {Zhang}, \citenamefont {Beloy}, \citenamefont {Hassan}, \citenamefont {McGrew}, \citenamefont {Chen}, \citenamefont {Siegel}, \citenamefont {Grogan},\ and\ \citenamefont {Ludlow}}]{Zhang2022-gp}%
    \BibitemOpen
    \bibfield  {author} {\bibinfo {author} {\bibfnamefont {X.}~\bibnamefont {Zhang}}, \bibinfo {author} {\bibfnamefont {K.}~\bibnamefont {Beloy}}, \bibinfo {author} {\bibfnamefont {Y.~S.}\ \bibnamefont {Hassan}}, \bibinfo {author} {\bibfnamefont {W.~F.}\ \bibnamefont {McGrew}}, \bibinfo {author} {\bibfnamefont {C.-C.}\ \bibnamefont {Chen}}, \bibinfo {author} {\bibfnamefont {J.~L.}\ \bibnamefont {Siegel}}, \bibinfo {author} {\bibfnamefont {T.}~\bibnamefont {Grogan}},\ and\ \bibinfo {author} {\bibfnamefont {A.~D.}\ \bibnamefont {Ludlow}},\ }\bibfield  {title} {\bibinfo {title} {{Subrecoil Clock-Transition Laser Cooling Enabling Shallow Optical Lattice Clocks}},\ }\href {https://doi.org/10.1103/PhysRevLett.129.113202} {\bibfield  {journal} {\bibinfo  {journal} {Phys. Rev. Lett.}\ }\textbf {\bibinfo {volume} {129}},\ \bibinfo {pages} {113202} (\bibinfo {year} {2022})}\BibitemShut {NoStop}%
  \bibitem [{\citenamefont {Ishiyama}\ \emph {et~al.}(2023)\citenamefont {Ishiyama}, \citenamefont {Ono}, \citenamefont {Takano}, \citenamefont {Sunaga},\ and\ \citenamefont {Takahashi}}]{Ishiyama2023-df}%
    \BibitemOpen
    \bibfield  {author} {\bibinfo {author} {\bibfnamefont {T.}~\bibnamefont {Ishiyama}}, \bibinfo {author} {\bibfnamefont {K.}~\bibnamefont {Ono}}, \bibinfo {author} {\bibfnamefont {T.}~\bibnamefont {Takano}}, \bibinfo {author} {\bibfnamefont {A.}~\bibnamefont {Sunaga}},\ and\ \bibinfo {author} {\bibfnamefont {Y.}~\bibnamefont {Takahashi}},\ }\bibfield  {title} {\bibinfo {title} {Observation of an inner-shell orbital clock transition in neutral ytterbium atoms},\ }\href {https://doi.org/10.1103/PhysRevLett.130.153402} {\bibfield  {journal} {\bibinfo  {journal} {Phys. Rev. Lett.}\ }\textbf {\bibinfo {volume} {130}},\ \bibinfo {pages} {153402} (\bibinfo {year} {2023})}\BibitemShut {NoStop}%
  \bibitem [{\citenamefont {Kawasaki}\ \emph {et~al.}(2024)\citenamefont {Kawasaki}, \citenamefont {Kobayashi}, \citenamefont {Nishiyama}, \citenamefont {Tanabe},\ and\ \citenamefont {Yasuda}}]{Kawasaki2024-qo}%
    \BibitemOpen
    \bibfield  {author} {\bibinfo {author} {\bibfnamefont {A.}~\bibnamefont {Kawasaki}}, \bibinfo {author} {\bibfnamefont {T.}~\bibnamefont {Kobayashi}}, \bibinfo {author} {\bibfnamefont {A.}~\bibnamefont {Nishiyama}}, \bibinfo {author} {\bibfnamefont {T.}~\bibnamefont {Tanabe}},\ and\ \bibinfo {author} {\bibfnamefont {M.}~\bibnamefont {Yasuda}},\ }\bibfield  {title} {\bibinfo {title} {Isotope-shift analysis with the $4f^{14}6s^2{}^1\mathrm{S}_0$ – $4f^{13}5d6s^2~(j=2)$ transition in ytterbium},\ }\href {https://doi.org/10.1103/physreva.109.062806} {\bibfield  {journal} {\bibinfo  {journal} {Phys. Rev. A}\ }\textbf {\bibinfo {volume} {109}},\ \bibinfo {pages} {062806} (\bibinfo {year} {2024})}\BibitemShut {NoStop}%
  \bibitem [{\citenamefont {Kato}\ \emph {et~al.}(2016)\citenamefont {Kato}, \citenamefont {Inaba}, \citenamefont {Sugawa}, \citenamefont {Shibata}, \citenamefont {Yamamoto}, \citenamefont {Yamashita},\ and\ \citenamefont {Takahashi}}]{Kato2016-ie}%
    \BibitemOpen
    \bibfield  {author} {\bibinfo {author} {\bibfnamefont {S.}~\bibnamefont {Kato}}, \bibinfo {author} {\bibfnamefont {K.}~\bibnamefont {Inaba}}, \bibinfo {author} {\bibfnamefont {S.}~\bibnamefont {Sugawa}}, \bibinfo {author} {\bibfnamefont {K.}~\bibnamefont {Shibata}}, \bibinfo {author} {\bibfnamefont {R.}~\bibnamefont {Yamamoto}}, \bibinfo {author} {\bibfnamefont {M.}~\bibnamefont {Yamashita}},\ and\ \bibinfo {author} {\bibfnamefont {Y.}~\bibnamefont {Takahashi}},\ }\bibfield  {title} {\bibinfo {title} {{Laser spectroscopic probing of coexisting superfluid and insulating states of an atomic {Bose-Hubbard} system}},\ }\href {https://doi.org/10.1038/ncomms11341} {\bibfield  {journal} {\bibinfo  {journal} {Nat. Commun.}\ }\textbf {\bibinfo {volume} {7}},\ \bibinfo {pages} {11341} (\bibinfo {year} {2016})}\BibitemShut {NoStop}%
  \bibitem [{\citenamefont {Taichenachev}\ \emph {et~al.}(2006)\citenamefont {Taichenachev}, \citenamefont {Yudin}, \citenamefont {Oates}, \citenamefont {Hoyt}, \citenamefont {Barber},\ and\ \citenamefont {Hollberg}}]{Taichenachev2006-ta}%
    \BibitemOpen
    \bibfield  {author} {\bibinfo {author} {\bibfnamefont {A.~V.}\ \bibnamefont {Taichenachev}}, \bibinfo {author} {\bibfnamefont {V.~I.}\ \bibnamefont {Yudin}}, \bibinfo {author} {\bibfnamefont {C.~W.}\ \bibnamefont {Oates}}, \bibinfo {author} {\bibfnamefont {C.~W.}\ \bibnamefont {Hoyt}}, \bibinfo {author} {\bibfnamefont {Z.~W.}\ \bibnamefont {Barber}},\ and\ \bibinfo {author} {\bibfnamefont {L.}~\bibnamefont {Hollberg}},\ }\bibfield  {title} {\bibinfo {title} {Magnetic field-induced spectroscopy of forbidden optical transitions with application to lattice-based optical atomic clocks},\ }\href {https://doi.org/10.1103/PhysRevLett.96.083001} {\bibfield  {journal} {\bibinfo  {journal} {Phys. Rev. Lett.}\ }\textbf {\bibinfo {volume} {96}},\ \bibinfo {pages} {083001} (\bibinfo {year} {2006})}\BibitemShut {NoStop}%
  \bibitem [{\citenamefont {Barber}\ \emph {et~al.}(2006)\citenamefont {Barber}, \citenamefont {Hoyt}, \citenamefont {Oates}, \citenamefont {Hollberg}, \citenamefont {Taichenachev},\ and\ \citenamefont {Yudin}}]{Barber2006-tu}%
    \BibitemOpen
    \bibfield  {author} {\bibinfo {author} {\bibfnamefont {Z.~W.}\ \bibnamefont {Barber}}, \bibinfo {author} {\bibfnamefont {C.~W.}\ \bibnamefont {Hoyt}}, \bibinfo {author} {\bibfnamefont {C.~W.}\ \bibnamefont {Oates}}, \bibinfo {author} {\bibfnamefont {L.}~\bibnamefont {Hollberg}}, \bibinfo {author} {\bibfnamefont {A.~V.}\ \bibnamefont {Taichenachev}},\ and\ \bibinfo {author} {\bibfnamefont {V.~I.}\ \bibnamefont {Yudin}},\ }\bibfield  {title} {\bibinfo {title} {Direct excitation of the forbidden clock transition in neutral ${}^{174}\mathrm{Yb}$ atoms confined to an optical lattice},\ }\href {https://doi.org/10.1103/PhysRevLett.96.083002} {\bibfield  {journal} {\bibinfo  {journal} {Phys. Rev. Lett.}\ }\textbf {\bibinfo {volume} {96}},\ \bibinfo {pages} {083002} (\bibinfo {year} {2006})}\BibitemShut {NoStop}%
  \bibitem [{\citenamefont {Fowler}\ \emph {et~al.}(2009)\citenamefont {Fowler}, \citenamefont {Stephens},\ and\ \citenamefont {Groszkowski}}]{Fowler2009-jf}%
    \BibitemOpen
    \bibfield  {author} {\bibinfo {author} {\bibfnamefont {A.~G.}\ \bibnamefont {Fowler}}, \bibinfo {author} {\bibfnamefont {A.~M.}\ \bibnamefont {Stephens}},\ and\ \bibinfo {author} {\bibfnamefont {P.}~\bibnamefont {Groszkowski}},\ }\bibfield  {title} {\bibinfo {title} {High-threshold universal quantum computation on the surface code},\ }\href {https://doi.org/10.1103/PhysRevA.80.052312} {\bibfield  {journal} {\bibinfo  {journal} {Phys. Rev. A}\ }\textbf {\bibinfo {volume} {80}},\ \bibinfo {pages} {052312} (\bibinfo {year} {2009})}\BibitemShut {NoStop}%
  \bibitem [{\citenamefont {Nikolov}\ \emph {et~al.}(2023)\citenamefont {Nikolov}, \citenamefont {Diamond-Hitchcock}, \citenamefont {Bass}, \citenamefont {Spong},\ and\ \citenamefont {Pritchard}}]{Nikolov2023-fv}%
    \BibitemOpen
    \bibfield  {author} {\bibinfo {author} {\bibfnamefont {B.}~\bibnamefont {Nikolov}}, \bibinfo {author} {\bibfnamefont {E.}~\bibnamefont {Diamond-Hitchcock}}, \bibinfo {author} {\bibfnamefont {J.}~\bibnamefont {Bass}}, \bibinfo {author} {\bibfnamefont {N.~L.~R.}\ \bibnamefont {Spong}},\ and\ \bibinfo {author} {\bibfnamefont {J.~D.}\ \bibnamefont {Pritchard}},\ }\bibfield  {title} {\bibinfo {title} {{Randomized Benchmarking Using Nondestructive Readout in a Two-Dimensional Atom Array}},\ }\href {https://doi.org/10.1103/PhysRevLett.131.030602} {\bibfield  {journal} {\bibinfo  {journal} {Phys. Rev. Lett.}\ }\textbf {\bibinfo {volume} {131}},\ \bibinfo {pages} {030602} (\bibinfo {year} {2023})}\BibitemShut {NoStop}%
  \bibitem [{\citenamefont {Holland}\ \emph {et~al.}(2023)\citenamefont {Holland}, \citenamefont {Lu},\ and\ \citenamefont {Cheuk}}]{Holland2023-uz}%
    \BibitemOpen
    \bibfield  {author} {\bibinfo {author} {\bibfnamefont {C.~M.}\ \bibnamefont {Holland}}, \bibinfo {author} {\bibfnamefont {Y.}~\bibnamefont {Lu}},\ and\ \bibinfo {author} {\bibfnamefont {L.~W.}\ \bibnamefont {Cheuk}},\ }\bibfield  {title} {\bibinfo {title} {On-demand entanglement of molecules in a reconfigurable optical tweezer array},\ }\href {https://doi.org/10.1126/science.adf4272} {\bibfield  {journal} {\bibinfo  {journal} {Science}\ }\textbf {\bibinfo {volume} {382}},\ \bibinfo {pages} {1143} (\bibinfo {year} {2023})}\BibitemShut {NoStop}%
  \bibitem [{\citenamefont {Ruttley}\ \emph {et~al.}(2024{\natexlab{a}})\citenamefont {Ruttley}, \citenamefont {Hepworth}, \citenamefont {Guttridge},\ and\ \citenamefont {Cornish}}]{Ruttley2024-vx}%
    \BibitemOpen
    \bibfield  {author} {\bibinfo {author} {\bibfnamefont {D.~K.}\ \bibnamefont {Ruttley}}, \bibinfo {author} {\bibfnamefont {T.~R.}\ \bibnamefont {Hepworth}}, \bibinfo {author} {\bibfnamefont {A.}~\bibnamefont {Guttridge}},\ and\ \bibinfo {author} {\bibfnamefont {S.~L.}\ \bibnamefont {Cornish}},\ }\bibfield  {title} {\bibinfo {title} {Long-lived entanglement of molecules in magic-wavelength optical tweezers},\ }\Eprint {https://arxiv.org/abs/2408.14904} {arXiv:2408.14904}  (\bibinfo {year} {2024}{\natexlab{a}})\BibitemShut {NoStop}%
  \bibitem [{\citenamefont {Ruttley}\ \emph {et~al.}(2024{\natexlab{b}})\citenamefont {Ruttley}, \citenamefont {Guttridge}, \citenamefont {Hepworth},\ and\ \citenamefont {Cornish}}]{Ruttley2024-iw}%
    \BibitemOpen
    \bibfield  {author} {\bibinfo {author} {\bibfnamefont {D.~K.}\ \bibnamefont {Ruttley}}, \bibinfo {author} {\bibfnamefont {A.}~\bibnamefont {Guttridge}}, \bibinfo {author} {\bibfnamefont {T.~R.}\ \bibnamefont {Hepworth}},\ and\ \bibinfo {author} {\bibfnamefont {S.~L.}\ \bibnamefont {Cornish}},\ }\bibfield  {title} {\bibinfo {title} {Enhanced quantum control of individual ultracold molecules using optical tweezer arrays},\ }\href {https://doi.org/10.1103/prxquantum.5.020333} {\bibfield  {journal} {\bibinfo  {journal} {PRX quantum}\ }\textbf {\bibinfo {volume} {5}},\ \bibinfo {pages} {020333} (\bibinfo {year} {2024}{\natexlab{b}})}\BibitemShut {NoStop}%
  \bibitem [{\citenamefont {Picard}\ \emph {et~al.}(2024)\citenamefont {Picard}, \citenamefont {Park}, \citenamefont {Patenotte}, \citenamefont {Gebretsadkan}, \citenamefont {Wellnitz}, \citenamefont {Rey},\ and\ \citenamefont {Ni}}]{Picard2024-kr}%
    \BibitemOpen
    \bibfield  {author} {\bibinfo {author} {\bibfnamefont {L.~R.~B.}\ \bibnamefont {Picard}}, \bibinfo {author} {\bibfnamefont {A.~J.}\ \bibnamefont {Park}}, \bibinfo {author} {\bibfnamefont {G.~E.}\ \bibnamefont {Patenotte}}, \bibinfo {author} {\bibfnamefont {S.}~\bibnamefont {Gebretsadkan}}, \bibinfo {author} {\bibfnamefont {D.}~\bibnamefont {Wellnitz}}, \bibinfo {author} {\bibfnamefont {A.~M.}\ \bibnamefont {Rey}},\ and\ \bibinfo {author} {\bibfnamefont {K.-K.}\ \bibnamefont {Ni}},\ }\bibfield  {title} {\bibinfo {title} {{Sub-millisecond Entanglement and {iSWAP} Gate between Molecular Qubits}},\ }\Eprint {https://arxiv.org/abs/2406.15345} {arXiv:2406.15345}  (\bibinfo {year} {2024})\BibitemShut {NoStop}%
  \bibitem [{\citenamefont {Holland}\ \emph {et~al.}(2024)\citenamefont {Holland}, \citenamefont {Lu}, \citenamefont {Li}, \citenamefont {Welsh},\ and\ \citenamefont {Cheuk}}]{Holland2024-wn}%
    \BibitemOpen
    \bibfield  {author} {\bibinfo {author} {\bibfnamefont {C.~M.}\ \bibnamefont {Holland}}, \bibinfo {author} {\bibfnamefont {Y.}~\bibnamefont {Lu}}, \bibinfo {author} {\bibfnamefont {S.~J.}\ \bibnamefont {Li}}, \bibinfo {author} {\bibfnamefont {C.~L.}\ \bibnamefont {Welsh}},\ and\ \bibinfo {author} {\bibfnamefont {L.~W.}\ \bibnamefont {Cheuk}},\ }\bibfield  {title} {\bibinfo {title} {{Demonstration of Erasure Conversion in a Molecular Tweezer Array}},\ }\Eprint {https://arxiv.org/abs/2406.02391} {arXiv:2406.02391}  (\bibinfo {year} {2024})\BibitemShut {NoStop}%
  \bibitem [{\citenamefont {Dzuba}\ \emph {et~al.}(2018)\citenamefont {Dzuba}, \citenamefont {Flambaum},\ and\ \citenamefont {Schiller}}]{Dzuba2018-tz}%
    \BibitemOpen
    \bibfield  {author} {\bibinfo {author} {\bibfnamefont {V.~A.}\ \bibnamefont {Dzuba}}, \bibinfo {author} {\bibfnamefont {V.~V.}\ \bibnamefont {Flambaum}},\ and\ \bibinfo {author} {\bibfnamefont {S.}~\bibnamefont {Schiller}},\ }\bibfield  {title} {\bibinfo {title} {Testing physics beyond the standard model through additional clock transitions in neutral ytterbium},\ }\href {https://doi.org/10.1103/PhysRevA.98.022501} {\bibfield  {journal} {\bibinfo  {journal} {Phys. Rev. A}\ }\textbf {\bibinfo {volume} {98}},\ \bibinfo {pages} {022501} (\bibinfo {year} {2018})}\BibitemShut {NoStop}%
  \bibitem [{\citenamefont {Cao}\ \emph {et~al.}(2024)\citenamefont {Cao}, \citenamefont {Eckner}, \citenamefont {Yelin}, \citenamefont {Young}, \citenamefont {Jandura}, \citenamefont {Yan}, \citenamefont {Kim}, \citenamefont {Pupillo}, \citenamefont {Ye}, \citenamefont {Oppong},\ and\ \citenamefont {Kaufman}}]{Cao2024-hp}%
    \BibitemOpen
    \bibfield  {author} {\bibinfo {author} {\bibfnamefont {A.}~\bibnamefont {Cao}}, \bibinfo {author} {\bibfnamefont {W.~J.}\ \bibnamefont {Eckner}}, \bibinfo {author} {\bibfnamefont {T.~L.}\ \bibnamefont {Yelin}}, \bibinfo {author} {\bibfnamefont {A.~W.}\ \bibnamefont {Young}}, \bibinfo {author} {\bibfnamefont {S.}~\bibnamefont {Jandura}}, \bibinfo {author} {\bibfnamefont {L.}~\bibnamefont {Yan}}, \bibinfo {author} {\bibfnamefont {K.}~\bibnamefont {Kim}}, \bibinfo {author} {\bibfnamefont {G.}~\bibnamefont {Pupillo}}, \bibinfo {author} {\bibfnamefont {J.}~\bibnamefont {Ye}}, \bibinfo {author} {\bibfnamefont {N.~D.}\ \bibnamefont {Oppong}},\ and\ \bibinfo {author} {\bibfnamefont {A.~M.}\ \bibnamefont {Kaufman}},\ }\bibfield  {title} {\bibinfo {title} {{Multi-qubit gates and 'Schr\"{o}dinger cat' states in an optical clock}},\ }\Eprint {https://arxiv.org/abs/2402.16289} {arXiv:2402.16289}  (\bibinfo {year} {2024})\BibitemShut {NoStop}%
  \bibitem [{\citenamefont {Unnikrishnan}\ \emph {et~al.}(2024)\citenamefont {Unnikrishnan}, \citenamefont {Ilzh\"ofer}, \citenamefont {Scholz}, \citenamefont {H\"olzl}, \citenamefont {G\"otzelmann}, \citenamefont {Gupta}, \citenamefont {Zhao}, \citenamefont {Krauter}, \citenamefont {Weber}, \citenamefont {Makki}, \citenamefont {B\"uchler}, \citenamefont {Pfau},\ and\ \citenamefont {Meinert}}]{Unnikrishnan2024-ec}%
    \BibitemOpen
    \bibfield  {author} {\bibinfo {author} {\bibfnamefont {G.}~\bibnamefont {Unnikrishnan}}, \bibinfo {author} {\bibfnamefont {P.}~\bibnamefont {Ilzh\"ofer}}, \bibinfo {author} {\bibfnamefont {A.}~\bibnamefont {Scholz}}, \bibinfo {author} {\bibfnamefont {C.}~\bibnamefont {H\"olzl}}, \bibinfo {author} {\bibfnamefont {A.}~\bibnamefont {G\"otzelmann}}, \bibinfo {author} {\bibfnamefont {R.~K.}\ \bibnamefont {Gupta}}, \bibinfo {author} {\bibfnamefont {J.}~\bibnamefont {Zhao}}, \bibinfo {author} {\bibfnamefont {J.}~\bibnamefont {Krauter}}, \bibinfo {author} {\bibfnamefont {S.}~\bibnamefont {Weber}}, \bibinfo {author} {\bibfnamefont {N.}~\bibnamefont {Makki}}, \bibinfo {author} {\bibfnamefont {H.~P.}\ \bibnamefont {B\"uchler}}, \bibinfo {author} {\bibfnamefont {T.}~\bibnamefont {Pfau}},\ and\ \bibinfo {author} {\bibfnamefont {F.}~\bibnamefont {Meinert}},\ }\bibfield  {title} {\bibinfo {title} {Coherent control of the fine-structure qubit in a single alkaline-earth atom},\ }\href
    {https://doi.org/10.1103/PhysRevLett.132.150606} {\bibfield  {journal} {\bibinfo  {journal} {Phys. Rev. Lett.}\ }\textbf {\bibinfo {volume} {132}},\ \bibinfo {pages} {150606} (\bibinfo {year} {2024})}\BibitemShut {NoStop}%
  \bibitem [{\citenamefont {Pucher}\ \emph {et~al.}(2024)\citenamefont {Pucher}, \citenamefont {Klüsener}, \citenamefont {Spriestersbach}, \citenamefont {Geiger}, \citenamefont {Schindewolf}, \citenamefont {Bloch},\ and\ \citenamefont {Blatt}}]{Pucher2024-pf}%
    \BibitemOpen
    \bibfield  {author} {\bibinfo {author} {\bibfnamefont {S.}~\bibnamefont {Pucher}}, \bibinfo {author} {\bibfnamefont {V.}~\bibnamefont {Klüsener}}, \bibinfo {author} {\bibfnamefont {F.}~\bibnamefont {Spriestersbach}}, \bibinfo {author} {\bibfnamefont {J.}~\bibnamefont {Geiger}}, \bibinfo {author} {\bibfnamefont {A.}~\bibnamefont {Schindewolf}}, \bibinfo {author} {\bibfnamefont {I.}~\bibnamefont {Bloch}},\ and\ \bibinfo {author} {\bibfnamefont {S.}~\bibnamefont {Blatt}},\ }\bibfield  {title} {\bibinfo {title} {Fine-structure qubit encoded in metastable strontium trapped in an optical lattice},\ }\href {https://doi.org/10.1103/PhysRevLett.132.150605} {\bibfield  {journal} {\bibinfo  {journal} {Phys. Rev. Lett.}\ }\textbf {\bibinfo {volume} {132}},\ \bibinfo {pages} {150605} (\bibinfo {year} {2024})}\BibitemShut {NoStop}%
  \bibitem [{\citenamefont {Jandura}\ and\ \citenamefont {Pupillo}(2022)}]{Jandura2022-ei}%
    \BibitemOpen
    \bibfield  {author} {\bibinfo {author} {\bibfnamefont {S.}~\bibnamefont {Jandura}}\ and\ \bibinfo {author} {\bibfnamefont {G.}~\bibnamefont {Pupillo}},\ }\bibfield  {title} {\bibinfo {title} {{Time-optimal two- and three-qubit gates for Rydberg atoms}},\ }\href {https://doi.org/10.22331/q-2022-05-13-712} {\bibfield  {journal} {\bibinfo  {journal} {Quantum}\ }\textbf {\bibinfo {volume} {6}},\ \bibinfo {pages} {712} (\bibinfo {year} {2022})}\BibitemShut {NoStop}%
  \bibitem [{\citenamefont {Ireland}\ \emph {et~al.}(2024)\citenamefont {Ireland}, \citenamefont {Walker},\ and\ \citenamefont {Pritchard}}]{Ireland2024-wa}%
    \BibitemOpen
    \bibfield  {author} {\bibinfo {author} {\bibfnamefont {P.~M.}\ \bibnamefont {Ireland}}, \bibinfo {author} {\bibfnamefont {D.~M.}\ \bibnamefont {Walker}},\ and\ \bibinfo {author} {\bibfnamefont {J.~D.}\ \bibnamefont {Pritchard}},\ }\bibfield  {title} {\bibinfo {title} {{Interspecies F\"{o}rster resonances for {Rb-Cs} Rydberg $d$-states for enhanced multi-qubit gate fidelities}},\ }\href {https://doi.org/10.1103/PhysRevResearch.6.013293} {\bibfield  {journal} {\bibinfo  {journal} {Phys. Rev. Res.}\ }\textbf {\bibinfo {volume} {6}},\ \bibinfo {pages} {013293} (\bibinfo {year} {2024})}\BibitemShut {NoStop}%
  \bibitem [{\citenamefont {Anand}\ \emph {et~al.}(2024)\citenamefont {Anand}, \citenamefont {Bradley}, \citenamefont {White}, \citenamefont {Ramesh}, \citenamefont {Singh},\ and\ \citenamefont {Bernien}}]{Anand2024-xl}%
    \BibitemOpen
    \bibfield  {author} {\bibinfo {author} {\bibfnamefont {S.}~\bibnamefont {Anand}}, \bibinfo {author} {\bibfnamefont {C.~E.}\ \bibnamefont {Bradley}}, \bibinfo {author} {\bibfnamefont {R.}~\bibnamefont {White}}, \bibinfo {author} {\bibfnamefont {V.}~\bibnamefont {Ramesh}}, \bibinfo {author} {\bibfnamefont {K.}~\bibnamefont {Singh}},\ and\ \bibinfo {author} {\bibfnamefont {H.}~\bibnamefont {Bernien}},\ }\bibfield  {title} {\bibinfo {title} {{A dual-species Rydberg array}},\ }\Eprint {https://arxiv.org/abs/2401.10325} {arXiv:2401.10325}  (\bibinfo {year} {2024})\BibitemShut {NoStop}%
  \bibitem [{\citenamefont {Nogrette}\ \emph {et~al.}(2014)\citenamefont {Nogrette}, \citenamefont {Labuhn}, \citenamefont {Ravets}, \citenamefont {Barredo}, \citenamefont {Béguin}, \citenamefont {Vernier}, \citenamefont {Lahaye},\ and\ \citenamefont {Browaeys}}]{Nogrette2014-zl}%
    \BibitemOpen
    \bibfield  {author} {\bibinfo {author} {\bibfnamefont {F.}~\bibnamefont {Nogrette}}, \bibinfo {author} {\bibfnamefont {H.}~\bibnamefont {Labuhn}}, \bibinfo {author} {\bibfnamefont {S.}~\bibnamefont {Ravets}}, \bibinfo {author} {\bibfnamefont {D.}~\bibnamefont {Barredo}}, \bibinfo {author} {\bibfnamefont {L.}~\bibnamefont {Béguin}}, \bibinfo {author} {\bibfnamefont {A.}~\bibnamefont {Vernier}}, \bibinfo {author} {\bibfnamefont {T.}~\bibnamefont {Lahaye}},\ and\ \bibinfo {author} {\bibfnamefont {A.}~\bibnamefont {Browaeys}},\ }\bibfield  {title} {\bibinfo {title} {Single-atom trapping in holographic {2D} arrays of microtraps with arbitrary geometries},\ }\href {https://doi.org/10.1103/physrevx.4.021034} {\bibfield  {journal} {\bibinfo  {journal} {Phys. Rev. X.}\ }\textbf {\bibinfo {volume} {4}},\ \bibinfo {pages} {021034} (\bibinfo {year} {2014})}\BibitemShut {NoStop}%
  \bibitem [{\citenamefont {Bergschneider}\ \emph {et~al.}(2018)\citenamefont {Bergschneider}, \citenamefont {Klinkhamer}, \citenamefont {Becher}, \citenamefont {Klemt}, \citenamefont {Zürn}, \citenamefont {Preiss},\ and\ \citenamefont {Jochim}}]{Bergschneider2018-ki}%
    \BibitemOpen
    \bibfield  {author} {\bibinfo {author} {\bibfnamefont {A.}~\bibnamefont {Bergschneider}}, \bibinfo {author} {\bibfnamefont {V.~M.}\ \bibnamefont {Klinkhamer}}, \bibinfo {author} {\bibfnamefont {J.~H.}\ \bibnamefont {Becher}}, \bibinfo {author} {\bibfnamefont {R.}~\bibnamefont {Klemt}}, \bibinfo {author} {\bibfnamefont {G.}~\bibnamefont {Zürn}}, \bibinfo {author} {\bibfnamefont {P.~M.}\ \bibnamefont {Preiss}},\ and\ \bibinfo {author} {\bibfnamefont {S.}~\bibnamefont {Jochim}},\ }\bibfield  {title} {\bibinfo {title} {Spin-resolved single-atom imaging of $^{6}${Li} in free space},\ }\href {https://doi.org/10.1103/PhysRevA.97.063613} {\bibfield  {journal} {\bibinfo  {journal} {Phys. Rev. A}\ }\textbf {\bibinfo {volume} {97}},\ \bibinfo {pages} {063613} (\bibinfo {year} {2018})}\BibitemShut {NoStop}%
  \bibitem [{\citenamefont {Jones}\ \emph {et~al.}(2006)\citenamefont {Jones}, \citenamefont {Tiesinga}, \citenamefont {Lett},\ and\ \citenamefont {Julienne}}]{Jones2006-oh}%
    \BibitemOpen
    \bibfield  {author} {\bibinfo {author} {\bibfnamefont {K.~M.}\ \bibnamefont {Jones}}, \bibinfo {author} {\bibfnamefont {E.}~\bibnamefont {Tiesinga}}, \bibinfo {author} {\bibfnamefont {P.~D.}\ \bibnamefont {Lett}},\ and\ \bibinfo {author} {\bibfnamefont {P.~S.}\ \bibnamefont {Julienne}},\ }\bibfield  {title} {\bibinfo {title} {{Ultracold photoassociation spectroscopy: Long-range molecules and atomic scattering}},\ }\href {https://doi.org/10.1103/RevModPhys.78.483} {\bibfield  {journal} {\bibinfo  {journal} {Rev. Mod. Phys.}\ }\textbf {\bibinfo {volume} {78}},\ \bibinfo {pages} {483} (\bibinfo {year} {2006})}\BibitemShut {NoStop}%
  \bibitem [{\citenamefont {Weiner}\ \emph {et~al.}(1999)\citenamefont {Weiner}, \citenamefont {Bagnato}, \citenamefont {Zilio},\ and\ \citenamefont {Julienne}}]{Weiner1999-tv}%
    \BibitemOpen
    \bibfield  {author} {\bibinfo {author} {\bibfnamefont {J.}~\bibnamefont {Weiner}}, \bibinfo {author} {\bibfnamefont {V.~S.}\ \bibnamefont {Bagnato}}, \bibinfo {author} {\bibfnamefont {S.}~\bibnamefont {Zilio}},\ and\ \bibinfo {author} {\bibfnamefont {P.~S.}\ \bibnamefont {Julienne}},\ }\bibfield  {title} {\bibinfo {title} {Experiments and theory in cold and ultracold collisions},\ }\href {https://doi.org/10.1103/RevModPhys.71.1} {\bibfield  {journal} {\bibinfo  {journal} {Rev. Mod. Phys.}\ }\textbf {\bibinfo {volume} {71}},\ \bibinfo {pages} {1} (\bibinfo {year} {1999})}\BibitemShut {NoStop}%
  \bibitem [{\citenamefont {Enomoto}\ \emph {et~al.}(2008)\citenamefont {Enomoto}, \citenamefont {Kitagawa}, \citenamefont {Tojo},\ and\ \citenamefont {Takahashi}}]{Enomoto2008-au}%
    \BibitemOpen
    \bibfield  {author} {\bibinfo {author} {\bibfnamefont {K.}~\bibnamefont {Enomoto}}, \bibinfo {author} {\bibfnamefont {M.}~\bibnamefont {Kitagawa}}, \bibinfo {author} {\bibfnamefont {S.}~\bibnamefont {Tojo}},\ and\ \bibinfo {author} {\bibfnamefont {Y.}~\bibnamefont {Takahashi}},\ }\bibfield  {title} {\bibinfo {title} {Hyperfine-structure-induced purely long-range molecules},\ }\href {https://doi.org/10.1103/PhysRevLett.100.123001} {\bibfield  {journal} {\bibinfo  {journal} {Phys. Rev. Lett.}\ }\textbf {\bibinfo {volume} {100}},\ \bibinfo {pages} {123001} (\bibinfo {year} {2008})}\BibitemShut {NoStop}%
  \bibitem [{\citenamefont {Blatt}(1967)}]{BLATT1967382}%
    \BibitemOpen
    \bibfield  {author} {\bibinfo {author} {\bibfnamefont {J.~M.}\ \bibnamefont {Blatt}},\ }\bibfield  {title} {\bibinfo {title} {{Practical points concerning the solution of the Schr\"{o}dinger equation}},\ }\href {https://doi.org/https://doi.org/10.1016/0021-9991(67)90046-0} {\bibfield  {journal} {\bibinfo  {journal} {Journal of Computational Physics}\ }\textbf {\bibinfo {volume} {1}},\ \bibinfo {pages} {382} (\bibinfo {year} {1967})}\BibitemShut {NoStop}%
  \bibitem [{\citenamefont {Borkowski}\ \emph {et~al.}(2011)\citenamefont {Borkowski}, \citenamefont {Ciury{\l}o}, \citenamefont {Julienne}, \citenamefont {Yamazaki}, \citenamefont {Hara}, \citenamefont {Enomoto}, \citenamefont {Taie}, \citenamefont {Sugawa}, \citenamefont {Takasu},\ and\ \citenamefont {Takahashi}}]{Borkowski2011-tq}%
    \BibitemOpen
    \bibfield  {author} {\bibinfo {author} {\bibfnamefont {M.}~\bibnamefont {Borkowski}}, \bibinfo {author} {\bibfnamefont {R.}~\bibnamefont {Ciury{\l}o}}, \bibinfo {author} {\bibfnamefont {P.~S.}\ \bibnamefont {Julienne}}, \bibinfo {author} {\bibfnamefont {R.}~\bibnamefont {Yamazaki}}, \bibinfo {author} {\bibfnamefont {H.}~\bibnamefont {Hara}}, \bibinfo {author} {\bibfnamefont {K.}~\bibnamefont {Enomoto}}, \bibinfo {author} {\bibfnamefont {S.}~\bibnamefont {Taie}}, \bibinfo {author} {\bibfnamefont {S.}~\bibnamefont {Sugawa}}, \bibinfo {author} {\bibfnamefont {Y.}~\bibnamefont {Takasu}},\ and\ \bibinfo {author} {\bibfnamefont {Y.}~\bibnamefont {Takahashi}},\ }\bibfield  {title} {\bibinfo {title} {Photoassociative production of ultracold heteronuclear ytterbium molecules},\ }\href {https://doi.org/10.1103/PhysRevA.84.030702} {\bibfield  {journal} {\bibinfo  {journal} {Phys. Rev. A}\ }\textbf {\bibinfo {volume} {84}},\ \bibinfo {pages} {030702(R)} (\bibinfo {year} {2011})}\BibitemShut {NoStop}%
  \bibitem [{\citenamefont {LeRoy}\ and\ \citenamefont {Bernstein}(1970)}]{LeRoy1970-lb}%
    \BibitemOpen
    \bibfield  {author} {\bibinfo {author} {\bibfnamefont {R.~J.}\ \bibnamefont {LeRoy}}\ and\ \bibinfo {author} {\bibfnamefont {R.~B.}\ \bibnamefont {Bernstein}},\ }\bibfield  {title} {\bibinfo {title} {Dissociation energy and long-range potential of diatomic molecules from vibrational spacings of higher levels},\ }\href {https://doi.org/10.1063/1.1673585} {\bibfield  {journal} {\bibinfo  {journal} {J. Chem. Phys.}\ }\textbf {\bibinfo {volume} {52}},\ \bibinfo {pages} {3869} (\bibinfo {year} {1970})}\BibitemShut {NoStop}%
  \bibitem [{\citenamefont {Brooks}\ \emph {et~al.}(2021)\citenamefont {Brooks}, \citenamefont {Spence}, \citenamefont {Guttridge}, \citenamefont {Alampounti}, \citenamefont {Rakonjac}, \citenamefont {McArd}, \citenamefont {Hutson},\ and\ \citenamefont {Cornish}}]{Brooks2021-ec}%
    \BibitemOpen
    \bibfield  {author} {\bibinfo {author} {\bibfnamefont {R.~V.}\ \bibnamefont {Brooks}}, \bibinfo {author} {\bibfnamefont {S.}~\bibnamefont {Spence}}, \bibinfo {author} {\bibfnamefont {A.}~\bibnamefont {Guttridge}}, \bibinfo {author} {\bibfnamefont {A.}~\bibnamefont {Alampounti}}, \bibinfo {author} {\bibfnamefont {A.}~\bibnamefont {Rakonjac}}, \bibinfo {author} {\bibfnamefont {L.~A.}\ \bibnamefont {McArd}}, \bibinfo {author} {\bibfnamefont {J.~M.}\ \bibnamefont {Hutson}},\ and\ \bibinfo {author} {\bibfnamefont {S.~L.}\ \bibnamefont {Cornish}},\ }\bibfield  {title} {\bibinfo {title} {Preparation of one {${}^{87}\mathrm{Rb}$} and one {${}^{133}\mathrm{Cs}$} atom in a single optical tweezer},\ }\href {https://doi.org/10.1088/1367-2630/ac0000} {\bibfield  {journal} {\bibinfo  {journal} {New J. Phys.}\ }\textbf {\bibinfo {volume} {23}},\ \bibinfo {pages} {065002} (\bibinfo {year} {2021})}\BibitemShut {NoStop}%
  \end{thebibliography}
\end{document}